\documentclass[review,numbers,sort&compress]{elsarticle}

\usepackage{lineno}
\usepackage[colorlinks,citecolor=blue,linktoc=all,linkcolor=cyan]{hyperref}
\usepackage{graphicx}

\usepackage[T1]{fontenc}
\usepackage{dsfont}               
\usepackage{mathrsfs}             
\usepackage{slashed}              
\usepackage{amsmath}
\usepackage{amssymb}
\usepackage{amsbsy}
\usepackage{amsfonts}
\usepackage{comment}
\usepackage{xspace}
\usepackage{orcidlink}
\usepackage{hyperref}
\usepackage{longtable}
\usepackage{float} 
\numberwithin{equation}{section}
\numberwithin{table}{section}
\numberwithin{figure}{section}
\journal{Progress in Particle and Nuclear Physics}

\usepackage{titlesec}
\usepackage{sectsty}
\titleformat{\section}{\normalfont\Large\bfseries}{\thesection}{1em}{}
\titleformat{\subsection}{\normalfont\large\bfseries}{\thesubsection}{1em}{}
\titleformat{\subsubsection}{\normalfont\normalsize\bfseries}{\thesubsubsection}{1em}{}

\begin{document}
%

\newcommand{\pp}           {pp\xspace}
\newcommand{\ppbar}        {\mbox{$\mathrm {p\overline{p}}$}\xspace}
\newcommand{\XeXe}         {\mbox{Xe--Xe}\xspace}
\newcommand{\PbPb}         {\mbox{Pb--Pb}\xspace}
\newcommand{\pA}           {\mbox{pA}\xspace}
\newcommand{\pPb}          {\mbox{p--Pb}\xspace}
\newcommand{\AuAu}         {\mbox{Au--Au}\xspace}
\newcommand{\dAu}          {\mbox{d--Au}\xspace}

\newcommand{\s}            {\ensuremath{\sqrt{s}}\xspace}
\newcommand{\snn}          {\ensuremath{\sqrt{s_{\mathrm{NN}}}}\xspace}
\newcommand{\pt}           {\ensuremath{p_{\rm T}}\xspace}
\newcommand{\meanpt}       {$\langle p_{\mathrm{T}}\rangle$\xspace}
\newcommand{\ycms}         {\ensuremath{y_{\rm CMS}}\xspace}
\newcommand{\ylab}         {\ensuremath{y_{\rm lab}}\xspace}
\newcommand{\etarange}[1]  {\mbox{$\left | \eta \right |~<~#1$}}
\newcommand{\yrange}[1]    {\mbox{$\left | y \right |~<~#1$}}
\newcommand{\dndy}         {\ensuremath{\mathrm{d}N_\mathrm{ch}/\mathrm{d}y}\xspace}
\newcommand{\dndeta}       {\ensuremath{\mathrm{d}N_\mathrm{ch}/\mathrm{d}\eta}\xspace}
\newcommand{\avdndeta}     {\ensuremath{\langle\dndeta\rangle}\xspace}
\newcommand{\avdndetam}    {\langle \mathrm{d}N_{\mathrm{ch}} / \mathrm{d}\eta\rangle_{|\eta| < 0.5}}
\newcommand{\dndetaqube}   {$\langle \mathrm{d}N_\mathrm{ch}/\mathrm{d}\eta \rangle^{1/3}_{|\eta|<0.5}$\xspace}
\newcommand{\dNdy}         {\ensuremath{\mathrm{d}N_\mathrm{ch}/\mathrm{d}y}\xspace}
\newcommand{\dNdpt}         {\ensuremath{\mathrm{d}N\mathrm{d}{p_{\rm T}}}\xspace}
\newcommand{\Npart}        {\ensuremath{N_\mathrm{part}}\xspace}
\newcommand{\Ncoll}        {\ensuremath{N_\mathrm{coll}}\xspace}
\newcommand{\dEdx}         {\ensuremath{\textrm{d}E/\textrm{d}x}\xspace}
\newcommand{\RpPb}         {\ensuremath{R_{\rm pPb}}\xspace}

\newcommand{\nineH}        {$\sqrt{s}~=~0.9$~Te\kern-.1emV\xspace}
\newcommand{\seven}        {$\sqrt{s}~=~7$~Te\kern-.1emV\xspace}
\newcommand{\twoH}         {$\sqrt{s}~=~0.2$~Te\kern-.1emV\xspace}
\newcommand{\twosevensix}  {$\sqrt{s}~=~2.76$~Te\kern-.1emV\xspace}
\newcommand{\five}         {$\sqrt{s}~=~5.02$~Te\kern-.1emV\xspace}
\newcommand{\twosevensixnn}{$\sqrt{s_{\mathrm{NN}}}~=~2.76$~Te\kern-.1emV\xspace}
\newcommand{\fivenn}       {$\sqrt{s_{\mathrm{NN}}}~=~5.02$~Te\kern-.1emV\xspace}
\newcommand{\LT}           {L{\'e}vy-Tsallis\xspace}
\newcommand{\GeVc}         {Ge\kern-.1emV/$c$\xspace}
\newcommand{\MeVc}         {Me\kern-.1emV/$c$\xspace}
\newcommand{\TeV}          {Te\kern-.1emV\xspace}
\newcommand{\GeV}          {Ge\kern-.1emV\xspace}
\newcommand{\MeV}          {Me\kern-.1emV\xspace}
\newcommand{\GeVmass}      {Ge\kern-.2emV/$c^2$\xspace}
\newcommand{\MeVmass}      {Me\kern-.2emV/$c^2$\xspace}
\newcommand{\lumi}         {\ensuremath{\mathcal{L}}\xspace}

\newcommand{\ITS}          {\rm{ITS}\xspace}
\newcommand{\TOF}          {\rm{TOF}\xspace}
\newcommand{\ZDC}          {\rm{ZDC}\xspace}
\newcommand{\ZDCs}         {\rm{ZDCs}\xspace}
\newcommand{\ZNA}          {\rm{ZNA}\xspace}
\newcommand{\ZNC}          {\rm{ZNC}\xspace}
\newcommand{\SPD}          {\rm{SPD}\xspace}
\newcommand{\SDD}          {\rm{SDD}\xspace}
\newcommand{\SSD}          {\rm{SSD}\xspace}
\newcommand{\TPC}          {\rm{TPC}\xspace}
\newcommand{\TRD}          {\rm{TRD}\xspace}
\newcommand{\VZERO}        {\rm{V0}\xspace}
\newcommand{\VZEROA}       {\rm{V0A}\xspace}
\newcommand{\VZEROC}       {\rm{V0C}\xspace}
\newcommand{\Vdecay} 	   {\ensuremath{V^{0}}\xspace}

\newcommand{\ee}           {\ensuremath{e^{+}e^{-}}} 
\newcommand{\pip}          {\ensuremath{\pi^{+}}\xspace}
\newcommand{\pim}          {\ensuremath{\pi^{-}}\xspace}
\newcommand{\pipm}          {\ensuremath{\pi^{\pm}}\xspace}
\newcommand{\kap}          {\ensuremath{\rm{K}^{+}}\xspace}
\newcommand{\kam}          {\ensuremath{\rm{K}^{-}}\xspace}
\newcommand{\pbar}         {\ensuremath{\rm\overline{p}}\xspace}
\newcommand{\kzero}        {\ensuremath{{\rm K}^{0}_{\rm{S}}}\xspace}
\newcommand{\lmb}          {\ensuremath{\Lambda}\xspace}
\newcommand{\almb}         {\ensuremath{\overline{\Lambda}}\xspace}
\newcommand{\Om}           {\ensuremath{\Omega^-}\xspace}
\newcommand{\Mo}           {\ensuremath{\overline{\Omega}^+}\xspace}
\newcommand{\X}            {\ensuremath{\Xi^-}\xspace}
\newcommand{\Ix}           {\ensuremath{\overline{\Xi}^+}\xspace}
\newcommand{\Xis}          {\ensuremath{\Xi^{\pm}}\xspace}
\newcommand{\Oms}          {\ensuremath{\Omega^{\pm}}\xspace}
\newcommand{\degree}       {\ensuremath{^{\rm o}}\xspace}

\newcommand{\CKS}          {K$^{*}$(892)$^{\pm}$}
\newcommand{\CKSshort}          {K$^{*\pm}$}
\newcommand{\NKSshort}          {K$^{*0}$}
\newcommand{\NKS}          {K$^{*}$(892)$^{0}$}
\newcommand{\KS}{K$\mathrm{^{0}_{S} }$}
\newcommand{\ENfive}        {$\sqrt{\it{s}_{\mathrm{NN}}} =$ 5.02 TeV\xspace}
\newcommand{\ENthirteen}    {$\sqrt{\it{s}} =$ 13 TeV\xspace}
\newcommand{\ENtwo}        {$\sqrt{\it{s}_{\mathrm{NN}}} =$ 2.76 TeV\xspace}
\newcommand{\CKSDC}     {K$^{*\pm} \rightarrow \mathrm{K^{0}_{S} \pi^{\pm}}$\xspace}
\newcommand{\CKSDClong}     {K$^{*}$(892)$^{\pm} \rightarrow \mathrm{K^{0}_{S} \pi^{\pm}}$\xspace}
\newcommand{\NKSDClong}     {K$^{*}$(892)$^{0} \rightarrow \mathrm{K^{\mp} \pi^{\pm}}$\xspace}
\newcommand{\CKSDCKL}     {K$^{*\pm} \rightarrow \mathrm{K^{0} \pi^{\pm}}$\xspace}
\newcommand{\RAA}{$R_{\mathrm{AA}}$\xspace}
\newcommand{\pion}          {$\mathrm{\pi}$}
\newcommand{\fone}          {f$_{1}$(1285)}
\newcommand{\foneshort}          {f$_{1}$}
\newcommand{\foneDC}     {f$_{1} (1285) \rightarrow \mathrm{K^{0}_{S} K^{\pm}\pi^{\mp}}$\xspace}
\newcommand{\Kstar}          {K$^{*}$$^{0}$\xspace}
\newcommand{\Ph}          {$\phi$\xspace}

\begin{frontmatter}

\title{Light-Flavour Resonance Production in High-Energy Heavy-Ion Collisions: An Experimental Review}

\author[prottayaddress]{Prottay Das}
\author[ajayaddress]{Ajay Kumar Dash}
\author[sandeepaddress]{Sandeep Dudi}
\author[prottayaddress]{Sourav Kundu}
\author[dmallikaddress]{Dukhishyam Mallick}
\author[bmohantyaddress]{Bedangadas Mohanty \corref{mycorrespondingauthor}}
\cortext[mycorrespondingauthor]{Corresponding author}
\ead{bedanga@niser.ac.in}
\author[nasimaddress]{Md Nasim}
\author[kishoraaddress]{Kishora Nayak}
\author[aswiniaddress]{Aswini Kumar Sahoo}
\author[aswiniaddress]{Subhash Singha}
\author[bmohantyaddress]{Ranbir Singh}


\address[prottayaddress]{CERN, Geneva, Switzerland}
\address[ajayaddress]{School of Earth $\&$ Planetary Sciences, National Institute of Science Education and Research, An OCC of Homi Bhabha National Institute, Jatni, Odisha, India}
\address[sandeepaddress]{Dipartimento di Fisica “E.R. Caianiello” dell’università degli Studi di Salerno and INFN of Napoli, gruppo collegato di Salerno, Italy}
\address[dmallikaddress]{INFN Sezione di Cagliari, Cagliari, Italy}
\address[bmohantyaddress]{School of Physical Sciences, National Institute of Science Education and Research Bhubaneswar, An OCC of Homi Bhabha National Institute, Jatni, Odisha, India }
\address[nasimaddress]{Department of Physical Sciences, Indian Institute of Science Education and Research, Berhampur, India}
\address[kishoraaddress]{Department of Physics, Panchayat College, Bargarh, India}
\address[aswiniaddress]{Institute of Modern Physics, Chinese Academy of Sciences, Huizhou, China}


\begin{abstract}
Resonances provide sensitivity to the late-stage dynamics of heavy-ion collisions, as their lifetimes are comparable to the duration of the hadronic phase. This review summarizes state-of-the-art measurements of light-flavour mesonic and baryonic resonances, including $\rho$(770), $K^{\star}$(892), $\phi$(1020), $\Delta$(1232), $\Lambda^{\star}$(1520), $\Sigma^{\star}$(1385) and $\Xi^\star$(1530), in pp, p–A and A–A collisions at SPS, RHIC and the LHC. Systematic trends in yields, mass and width modifications, transverse-momentum spectra, nuclear modification factors, and particle ratios reveal the interplay of re-scattering and regeneration, medium-induced suppression, and the development of collective dynamics with increasing system size and multiplicity.

Anisotropic-flow results confirm the coupling of resonances to the expanding medium, while recent vector-meson spin-alignment measurements offer fresh insights into hadronization mechanisms and local fields. Ultra-peripheral collisions provide vacuum-like baselines for isolating in-medium effects. Emerging opportunities for charm-resonance studies in upcoming high-luminosity experiments are also outlined. Together, these advances demonstrate the important role of resonance measurements in constraining the space–time evolution of strongly interacting matter.

\end{abstract}


\end{frontmatter}

\newpage

\thispagestyle{empty}
\tableofcontents


\section{Introduction}\label{intro}
The study of hadronic resonances in high-energy heavy-ion collisions has emerged as an important tool to probe the properties of strongly interacting matter under extreme conditions. Governed by the fundamental theory of Quantum Chromodynamics (QCD), the hot and dense medium formed in such collisions—commonly referred to as the quark-gluon plasma (QGP)—undergoes a complex evolution that spans partonic and hadronic stages. Hadronic resonances, due to their short lifetimes and multiple decay channels, are abundantly produced throughout this evolution and provide essential insight into the space-time dynamics of the system~\cite{NA49:2000jee,NA49:2011bfu,NA49:2008goy,DELPHI:1999res,Gavillet:2001ym,ALICE:2014jbq,ALICE:2017ban,STAR:2004bgh,Adams:2005dq}.

Unlike stable hadrons, resonances decay within the medium and their decay products are subject to subsequent interactions. This makes their measurable properties—such as mass, width, and yields—highly sensitive to the surrounding environment. Consequently, resonances act as dynamic probes of the hadronic phase, providing information on key phenomena such as re-scattering, regeneration, chemical and kinetic freeze-out, and collective behavior \cite{ALICE:2019xyr,ALICE:2023ifn,ALICE:2023edr,STAR:2003vqj,STAR:2022sir,Knospe:2015nva,Manninen:2008mg,Vovchenko:2019kes,Heinz:2007in}. Their comparative analysis with stable hadrons offers a means to constrain the duration and characteristics of the hadronic phase.

Among light-flavor resonances, particles such as $\rho$(770), $K^{*}$(892), $\Delta$(1232), $\Lambda^{*}$(1520) and $\phi$(1020) are particularly interesting. Each possesses distinct lifetimes, decay modes, and interaction cross sections, allowing physicists to map different stages of the system's evolution  \cite{STAR:2008bgi,STAR:2004yym,PHENIX:2014kia,PHENIX:2022hku,PHENIX:2022rvg,ALICE:2018qdv,Steinheimer:2017vju}. The $\phi$ meson, for instance, has a long lifetime and minimal hadronic interaction, making it a clean messenger of early-stage dynamics. In contrast, short-lived resonances such as \Kstar are more reflective of late-stage hadronic effects, offering a window into re-scattering and regeneration processes~\cite{ALICE:2021ptz,Markert:2005jv}.

Strange resonances play an additional role in the investigation of strangeness enhancement—one of the first proposed signatures of QGP formation \cite{STAR:2019bjj,STAR:2008inc,PHENIX:2004spo}. Resonances such as $K^{*}$, $\Lambda^{*}$, $\phi$, and $\Sigma$, which contain strange quarks, provide information on the production, transport, and equilibration of strangeness within the medium~\cite{STAR:2002npn,STAR:2010avo,NA49:2008goy,ALICE:2018ewo,ALICE:2022zuc}. Their modification patterns across system size and centrality serve as sensitive indicators of thermalization and hadronic dynamics in the strange sector.

Beyond hadronic interactions, ultra-peripheral collisions (UPCs) open a new frontier in resonance production. In these events, the electromagnetic fields of colliding ions induce photonuclear interactions, allowing for coherent photoproduction of vector mesons such as $K^{*}$, $\phi$ and $\rho^{0}$ ~\cite{H1:2009bga,STAR:2007elq,ALICE:2020ugp,CMS:2025lsm,CMS:2019awk,LHCb:2025fzk,H1:2009cml,THMF}. The clean environment of UPCs—largely free from strong final-state interactions—enables precision studies of resonance properties and probes the gluon structure of nuclei at low Bjorken-x~\cite{H1:2020lzc,ZEUS:1997rof}. Recent results from the ALICE experiment demonstrate the potential of this complementary approach in advancing our understanding of resonance dynamics in a different QCD regime ~\cite{ALICE:2018qdv,ALICE:2024upcKpm}.

Over the past decade, experiments at SPS ~\cite{NA49:2000jee,NA49:2011bfu,NA49:2008goy}, RHIC~\cite{PHENIX:2010bqp,STAR:2003pjh,STAR:2003wqp,STAR:2002svs,STAR:2004bgh}, and LHC  ~\cite{ALICE:2022wpn,ALICE:2022uac,ALICE:2016fzo,ALICE:2018pal} have built a rich dataset spanning a variety of collision energies and system sizes. These results reveal systematic trends in resonance yields, mass shifts, and width modifications, as well as variations in resonance-to-stable particle ratios. They provide indispensable benchmarks for theoretical models, including thermal fits ~\cite{Teaney:2002aj,Motornenko:2019jha,Schnedermann:1993ws,Stachel:2013zma} and transport simulations~\cite{Oliinychenko:2021enj}, that aim to describe the freeze-out conditions and hadronic interactions.

In addition to yields and spectra, resonances also contribute to the study of collective flow, a hallmark of QGP formation. The anisotropic flow of resonances—especially the $\phi$ meson—has been used to extract information on partonic collectivity, initial-state fluctuations, and the medium’s viscosity~\cite{Teaney:2000cw,Snellings:2011sz}. Observables such as directed ($v_{1}$), elliptic ($v_{2}$), and triangular ($v_{3}$) flow serve to map the early-time geometry and the dynamical response of the system.

A particularly novel aspect of resonance studies is spin alignment ~\cite{ALICE:2019aid,STAR:2022fan,Jiang:2016woz,Yang:2017sdk,Kumar:2023ghs}. Vector mesons such as $K^{*0} ~\text{and}~\phi$ offer an opportunity to probe the spin degrees of freedom in a rotating QCD medium. Spin alignment measurements are sensitive to the vorticity, magnetic fields, and polarization mechanisms present in non-central collisions and provide constraints on hadronization models incorporating spin-orbit couplings \cite{Becattini:2016gvu,Zhang:2023ppo,local1,local2,Muller:2021hpe,Lv:2024uev}.

This review aims to present a detailed and cohesive overview of the experimental landscape of light-flavour resonance production in high-energy heavy-ion collisions. We explore reconstruction techniques, analyze key observables across systems and energies, and discuss interpretations within statistical and dynamical models. Special attention is given to recent developments in flow, spin alignment, and UPC measurements, which together deepen our understanding of hadronic evolution and the broader QCD phase structure.

\section{Resonance Reconstruction in Experiments}\label{second}

\subsection{Resonance Introduction}
Resonances are short-lived hadronic states with lifetimes on the order of $10^{-23}$ seconds, typically comparable to the hadronization timescale in high-energy collisions. Due to their ephemeral nature, resonances decay promptly after their formation and cannot be detected directly. Instead, their presence is inferred by reconstructing their decay products using invariant mass techniques.

Modern high-energy experiments—such as those conducted at RHIC and the LHC—are equipped to identify stable or long-lived particles, including pions, kaons, protons, electrons, and muons. Resonances are reconstructed indirectly through their characteristic decay topologies and kinematic properties. Table~\ref{tab_resonance} summarizes the key resonances studied in such experiments, along with their dominant decay channels, branching ratios, and lifetimes. Table~\ref{tab_resonance1} summarizes these resonance measurements in various experiments. 
\begin{table}[!hbt]
\begin{center}
  \begin{tabular}{ |c|c|c|c|c| }
  \hline
  \textbf{Resonances}  &  \textbf{Decay channel}  &  \textbf{Branching ratio}  &  \textbf{Lifetime ($\tau$) [fm/c]}  \\
  \hline
   $\rho^{0}$ (770) & $\pi^{+} \pi^{-}$  &  1 & 1.1 \\
   \hline
      $\Delta$ (1232) & $p\pi$  & 1 & 1.6 \\
      \hline
      $f^{0}$ (980) & $\pi^{+} \pi^{-}$   & 2/3 & 2.6\\
      \hline
      $K^{*0}$ (892) & $K\pi$  & 2/3  & 4\\
      \hline
      $\Sigma^{*}$ (1385) & $\Lambda \pi$  & 0.88 & 5.5\\
      \hline
      $\Lambda^{*}$ (1520) & $pK$  & 0.45 & 12.6\\
      \hline
    $\Xi^{0}$ (1530) & ($\Lambda\pi^{+})\pi^{-}$  & 0.42 & 22 \\
         \hline
      $\phi$ (1020) & $K^{+}K^{-}$  & 0.49& 45\\
      \hline

  \hline
  \end{tabular}
\end{center}
  \caption{Properties of resonances studied in high-energy experiments: decay channels, branching ratios, and lifetimes~\cite{ParticleDataGroup:2024cfk}.}
  \label{tab_resonance}
  
\end{table}

\begin{table}[!hbt]
\begin{center}
\resizebox{\textwidth}{!}{%
  \begin{tabular}{|p{2.3cm}|p{3.7cm}|p{4.0cm}|p{4.3cm}|}
  \hline
  \textbf{Resonance}  &  \textbf{Experiment}  &  \textbf{System}  &  \textbf{Ref.}  \\
  \hline
   $\rho^{0}$ (770) & ALICE, STAR, DELPHI, HERA   & \pp, \pPb, \PbPb, \AuAu  & 
   ~\cite{DELPHI:1999hkl, H1:2009bga} \\
   \hline
   $\Delta$ (1232) & STAR & \pp, \dAu & 
   ~\cite{STAR:2008twt} \\
   \hline
   $f^{0}$ (980) & ALICE, DELPHI   & \pp & 
   ~\cite{ALICE:2022qnb, DELPHI:1999hkl} \\
   \hline
   $K^{*0}$ (896) & ALICE, STAR, NA49,  PHENIX, HERA  & \pp, \pPb, \PbPb, \AuAu  & 
   ~\cite{ALICE:2017ban, ALICE:2021ptz, ALICE:2019etb, ALICE:2016sak, STAR:2002npn, STAR:2004bgh, STAR:2010avo, PHENIX:2014kia, NA49:2011bfu, H1:2009bga} \\
   \hline
   $\Sigma^{*}$ (1385) & ALICE, STAR  & \pp, \pPb, \PbPb, \dAu & 
   ~\cite{ALICE:2023egx, ALICE:2017pgw, ALICE:2022zuc} \\
   \hline
   $\Lambda^{*}$ (1520) & ALICE, STAR  & \pp, \pPb, \PbPb, \dAu & 
   ~\cite{ALICE:2018ewo, ALICE:2020lam} \\
   \hline
   $\Xi^{0}$ (1530) &  ALICE  & \pp, \pPb & 
   ~\cite{ALICE:2023egx, ALICE:2017pgw} \\
   \hline
   $\phi$ (1020) & ALICE, STAR, NA49, PHENIX, HERA  & \pp, \pPb, \PbPb, \AuAu & 
   ~\cite{ALICE:2017ban,ALICE:2021ptz,ALICE:2019etb,ALICE:2016sak,STAR:2008inc,STAR:2008bgi,PHENIX:2022rvg, PHENIX:2022hku,PHENIX:2004spo,NA49:2008goy,NA49:2000jee,H1:2009bga} \\
   \hline
  \end{tabular}%
} 
\end{center}
  \caption{Summary of resonances studied across different experiments.}
  \label{tab_resonance1}
\end{table}

\subsection{Reconstruction Techniques}\label{chap2: sub:first} Resonances are reconstructed by calculating the invariant mass of their decay products, using the energy and momentum vectors of the daughter particles:
\begin{equation}
M_{\rm inv} = \sqrt{(E_1 + E_2)^2 - \left| \vec{p}_1 + \vec{p}_2 \right|^2}
\end{equation}
where $E_{1,2}$ and $\vec{p}_{1,2}$ denote the energies and momenta of the decay daughters.

\begin{figure}[!hbt]
	\centering
	\includegraphics[height=0.5\textwidth]{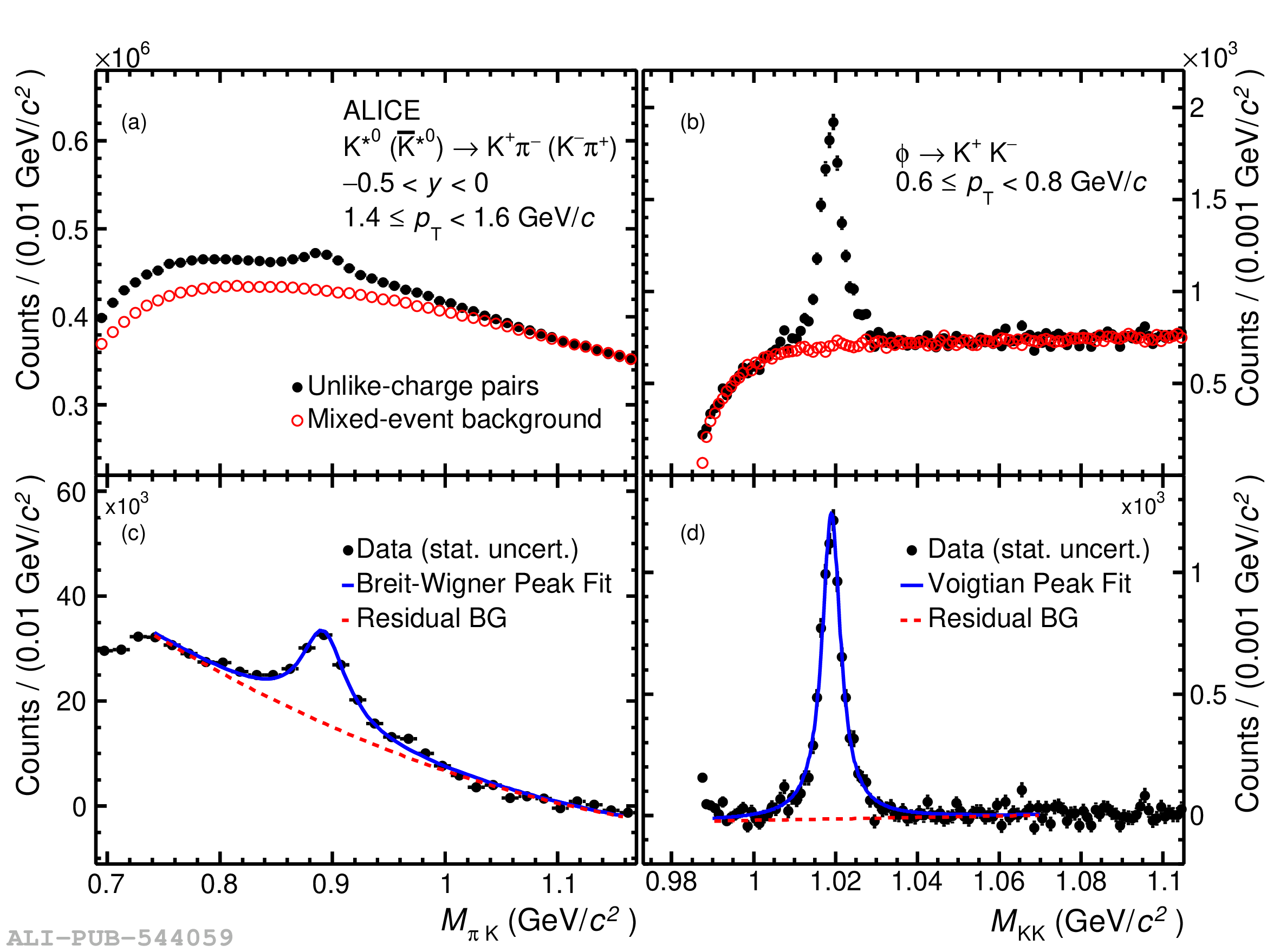}
	\caption{\label{fig:fig2_1} 
Invariant mass distributions for $K^{*0}$ and $\phi$ mesons in p--Pb collisions at $\sqrt{s_{\rm{NN}}} = 8.16$ TeV ~\cite{ALICE:2021rpa}. Panels (a, b): same-event distributions (black) overlaid with normalized mixed-event backgrounds (red). Panels (c, d): background-subtracted spectra fitted with a Breit--Wigner (for $K^{*0}$) and a Voigtian function (for $\phi$), respectively. The residual background is described by a second-order polynomial.
}
\end{figure}

For example, the hadronic decays of $K^{*0} \to K^{\pm} \pi^{\mp}$ and $\phi \to K^{+} K^{-}$ are reconstructed by forming unlike-sign pairs from the same event. Figure~\ref{fig:fig2_1} (upper panels) shows the resulting invariant mass distributions for these decay channels in \pPb collisions at $\sqrt{s_{\rm{NN}}} = 8.16$ TeV ~\cite{ALICE:2021rpa}. However, these distributions also contain a significant combinatorial background arising from uncorrelated particle pairs. Several standard techniques are employed to estimate and subtract this background:
\subsubsection{Mixed-event method:} Hadrons from different events—matched in multiplicity and vertex position—are paired to construct the combinatorial background. Typically, each event is mixed with 5–10 others to enhance statistics. The mixed-event distribution is normalized to the same-event distribution in invariant mass regions far from the expected resonance peak.
\subsubsection{Like-sign method:} In this approach, same-charge pairs ($K^{+}K^{+}$, $K^{-}K^{-}$ for $\phi$; $K^{+}\pi^{+}$, $K^{-}\pi^{-}$ for $K^{*0}$) are formed within the same event. The background distribution is normalized using the geometric mean: $2\sqrt{N^{++} \times N^{--}}$.
\subsubsection{Single track rotation method:} One of the decay daughters is rotated by an angle $\sim \pi$ in the transverse plane, effectively destroying the decay correlation while preserving the event's global kinematics. The resulting invariant mass distribution is built from multiple angular rotations to suppress statistical fluctuations.

Even after combinatorial background subtraction, residual correlated background may remain, arising from misidentified particles, decay chains of other resonances, or jet correlations. This is typically modeled using a low-order polynomial, informed by full-event Monte Carlo simulations.

\begin{figure}[!hbt]
	\centering
	\includegraphics[height=0.55\textwidth]{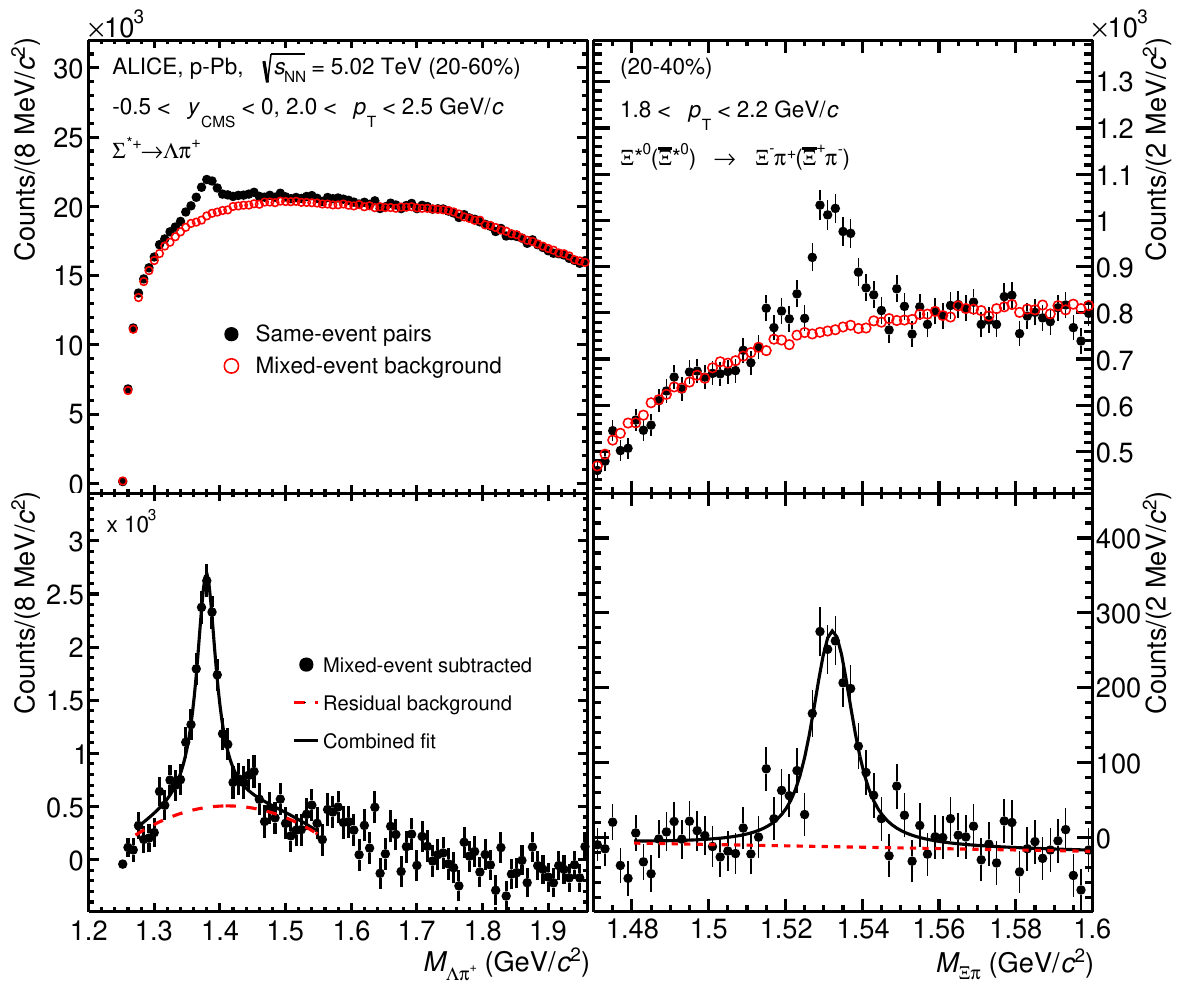}
	\caption{\label{fig:fig2_11} Invariant mass distributions for $\Lambda\pi^{+}$ and $\Xi^{\mp}\pi^{\pm}$ in p--Pb collisions at $\sqrt{s_{\rm{NN}}} = 5.02$ TeV ~\cite{ALICE:2017pgw}. Upper panels: same-event spectra with overlaid mixed-event backgrounds. Lower panels: background-subtracted signals fitted with Voigtian (for $\Sigma^{*+}$) and Breit--Wigner (for $\Xi^{*0}$) functions plus a second-order polynomial.
}
\end{figure}

Figure~\ref{fig:fig2_11} illustrates a similar procedure for $\Sigma^{*+}$ and $\Xi^{*0}$ resonances, showing the effectiveness of mixed-event subtraction and fitting using Breit--Wigner or Voigtian functions ~\cite{ALICE:2017pgw}.
\subsection{Hadronic vs. Leptonic Decay Channels}\label{chap2:subsub:first}
While resonances can decay via both hadronic and leptonic channels, most high-statistics measurements in STAR and ALICE utilize hadronic decays due to their higher branching ratios and detection efficiencies. Reconstruction of short-lived resonances such as the $\rho$ and $\phi$ mesons in leptonic decay channels
offers a unique advantage, since the electromagnetic decay products undergo negligible final-state hadronic interactions, allowing a direct probe of in-medium spectral functions and sensitivity to chiral symmetry restoration. Tche $\phi$ and $\omega$ mesons, which have been reconstructed in the $e^+e^-$ channel at midrapidity in Au+Au collisions at $\sqrt{s_{\rm{NN}}} = 200$ GeV (STAR)~\cite{STAR_PRC_RL}, and $\phi$ in the $\mu^+\mu^-$ channel at forward rapidity in pp, \pPb, and \PbPb collisions at the LHC (ALICE)~\cite{ALICE:2017_phif, ALICE:2021_phif}.

\subsection{Width and Mass of Resonances in pp, pA, and AA Collisions} \label{chap2: sub:second}
Systematic measurements of resonance mass and width in various collision systems provide insight into the hadronic phase and possible in-medium effects. Modifications to these parameters can reflect re-scattering, regeneration, and chiral symmetry restoration effects~\cite{GREEN198399} in the medium .
The invariant mass spectra—after background subtraction—are often fitted using a relativistic Breit--Wigner function, typically multiplied by a phase space factor, and supplemented with a polynomial to account for residual background. The resonance mass and width are extracted as free parameters in the fit.

\begin{figure}[!hbt]
	\centering
	\includegraphics[height=0.4\textwidth]{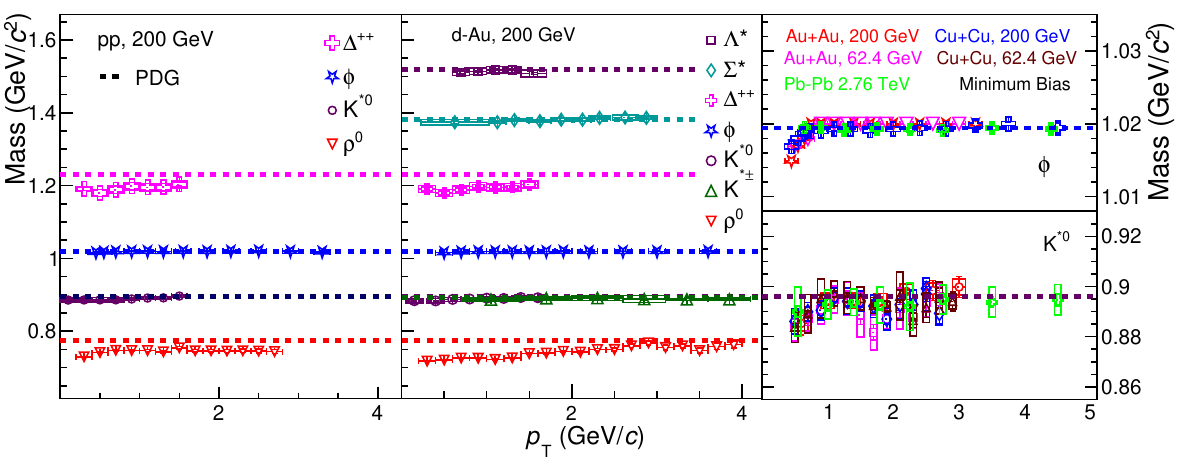}
	\caption{\label{fig:fig2_2} Mass as function $p_{\rm T}$, at midrapidity for various resonance particles across different collision systems and energies ~\cite{STAR:2004bgh, STAR:2008twt, ALICE:2014jbq, STAR:2010avo}.}
\end{figure}

\begin{figure}[!hbt]
	\centering
	\includegraphics[height=0.4\textwidth]{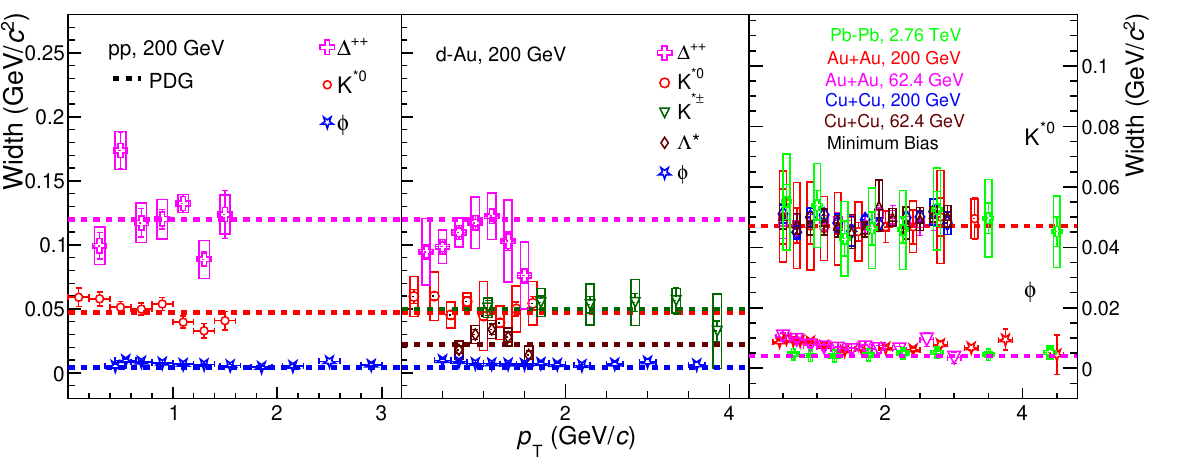}
	\caption{\label{fig:fig2_3} Transverse momentum dependence of resonance widths at midrapidity across different systems and energies ~\cite{STAR:2004bgh, STAR:2008twt, ALICE:2014jbq, STAR:2010avo}.}
\end{figure}

Figures~\ref{fig:fig2_2} and~\ref{fig:fig2_3} show the transverse momentum dependence of the measured mass and width of a few selected resonances in \pp, \pPb, and \PbPb collisions~\cite{STAR:2004bgh, STAR:2008twt, ALICE:2014jbq, STAR:2010avo}. The PDG~\cite{ParticleDataGroup:2022pth} vacuum values are shown as dashed lines.

Most observed mass shifts (especially for $\rho^0$ and $\Delta^{++}$) are dominated by phase-space distortions, interference, and reconstruction biases, not clean in-medium spectral modifications~\cite{STAR:2008twt}. Width measurements are largely resolution-limited, except in rare cases. Current hadronic decay measurements provide limited direct sensitivity to chiral restoration via mass/width modifications, motivating dilepton and UPC baselines.

In central A--A collisions, notable modifications in the mass of $K^{*0}$ are observed. At low $p_{\rm T}$ ($<1$ GeV/$c$), the mass of the $K^{*}$ meson exhibits a systematic downward shift, whereas its width remains largely unaltered. The $\phi$ meson mass is consistent with PDG at LHC energies but shows suppression at RHIC, particularly at low $p_{\rm T}$—likely dominated by detector resolution rather than medium effects. For $p_{\rm T} > 1$ GeV/$c$, both mass and width values for most resonances converge toward PDG values, indicating reduced interaction with the hadronic medium.
No strong dependence on beam energy or system size is evident in these observables within current experimental uncertainties.

\subsection{Yield of Resonances in pp, pA, AA Collisions}\label{chap2: sub:second}
The STAR (RHIC) and ALICE (LHC) collaborations have conducted systematic measurements of short-lived hadronic resonance yields in a broad range of collision systems— proton-proton (pp), proton--nucleus (p--A), and nucleus--nucleus (A--A). These measurements serve multiple purposes: they provide baselines from elementary collisions (pp), probe cold nuclear matter effects in p--A, and explore medium-induced modifications in the hot and dense environment created in A--A collisions. Resonance yields thus offer valuable insights into hadronization, re-scattering, and regeneration processes during the hadronic phase.

Resonances are reconstructed via their dominant hadronic decay channels, utilizing the momentum and particle identification information of the decay daughters. The invariant mass spectra—after subtracting the combinatorial background—are fitted with a resonance signal shape (typically a relativistic/non-relativistic Breit--Wigner or Voigtian function) superimposed on a residual background (modeled by a low-order polynomial). The signal yield is extracted from the integral of the fitted peak.

To obtain physical yields, the raw signal is corrected for:
\begin{itemize}
    \item Geometrical acceptance of the detector,
    \item Reconstruction and selection efficiencies (often evaluated via Monte Carlo simulations passed through GEANT),
    \item Branching ratios of the decay channel used.
\end{itemize}

The resulting yields are typically reported as transverse momentum ($p_{\rm T}$) differential spectra, $\mathrm{d}^{2}N/\mathrm{d}p_{\rm T}\mathrm{d}y$, for various event activity classes (centrality in A--A or multiplicity in pp/p--A), within a fixed rapidity interval (usually midrapidity).

To obtain the total midrapidity yield ($\mathrm{d}N/\mathrm{d}y$), the measured $p_{\rm T}$ spectra are integrated over the accessible range, and extrapolated to unmeasured low-$p_{\rm T}$ and high-$p_{\rm T}$ regions using fitting functions such as:
\begin{itemize}
    \item Boltzmann–Gibbs (thermal) distribution,
    \item Lévy–Tsallis function (suitable for non-thermal tails)~\cite{Tsallis:1987eu},
    \item Blast-Wave parameterization (for collective expansion scenarios)~\cite{Schnedermann:1993ws}.
\end{itemize}
The extrapolated fraction, particularly at low $p_{\rm T}$, can contribute significantly to the total yield—up to 30–40\% in central A--A collisions.

\begin{figure}[H]
	\centering
	\includegraphics[height=0.7\textwidth]{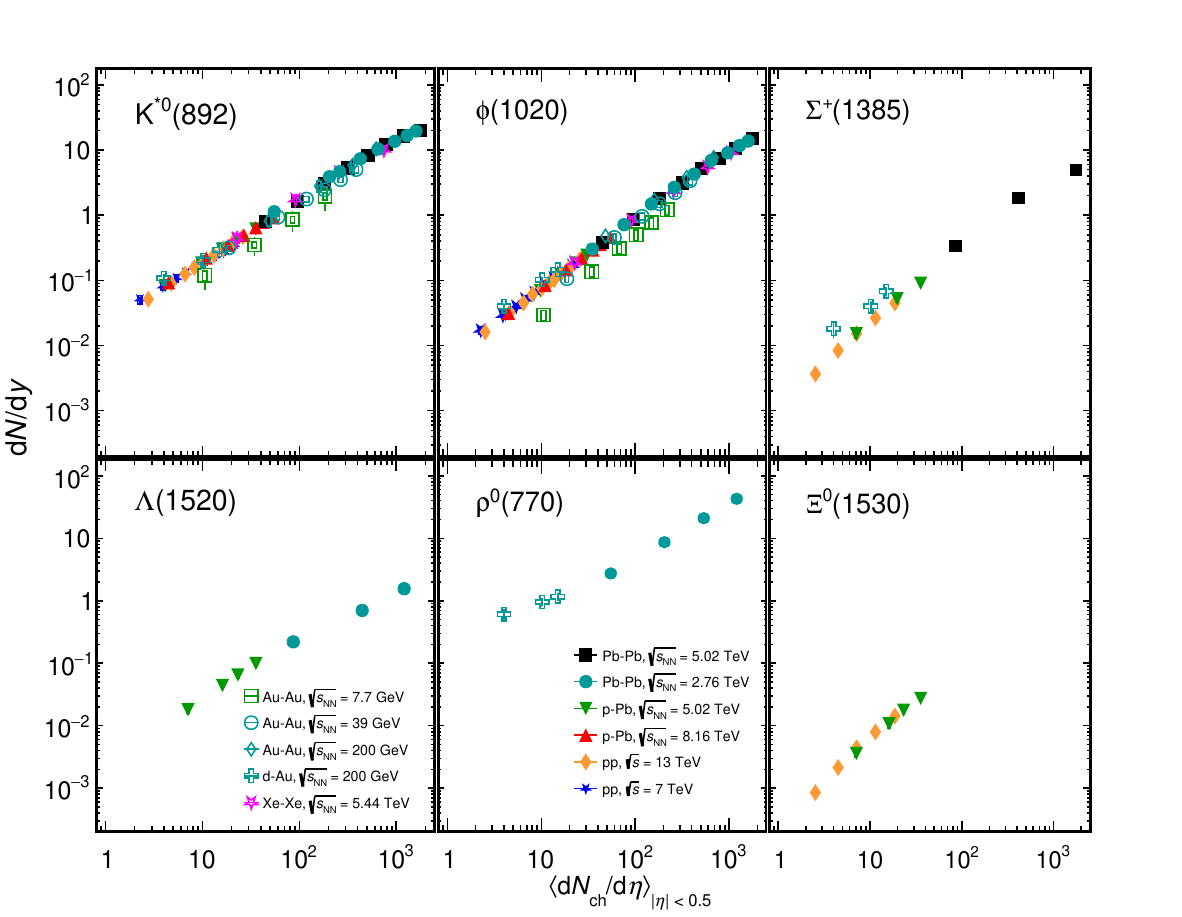}
	\caption{\label{fig:fig2_4} $p_{\rm T}$-integrated yield ($\mathrm{d}N/\mathrm{d}y$) as a function of the charged-particle multiplicity density at midrapidity for various resonance species in different collision systems and energies ~\cite{ALICE:2017ban, ALICE:2021ptz, ALICE:2019etb, ALICE:2016sak, ALICE:2021rpa, ALICE:2023egx, ALICE:2017pgw, ALICE:2022zuc, ALICE:2018ewo, ALICE:2020lam, ALICE:2023edr, ALICE:2021_xexe}.}
\end{figure}
Figure~\ref{fig:fig2_4} shows the $p_{\rm T}$-integrated midrapidity yields ($\mathrm{d}N/\mathrm{d}y$)~\cite{ALICE:2017ban, ALICE:2021ptz, ALICE:2019etb, ALICE:2016sak, ALICE:2021rpa, ALICE:2023egx, ALICE:2017pgw, ALICE:2022zuc, ALICE:2018ewo, ALICE:2020lam} of different resonances as a function of the average charged-particle multiplicity density ($\langle\mathrm{d}N_{\rm ch}/\mathrm{d}\eta\rangle$) at midrapidity, across a wide range of collision systems and beam energies.

A clear scaling behavior is observed: Resonance yields increase monotonically with event multiplicity. Remarkably, for a fixed value of $\langle\mathrm{d}N_{\rm ch}/\mathrm{d}\eta\rangle$, the yields of most resonances are consistent across different collision systems (pp, p--A, and A--A) and center-of-mass energies. This apparent universality suggests that the final-state multiplicity is a more fundamental variable governing resonance production than the initial-state system size or energy.

An exception is observed in the case of $\Sigma^{+}$ production in Au--Au collisions at $\sqrt{s_{\rm NN}} = 7.7$ GeV and in d--Au collisions at $\sqrt{s_{\rm NN}} = 200$ GeV, where the yields deviate from the scaling trend. These deviations may point to underlying nuclear effects, or non-trivial changes in hadronization dynamics at lower energies.

\subsection{Mean Transverse Momentum (\texorpdfstring{$\langle p_{\rm T} \rangle$}{<pT>}) of Resonances in pp, pA, and AA Collisions}\label{chap2: sub:second}
The mean transverse momentum (\meanpt) of resonances such as K$^{*0}$, $\phi$, $\Delta$, $\rho$, $\Lambda^{*}$, and $\Sigma^{*}$ has been systematically measured by the STAR (RHIC) and ALICE (LHC) collaborations in pp, p--A, and A--A collisions. The \meanpt is extracted from corrected transverse momentum spectra and provides an integrated measure of the spectral shape, offering insight into particle production dynamics, radial flow, and possible medium modifications.

For each centrality (in A--A) or multiplicity (in pp and p--A) class, \meanpt is computed by integrating the fully corrected $p_{\rm T}$ spectrum—combining the measured and extrapolated regions—weighted by $p_{\rm T}$. Higher \meanpt values typically correspond to harder spectra, which can originate from increased radial flow, parton energy loss, or changes in hadronization mechanisms.

\begin{figure}[!hbt]
	\centering
	\includegraphics[height=0.7\textwidth]{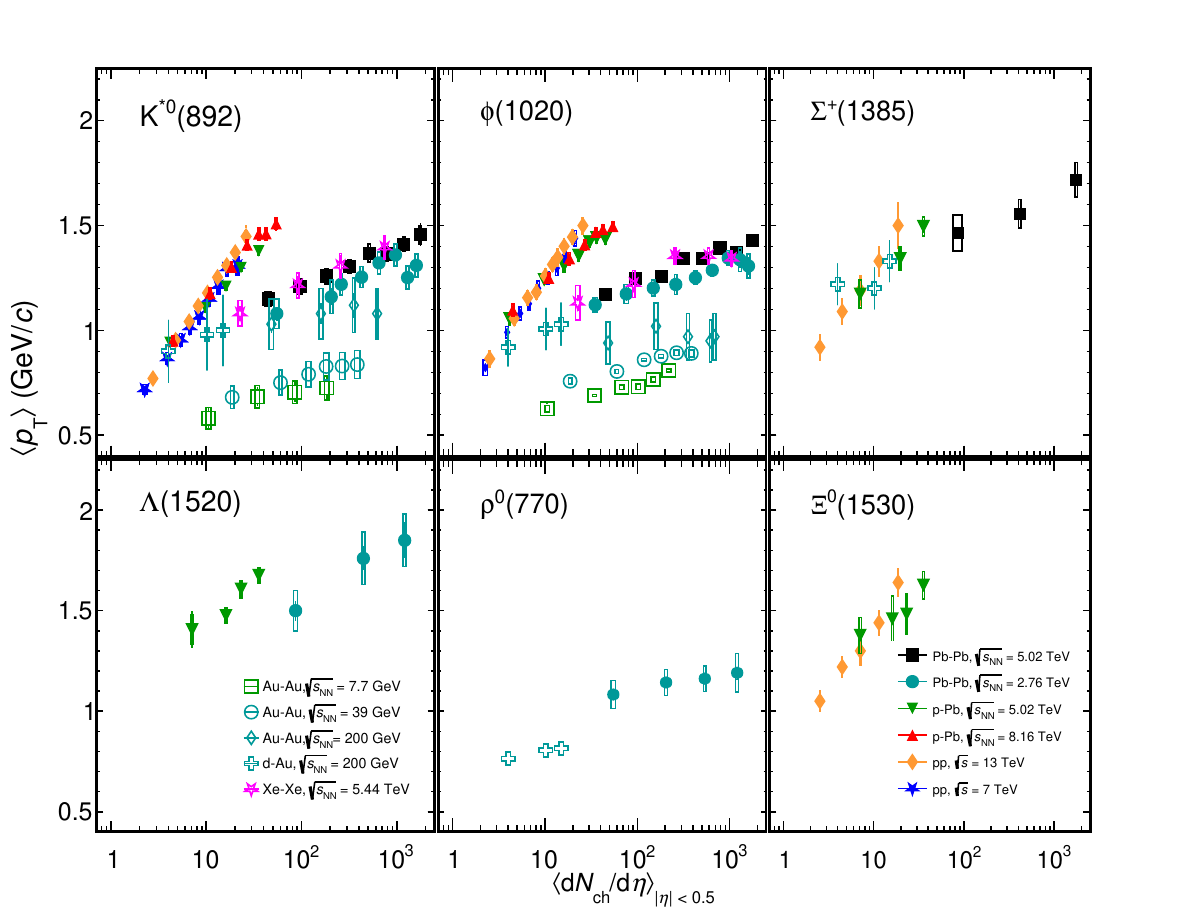}
	\caption{\label{fig:fig2_5} Mean transverse momentum ($\langle p_{\rm T} \rangle$) as a function of charged-particle multiplicity density at midrapidity for various resonances in different collision systems and energies ~\cite{ALICE:2017ban, ALICE:2021ptz, ALICE:2019etb, ALICE:2016sak, ALICE:2021rpa, ALICE:2023egx, ALICE:2017pgw, ALICE:2022zuc, ALICE:2018ewo, ALICE:2020lam, ALICE:2023edr, ALICE:2021_xexe}.}
\end{figure}
Figure~\ref{fig:fig2_5} shows the \meanpt of K$^{*0}$, $\phi$, $\Sigma^{+}$, $\Lambda^{*}$, $\rho$, and $\Xi^{*0}$ as a function of average midrapidity charged-particle multiplicity ($\langle \mathrm{d}N_{\rm ch}/\mathrm{d}\eta \rangle$) in various collision systems and energies ~\cite{ALICE:2017ban, ALICE:2021ptz, ALICE:2019etb, ALICE:2016sak, ALICE:2021rpa, ALICE:2023egx, ALICE:2017pgw, ALICE:2022zuc, ALICE:2018ewo, ALICE:2020lam}. A common trend is observed: \meanpt increases with increasing event multiplicity across all systems and resonance species.

Interestingly, the rate of increase is steeper in small systems such as \pp compared to p--A and A--A. At a fixed $\langle \mathrm{d}N_{\rm ch}/\mathrm{d}\eta \rangle$, \meanpt values in pp and p--A collisions are systematically higher than those in A--A collisions, particularly for $\langle \mathrm{d}N_{\rm ch}/\mathrm{d}\eta \rangle > 10$. For example, at $\langle \mathrm{d}N_{\rm ch}/\mathrm{d}\eta \rangle \approx 40$, the \meanpt of K$^{*0}$ and $\phi$ mesons is significantly higher in pp and p--Pb than in Pb--Pb collisions, despite comparable multiplicities. This indicates that event geometry and the underlying particle production mechanisms differ substantially between systems.

In contrast to A--A collisions, the approximate hydrodynamic mass ordering of \meanpt is not observed in high-multiplicity pp collisions, where mesons such as K$^{*0}$ and $\phi$ often have larger \meanpt than protons of similar mass. This behavior indicates that non-collective mechanisms, such as multiple parton interactions, minijets, and color reconnection, play an important role in shaping the spectral hardening in small systems~\cite{ALICE:2013wgn}. A study of charged hadrons in p--Pb collisions~\cite{ALICE:2013wgn} showed a similar multiplicity dependence of \meanpt, transitioning from a pp-like trend at low multiplicity to a saturation-like behavior at high multiplicity, reminiscent of Pb--Pb collisions.

In \pPb collisions at $\sqrt{s_{\mathrm{NN}}} = 5.02$ and $8.16$ TeV, the \meanpt of resonances at low multiplicity is comparable to that in pp for similar $\langle \mathrm{d}N_{\rm ch}/\mathrm{d}\eta \rangle$. However, at higher multiplicities, the \meanpt values in \pPb become systematically lower than in pp. This crossover again emphasizes the role of system geometry and initial-state configurations.

In \PbPb collisions, the increase of \meanpt with centrality (or multiplicity) is generally attributed to the development of strong radial flow in the thermalized medium. In central collisions, the rise of \meanpt tends to saturate or flatten, suggesting that the system has reached a collective expansion limit. The overall qualitative features of \meanpt evolution with $\langle \mathrm{d}N_{\rm ch}/\mathrm{d}\eta \rangle$ are similar at RHIC and LHC energies, despite differences in absolute values.

\begin{figure}[!hbt]
	\centering
	\includegraphics[height=0.5\textwidth]{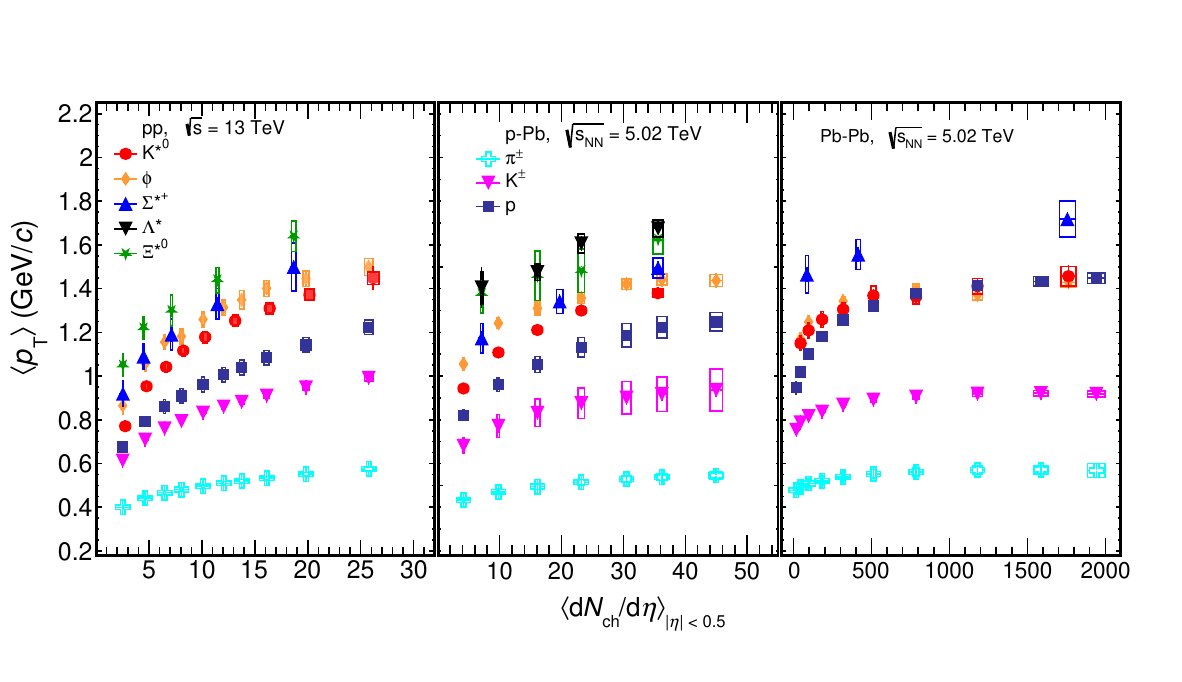}
	\caption{\label{fig:fig2_6} Comparison of mean transverse momentum for resonances and stable hadrons ($\pi^{\pm}$, K$^{\pm}$, p/$\bar{\mathrm{p}}$) as a function of charged-particle multiplicity density in pp (left), p--Pb (middle), and Pb--Pb (right) collisions~\cite{ALICE:2023egx, ALICE:2017pgw, ALICE:2019etb, ALICE:2016sak, ALICE:2021ptz, ALICE:2020lam, ALICE:2013wgn}}.
\end{figure}

Figure~\ref{fig:fig2_6} compares \meanpt of resonances with those of stable hadrons in \pp, \pPb and \PbPb collisions~\cite{ALICE:2023egx, ALICE:2017pgw, ALICE:2019etb, ALICE:2016sak, ALICE:2021ptz, ALICE:2020lam, ALICE:2013wgn}. For all particles, \meanpt increases with multiplicity, and the rise is steeper for more massive hadrons. In peripheral and intermediate-multiplicity events (pp and p--Pb), the $\phi$ and K$^{*0}$ mesons exhibit higher \meanpt than protons, despite similar masses. This mass ordering violation is not observed in central Pb--Pb collisions.

In central Pb--Pb collisions ($\langle \mathrm{d}N_{\rm ch}/\mathrm{d}\eta \rangle > 300$), the \meanpt values of K$^{*0}$, $\phi$, and protons become comparable within uncertainties, consistent with hydrodynamic predictions that \meanpt should scale primarily with hadron mass in a collective medium. In contrast, for $\langle \mathrm{d}N_{\rm ch}/\mathrm{d}\eta \rangle < 300$, the \meanpt of protons falls below that of the K$^{*0}$ and $\phi$, indicating a breakdown of mass ordering—possibly due to differing interactions in the hadronic phase or resonance regeneration effects. While collective radial expansion in large collision systems typically leads to a monotonic mass
ordering of \meanpt, this hierarchy can be violated when hadronization dynamics significantly influence particle production. In particular, recent theoretical studies within the EPOS framework have shown that the coexistence of a dense core region, hadronizing collectively, and a dilute corona component, dominated by string fragmentation, can modify the transverse-momentum distributions in a species-dependent manner. The relative contribution of core and corona particle
production depends on the local parton density and quark content, and can lead to deviations from the simple mass dependence expected from hydrodynamic flow alone. Such effects result in a breakdown of mass ordering in \meanpt
, highlighting the importance of hadronization mechanisms in shaping the observed momentum spectra ~\cite{epos4:model}.

\begin{figure}[!hbt]
	\centering
	\includegraphics[height=0.49\textwidth]{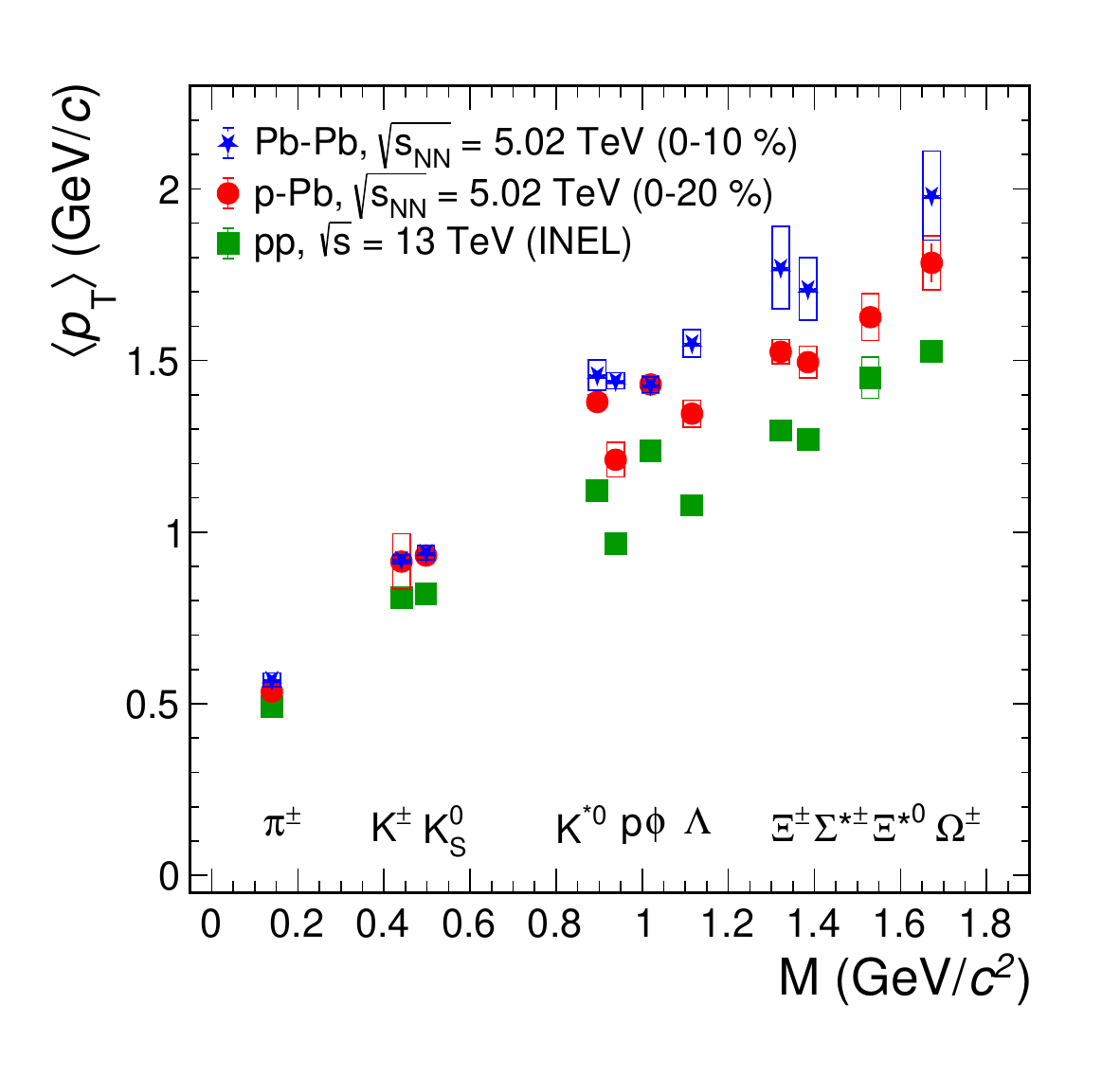}
	\caption{\label{fig:fig2_7} Mean transverse momentum as a function of particle mass in minimum-bias pp collisions at $\sqrt{s} = 13$ TeV, 0--20\% p--Pb collisions at $\sqrt{s_{\mathrm{NN}}} = 5.02$ TeV ~\cite{ALICE:2016sak, ALICE:2013wgn, ALICE:2017pgw, ALICE:2016mls}, and 0--10\% \PbPb collisions at $\sqrt{s_{\mathrm{NN}}} = 5.02$ TeV ~\cite{ALICE:2019hno, ALICE:2021ptz, ALICE:2022zuc, alice_PbPb2025:data}. }
\end{figure}
Figure~\ref{fig:fig2_7} shows \meanpt as a function of particle mass for three collision systems: minimum-bias pp at $\sqrt{s} = 13$ TeV, 0--20\% p--Pb at $\sqrt{s_{\mathrm{NN}}} = 5.02$ TeV, and 0--10\% Pb--Pb at the same energy. A general trend of increasing \meanpt with particle mass is observed in all systems. However, in pp and p--Pb collisions, the \meanpt of mesons such as K$^{*0}$ and $\phi$ is systematically higher than that of baryons like protons and $\Lambda$, despite comparable masses. This suggests a breaking of mass-scaling and points to different hadronization or final-state dynamics for mesons and baryons in small systems.

In central Pb--Pb collisions, the \meanpt values of all hadrons—including resonances and stable baryons—appear to follow a unified mass-dependent trend, consistent with radial flow effects in a collectively expanding medium. The \meanpt values in \pPb collisions generally lie between those of pp and Pb--Pb, indicating an intermediate level of collective behavior or spectral hardening.

\section{Physics from Resonance Production in Heavy-Ion Collisions}\label{third}

Hadronic resonances, owing to their relatively short lifetimes and sensitivity to the surrounding medium, serve as crucial probes of the space--time evolution of the matter created in relativistic heavy-ion collisions. Unlike stable hadrons, resonances decay inside the fireball and can suffer interactions in
the hadronic phase. Their yields, mass distributions, and flow patterns collectively encode rich information about freeze-out dynamics, the duration of
the hadronic phase, chiral symmetry restoration, and possible exotic QCD states. This section elaborates on the diverse physics that can be accessed
through systematic studies of resonance production.

\textbf{Hadronic Phase and Resonance Probes}

Hydrodynamic models have proven remarkably successful in describing the evolution of the quark–gluon plasma (QGP), under the assumptions of local thermal equilibrium and well-defined initial conditions~\cite{Blaizot:1987cc, Teaney:2009qa, Miller:2007ri, Schenke:2019pmk, Liu:2015nwa, Schenke:2012wb, Schenke:2010rr}. As the system expands and cools, it eventually reaches the hadronization temperature of approximately 155 MeV, where the deconfined partonic matter undergoes a transition into color-neutral hadrons—a process known as hadronization~\cite{Greco:2003xt, Greco:2003mm, Fries:2003vb, Fries:2003kq}. Following hadronization, the system enters the chemical freeze-out stage~\cite{Teaney:2002aj}, at which point inelastic collisions cease and the relative abundances of hadron species become fixed~\cite{Manninen:2008mg, Heinz:2007in}. However, elastic interactions among hadrons can still persist and influence the transverse momentum spectra until the system reaches kinetic freeze-out~\cite{Heinz:2007in}, characterized by the point at which
the mean free path of particles exceeds the system size. Beyond this stage, particles decouple and steam freely to the detectors.

The evolution from hadronization to kinetic freeze-out is commonly referred to as the hadronic phase, during which the system exists as a dense and
strongly interacting hadron gas~\cite{Steinheimer:2017vju}. Since the hadronization temperature is close to the chemical freeze-out temperature, the interval between chemical and kinetic freeze-out provides a useful experimental proxy for studying this phase. In this regime, the hadrons interact primarily through elastic and pseudo-elastic collisions, making it accessible via hadronic resonance probes.
Two competing mechanisms—rescattering and regeneration—govern the observable yields of resonances~\cite{Markert:2005jv,Singha:2015fia,Sahoo:2023dkv,Sahoo:2023rko}. In rescattering, the decay products of a resonance undergo elastic or pseudo-elastic interactions (via intermediate resonance states), scrambling the correlation information needed to reconstruct the parent resonance. For instance, a pion from a \NKSshort\xspace decay might scatter with another pion in the medium via a reaction like $\pi^-\pi^+ \rightarrow \rho^0 \rightarrow \pi^-\pi^+$, rendering the original \NKSshort\xspace undetectable. Conversely, regeneration can occur when two hadrons in the medium recombine to form a resonance, such as $\mathrm{K^{\mp}\pi^{\pm}} \rightarrow$ \NKSshort~$\rightarrow \mathrm{K^{\mp}\pi^{\pm}}$, effectively replenishing the resonance yield.

The relative importance of rescattering versus regeneration depends on several factors: the duration of the hadronic phase, the density and temperature of the medium, the hadronic cross sections of decay products, and the intrinsic lifetime of the resonance. Resonances with lifetimes comparable
to or shorter than the hadronic phase (e.g., \NKSshort, $\rho$) are particularly sensitive to these effects, while longer-lived resonances like the $\phi$ tend to escape unaffected, offering a chronometric handle on the hadronic lifetime. In addition to modifying yields, medium interactions can alter the mass
and width of resonances reconstructed via hadronic decays. These in-medium modifications lead to distorted line shapes in invariant mass spectra and
carry valuable information about the strength and duration of hadronic interactions. 

Experimentally, this physics is accessed by analyzing yield ratios of resonances to stable hadrons, such as \NKSshort/K or $\Lambda$(1520)/$\Lambda$ across different transverse momenta and collision centralities (or system sizes). Comparisons with thermal and transport model predictions, and with reference measurements from proton–proton collisions, where no extended hadronic phase is expected, provide insights into the properties and evolution of the hadronic
phase in heavy-ion collisions.

At higher transverse momentum, resonance production is influenced by partonic energy loss in the dense QGP medium~\cite{ALICE:2017ban, ALICE:2021ptz}. The degree of suppression depends on the initial energy density, the QGP lifetime, the parton path length through the medium, and parton flavor. This suppression
is quantified using the nuclear modification factor $R_{\mathrm{AA/AB}}$\cite{PHENIX:2001hpc}, defined as
\begin{equation}
     R_{AA/AB} =
\frac{d^2N_{AA/AB}/dp_T dy}
{\langle T_{AA/AB} \rangle \, d^2\sigma^{\mathrm{INEL}}_{pp}/dp_T dy}
\end{equation}

Here, $\frac{d^2N^{\text{AA/AB}}}{dp_T dy}$ is the differential yield of the particle species in nucleus–nucleus collisions, and  $\frac{d^2\sigma^{\text{pp}}_{\text{INEL}}}{dp_T dy}$ is the inelastic cross section in minimum bias proton-proton collisions. The factor $\langle T_{AA/AB} \rangle$ is the nuclear overlap function, proportional to the number of binary nucleon–nucleon collisions. Values of $R_{\rm AA} < 1$ at high $p_{\rm T}$ reflect suppression due to final-state effects like partonic energy loss~\cite{STAR:2017ieb}, nuclear shadowing~\cite{Armesto:2006ph}, or other mechanisms including Cronin enhancement or medium-induced broadening~\cite{ALICE:2016fzo}. Measuring $R_{\rm AA}$ for resonances thus provides complementary insights into both early-time partonic dynamics and late-time hadronic interactions.

Experimental results from STAR~\cite{STAR:2008bgi} and ALICE~\cite{ALICE:2019xyr} have established clear signatures of in-medium modifications and resonance suppression. These observations highlight the utility of short-lived resonances as chronometers of the hadronic phase and offer stringent benchmarks for hydrodynamic and transport model descriptions.

\subsection{Rescattering and Regeneration}\label{chap3:sub:first}

\begin{figure}[!hbt]
	\centering
	\includegraphics[height=0.5\textwidth]{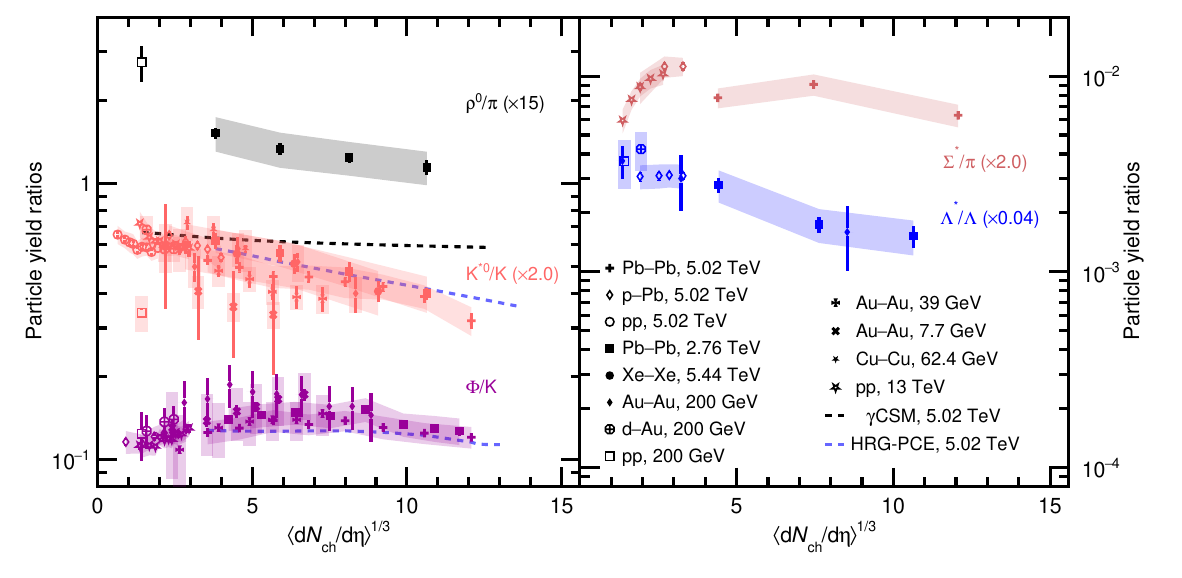}
	\caption{\label{fig:fig1p} $p_{\rm T}$-integrated yield ratios of mesonic (left) and baryonic (right) resonances to their corresponding stable hadrons as a function of charged-particle multiplicity density, expressed as \dndetaqube, at midrapidity. The data points are compiled from a wide range of collision systems and energies, as reported in Refs.~\cite{ALICE:2014jbq, ALICE:2023edr, ALICE:2017ban, ALICE:2018qdv, ALICE:2018ewo, ALICE:2021ptz, ALICE:2019xyr, ALICE:2022zuc, ALICE:2023ifn, ALICE:2023egx, ALICE:2020jsh, ALICE:2018pal, ALICE:2022uac, ALICE:2017pgw, ALICE:2016sak, Abelev:2013xaa, STAR:2003vqj, STAR:2004yym, STAR:2004bgh, STAR:2008twt, STAR:2008inc, STAR:2010avo, STAR:2022sir, STAR:2019bjj}. The measured \NKSshort/K ratio is compared with predictions from a thermal model ($\gamma_{\rm{S}}$CSM, black dotted line) and the HRG model with partial chemical equilibrium (HRG-PCE, blue dotted line).}
\end{figure}

Resonance production has been extensively investigated across a wide range of collision systems and energies, from the SPS to RHIC and the LHC. Figure~\ref{fig:fig1p} presents the $p_{\mathrm{T}}$-integrated yield ratios of several resonances to their corresponding long-lived hadrons as a function of \dndetaqube, which serves as a proxy for the system size. The choice of \dndetaqube is motivated by femtoscopic studies demonstrating a linear scaling of HBT radii with this quantity~\cite{ALICE:2011dyt}, underscoring its relevance for characterizing the spatial extent of the particle-emitting source. The left panel of Fig.~\ref{fig:fig1p} shows mesonic resonances, while the right panel focuses on baryonic species.

A pronounced, system-size--dependent suppression is observed for short-lived mesonic resonances such as $\rho^{0}/\pi$ and \NKSshort/K with increasing charged-particle multiplicity. This behavior is commonly attributed to hadronic rescattering occurring after chemical freeze-out. Since the lifetimes of these resonances are comparable to the duration of the hadronic phase, their decay daughters can undergo further interactions before kinetic freeze-out. Such interactions distort the decay kinematics, thereby reducing the probability of invariant-mass reconstruction and leading to a suppression of the measured resonance yields.
In contrast, the $\phi$ meson, with a lifetime of approximately 45~fm/$c$, decays predominantly outside the hadronic medium. Its decay products therefore experience minimal rescattering, and regeneration via kaon coalescence may even partially compensate for losses. As a result, the $\phi$/K yield ratio exhibits little to no dependence on charged-particle multiplicity, highlighting the relative insensitivity of the $\phi$ meson to late-stage hadronic interactions.

Beyond lifetime considerations, the observed suppression hierarchy among the
$\rho^{0}$, \NKSshort, and $\phi$ resonances can be understood in terms of the resonance-driven interaction strengths of their decay daughters in the hadronic phase. Microscopic transport and hadron resonance based models indicate that low-energy hadronic scattering proceeds predominantly through intermediate resonances, resulting in a hierarchy of elastic and pseudo-elastic reaction probabilities. $\rho^{0}$ decay daughters rescatter most efficiently via the broad $\rho(770)$ resonance in the $\pi\pi$ channel~\cite{Pelaez:2015qba}, while $\pi K$ scattering through the $K^{*}(892)$~\cite{Estabrooks:1977xe,Aston:1987ir}. In contrast, coupled-channel analyses show that $K\bar{K}$ interactions lack broad low-energy resonant enhancement outside the narrow $\phi(1020)$ pole, implying limited rescattering of $\phi$ decay daughters in the hadronic medium. In baryon-rich environments, meson--baryon channels such as $\pi N \!\leftrightarrow\! \Delta(1232)$ and $\pi\Lambda \!\leftrightarrow\! \Sigma^{*}$ can further deplete reconstructible baryonic resonances. Taken together, this hierarchy of resonance-mediated interaction strengths—strongest for $\pi\pi$, intermediate for $\pi K$, and weakest for $K\bar{K}$—provides the microscopic basis for the ordered suppression pattern observed across collision systems and multiplicities.

When combined with the markedly different lifetimes of these resonances—$\tau \sim 1$~fm/$c$ for the $\rho^{0}$, $\tau \sim 4$~fm/$c$ for the \NKSshort, and $\tau \sim 45$~fm/$c$ for the $\phi$—this hierarchy of interaction strengths naturally explains the strong suppression observed for $\rho^{0}/\pi$ and \NKSshort/K, in contrast to the near-constancy of the $\phi$/K ratio across collision systems and multiplicities. A similar suppression pattern is observed for baryonic resonances such as the $\Lambda(1520)/\Lambda$ ratio across collision systems and energies. The decay daughters of the $\Lambda(1520)$ interact via meson--baryon channels ($\pi N$, $K N$, $\pi \Lambda$) that are known to exhibit large cross sections, comparable to or exceeding those of the $\pi K$ system, making baryonic resonances particularly sensitive to hadronic rescattering. In contrast, the $\Sigma^{*}/\pi$ ratio does not exhibit a clear system-size dependence within current experimental uncertainties. Resonances with short lifetimes and strongly interacting decay products decay predominantly inside the hadronic medium and are therefore most susceptible to rescattering effects.

A smooth evolution of all resonance-to-stable particle ratios is observed across collision systems and energies, despite differences in initial conditions. This universality indicates that final-state hadronic interactions governed primarily by the particle density play a dominant role in shaping resonance yields. It is important to emphasize, however, that the observed suppression does not serve as a direct measurement of the hadronic phase lifetime. Instead, the reduced reconstructible yields reflect the effective hadronic interaction history experienced by the resonance decay daughters, shaped by decay timing, medium density, rescattering probabilities, and possible regeneration. One striking example is the $\Lambda(1520)$, which, despite its relatively long vacuum lifetime, exhibits strong suppression in central heavy-ion collisions. This highlights that resonance suppression is not simply determined by the lifetime but results from a complex balance of hadronic rescattering, regeneration, and feed-down contributions. Hybrid approaches combining hydrodynamics with a hadronic transport afterburner, as well as microscopic transport and Hadron Resonance Gas based frameworks, consistently indicate that the final observable resonance yields are governed not by the vacuum lifetime alone, but by the interaction strength of the decay products, the availability of regeneration channels, and the overall dynamical evolution of the hadronic medium. Consequently, while resonance suppression reflects the presence and extent of the hadronic phase, quantitative constraints on its duration require interpretation within such dynamical modeling frameworks.

To further quantify these effects, the \NKSshort/K ratio is compared with predictions from two theoretical approaches: the strangeness canonical statistical model with a suppression factor ($\gamma_{\mathrm{S}}$CSM)~\cite{Vovchenko:2019kes} and the Hadron Resonance Gas model with Partial Chemical Equilibrium (HRG-PCE)~\cite{Motornenko:2019jha}. Conventional statistical hadronization models assume thermal and chemical equilibrium at chemical freeze-out, with conserved baryon number, strangeness, and electric charge within a fixed correlation volume. While the $\gamma_{\mathrm{S}}$CSM successfully reproduces yields of most light-flavor hadrons, it significantly overpredicts the \NKSshort/K ratio in high-multiplicity events, where hadronic rescattering is expected to be strongest.

In contrast, the HRG-PCE model incorporates partial chemical equilibrium during the hadronic phase, in which the yields of short-lived resonances continue to evolve after chemical freeze-out. This framework provides a markedly improved description of the \NKSshort/K data and also reproduces the observed flat behavior of the $\phi$/K ratio.
At the microscopic level, an important refinement of this picture is that the same pseudo-elastic hadronic interactions which deplete reconstructable resonances can also regenerate them. Reactions such as $\pi\pi \leftrightarrow \rho$, $\pi K \leftrightarrow$ \NKSshort, and $\pi N \leftrightarrow \Delta$, as well as $\pi \Lambda \leftrightarrow \Sigma^{*}$, dominate the meson--meson and meson--baryon scattering rates in the hadronic phase and, to good approximation, obey the law of mass action. Within the HRG-PCE framework and in hadronic transport calculations, these channels dynamically maintain near kinetic equilibrium between short-lived resonances and their decay products down to temperatures close to kinetic freeze-out. A key feature of the HRG-PCE scenario is the emergence of species-dependent effective chemical potentials below chemical freeze-out, reflecting the conservation of particle yields in this stage rather than full chemical equilibrium. As the system cools, these chemical potentials grow and directly influence pseudo-elastic reaction rates, making regeneration strongly temperature- and density-dependent. Thus, regeneration rates evolve throughout the hadronic phase, making the suppression pattern sensitive not only to resonance lifetimes and hadronic cross sections, but also to the changing chemical environment, which is particularly important for low-energy RHIC and SPS conditions where the chemical potentials are large. The experimentally measured short-lived resonances to stable hadron ratios should therefore be interpreted as the net outcome of two competing effects: the loss of reconstructable resonances due to daughter rescattering and their continuous regeneration via pseudo-elastic hadronic interactions. Consequently, the observed suppression does not represent a simple survival probability of primordial resonances formed at chemical freeze-out, but rather encodes the dynamical balance between absorption and regeneration throughout the hadronic phase.

\begin{figure}[!hbt]
	\centering
	\includegraphics[height=0.5\textwidth]{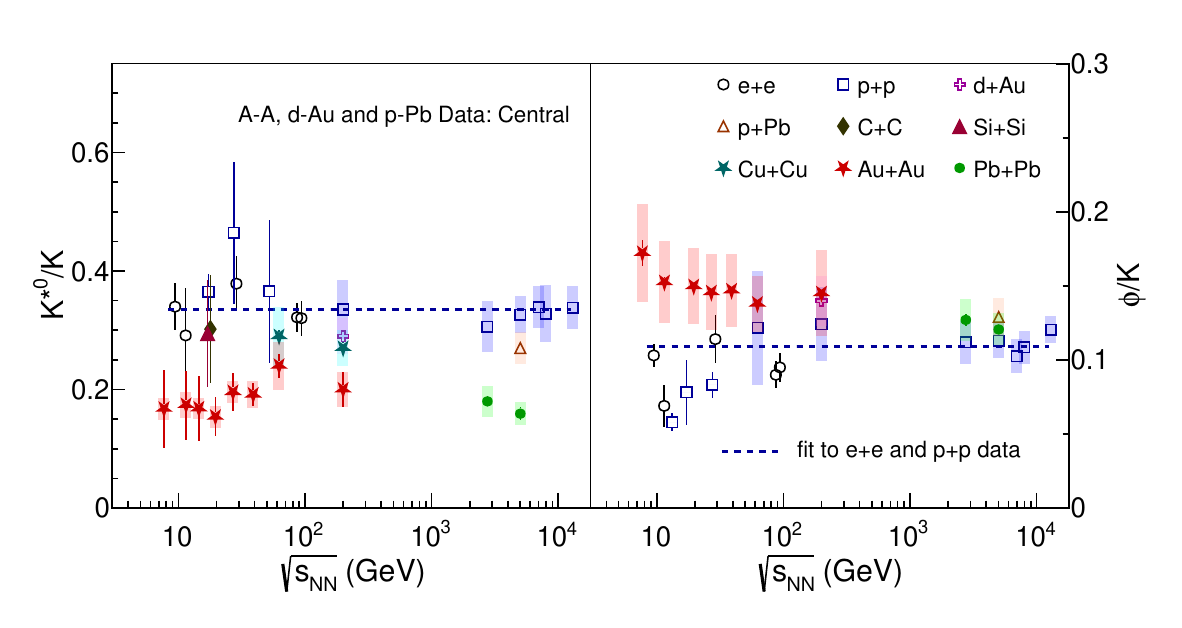}
	\caption{\label{fig:fig2p} $p_{\rm T}$-integrated yield ratios of \NKSshort/K (left) and $\phi$/K (right) at midrapidity as a function of collision energy. All data points are compiled from Refs.~\cite{ALICE:2014jbq, ALICE:2023edr, ALICE:2017ban, ALICE:2018qdv, ALICE:2018ewo, ALICE:2021ptz, ALICE:2019xyr, ALICE:2022zuc, ALICE:2023ifn, ALICE:2023egx, ALICE:2020jsh, ALICE:2018pal, ALICE:2022uac, ALICE:2017pgw, ALICE:2016sak, STAR:2003vqj, STAR:2004yym, STAR:2004bgh, STAR:2008twt, STAR:2008inc, STAR:2010avo, STAR:2022sir, STAR:2019bjj}.}
\end{figure}

To further elucidate system-size effects, Fig.~\ref{fig:fig2p} presents the \NKSshort/K and \( \phi/K \) yield ratios in central collisions as a function of collision energy. The left panel reveals that the \NKSshort/K ratio in heavy-ion collisions is consistently lower than in pp or e$^+$e$^-$ collisions, with p–Pb data falling in between—mirroring the expected hierarchy of system sizes. This pattern supports the interpretation that rescattering is more pronounced in larger systems. The higher ratio observed in Cu–Cu compared to Au–Au collisions aligns with the expectation from reduced hadronic phase density in smaller systems.

A zeroth-order polynomial fit to the pp and $\text{e}^{+}\text{e}^{-}$ data establishes a baseline, and the heavy-ion data point is found to lie approximately 3.5 standard deviations below this reference, quantifying the suppression and strengthening the argument for significant hadronic phase effects.

The right panel of Fig.~\ref{fig:fig2p} shows that the \( \phi/K \) ratio is largely independent of system size and energy, with a slight enhancement at lower energies in Au–Au collisions likely due to canonical suppression of kaon production rather than rescattering.

\begin{figure}[!hbt]
	\centering
	\includegraphics[height=0.35\textwidth]{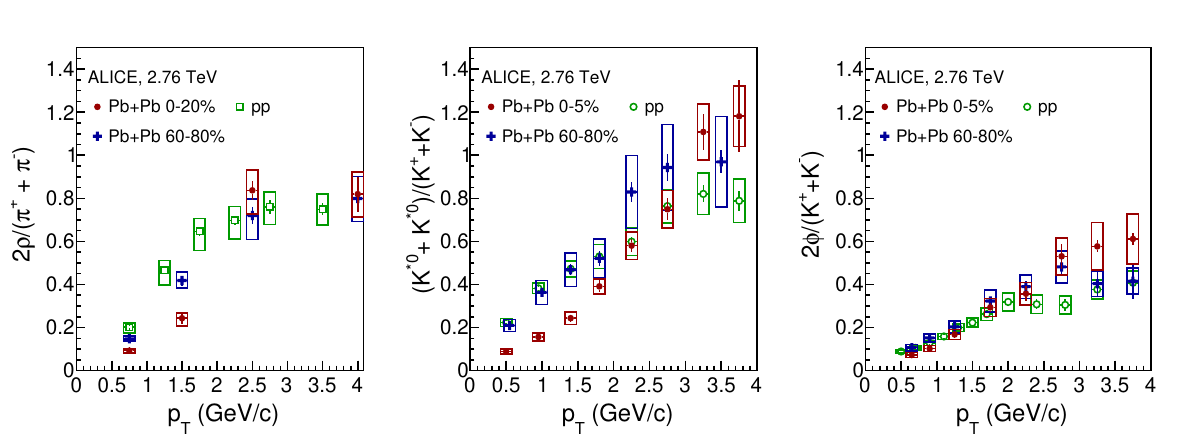}
	\caption{\label{fig:fig3p} $p_{\rm T}$-differential yield ratios of $\rho^{0}$/$\pi$ (left), \NKSshort/K (middle), and $\phi$/K (right) at midrapidity in Pb–Pb collisions at \snn = 2.76 TeV. Results are shown for central (red markers) and peripheral (blue markers) Pb--Pb collisions, along with minimum-bias pp collisions (green markers). Data are taken from Refs.~\cite{ALICE:2014jbq, ALICE:2018qdv}}
\end{figure}

To examine the transverse momentum dependence of rescattering, Fig.~\ref{fig:fig3p} displays the \( p_{\mathrm{T}} \)-differential ratios of \( \rho^{0}/\pi \), \NKSshort/K, and \( \phi/K \) in Pb–Pb collisions at \( \sqrt{s_{\mathrm{NN}}} = 2.76 \) TeV. Central events (in red) show a marked suppression of \( \rho^{0}/\pi \) and \NKSshort/K at low \( p_{\mathrm{T}} \) (\(< 2\) GeV/\(c\)) relative to peripheral and pp collisions, consistent with rescattering effects predominantly affecting soft particles. In contrast, the \( \phi/K \) ratio remains flat across all systems, further affirming its minimal sensitivity to hadronic reinteraction.

To quantitatively assess the observed suppression, a Boltzmann-Gibbs blast-wave model is employed to generate reference \( p_{\mathrm{T}} \) spectra for \NKSshort and \( \phi \) mesons at kinetic freeze-out, assuming no rescattering. Model parameters—kinetic freeze-out temperature \( T_{\mathrm{kin}} \), surface velocity \( \beta_{s} \), and velocity profile exponent \( n \)—are extracted from fits to the pion, kaon, and proton spectra~\cite{ALICE:2013mez}. The resonance yields are normalized to thermal model expectations at \( T_{\mathrm{ch}} = 155 \) MeV~\cite{Stachel:2013zma}, representing yields in the absence of post-hadronization effects.

\begin{figure}[!hbt]
	\centering
	\includegraphics[height=0.5\textwidth]{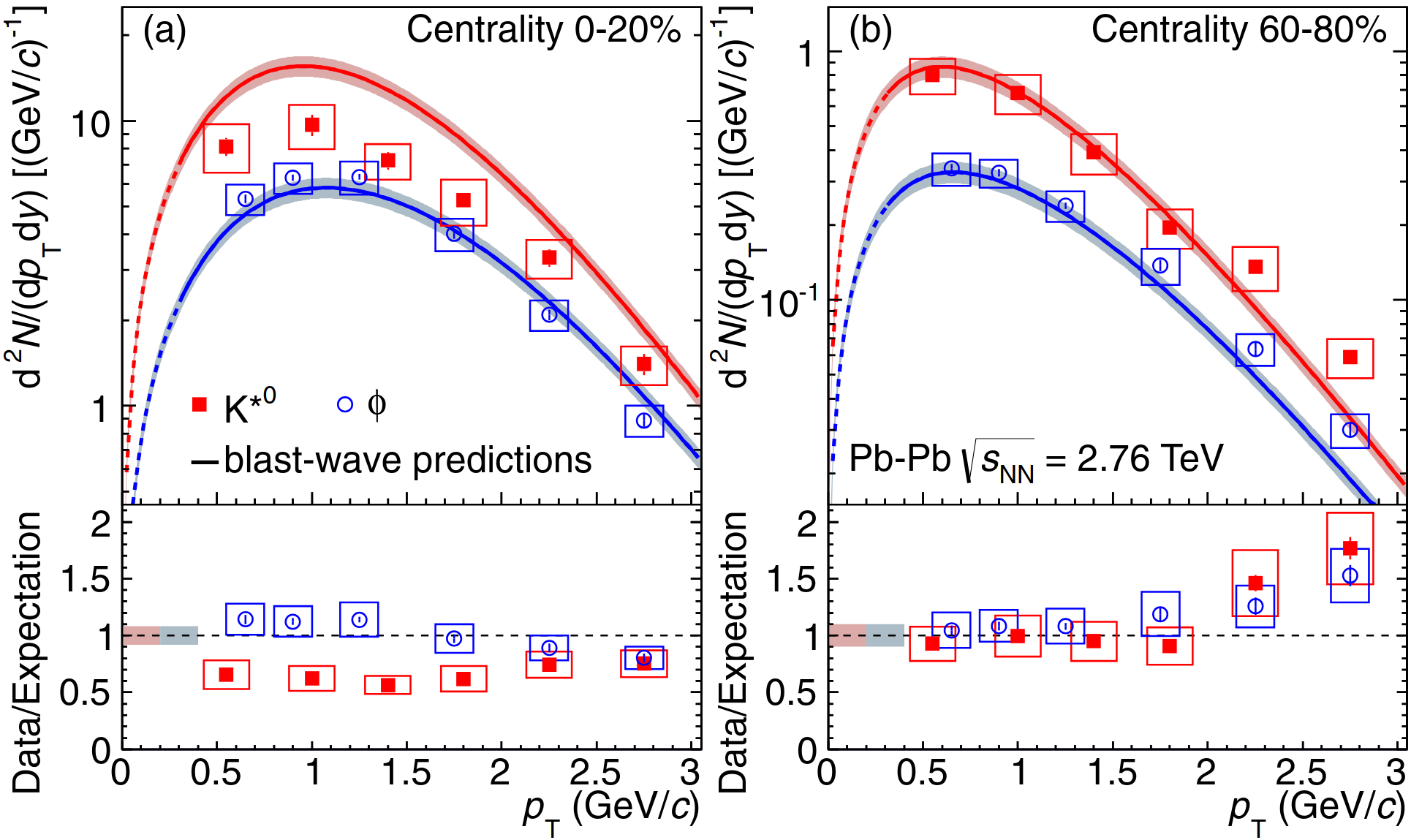}
	\caption{\label{fig:fig4p} \pt distributions of \NKSshort and $\phi$ mesons in Pb--Pb collisions at $\sqrt{s_{\mathrm{NN}}} = 2.76$ TeV for central (left) and peripheral (right) collisions~\cite{ALICE:2014jbq}. The expected distributions are calculated using Boltzmann-Gibbs blast-wave functions~\cite{Schnedermann:1993ws}, with parameters extracted from simultaneous fits to the $p_{\rm T}$ spectra of pions, kaons, and (anti)protons~\cite{ALICE:2013mez} (see text for details). Shaded bands represent the uncertainties associated with the blast-wave fit parameters. The lower panels display the ratios of the measured resonance yields to the model expectations..}
\end{figure}

Figure~\ref{fig:fig4p} compares the measured and model spectra. For the \( \phi \) meson, the measured-to-expected ratio remains near unity across \( p_{\mathrm{T}} \), indicating negligible suppression. In contrast, the \NKSshort yield is significantly lower than the model expectation in central events, with an average suppression factor of \( 0.6 \pm 0.1 \) below \( p_{\mathrm{T}} = 3 \) GeV/\(c\), corresponding to a \( \sim 4\sigma \) deviation. This suppression is absent in peripheral events and at higher \( p_{\mathrm{T}} \), where reinteraction effects are reduced.

These results provide compelling evidence that rescattering in the hadronic phase plays a decisive role in attenuating short-lived resonance yields, particularly in central heavy-ion collisions and at low transverse momentum, while leaving longer-lived resonances largely unaffected.

\begin{figure}[!hbt]
	\centering
	\includegraphics[height=0.38\textwidth]{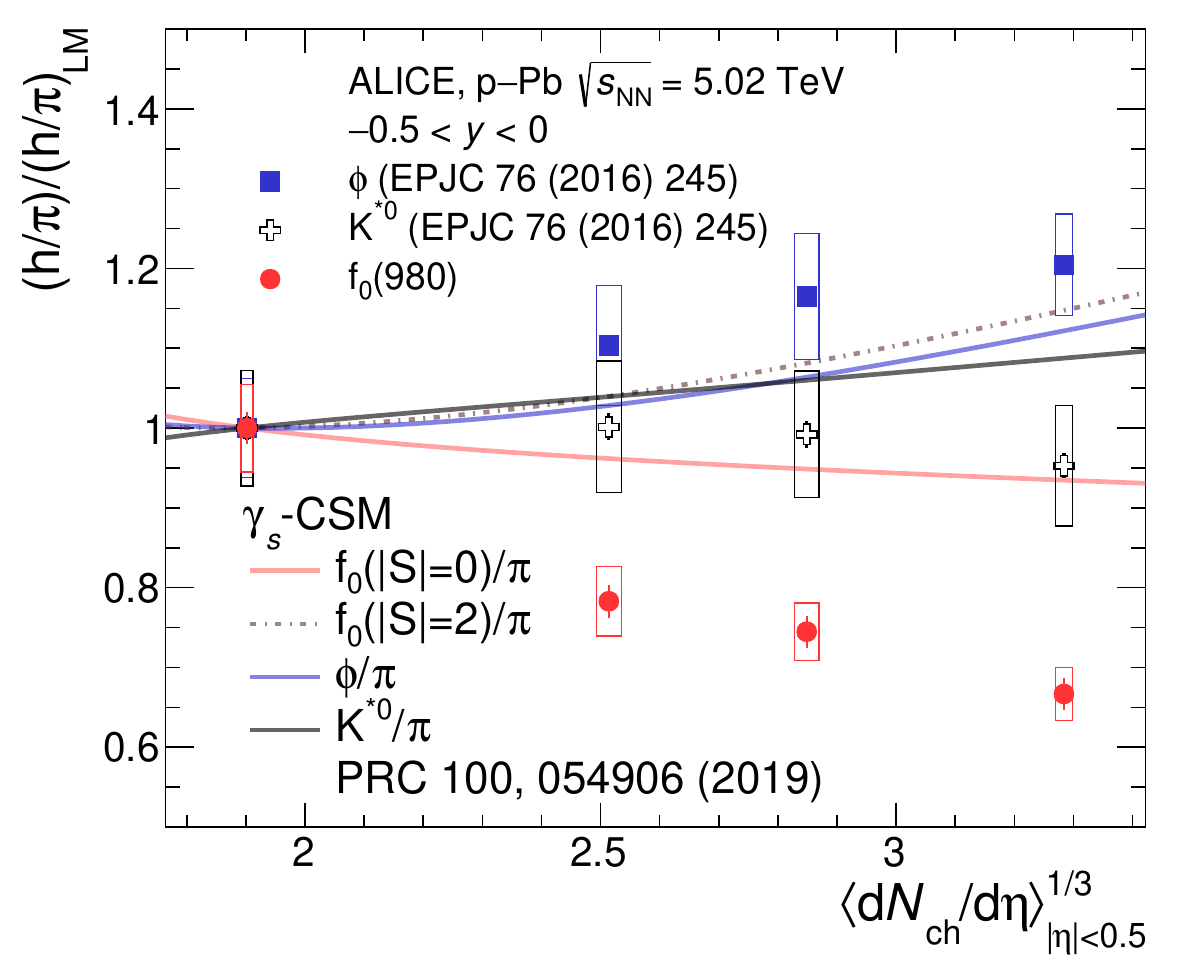}
	\includegraphics[height=0.38\textwidth]{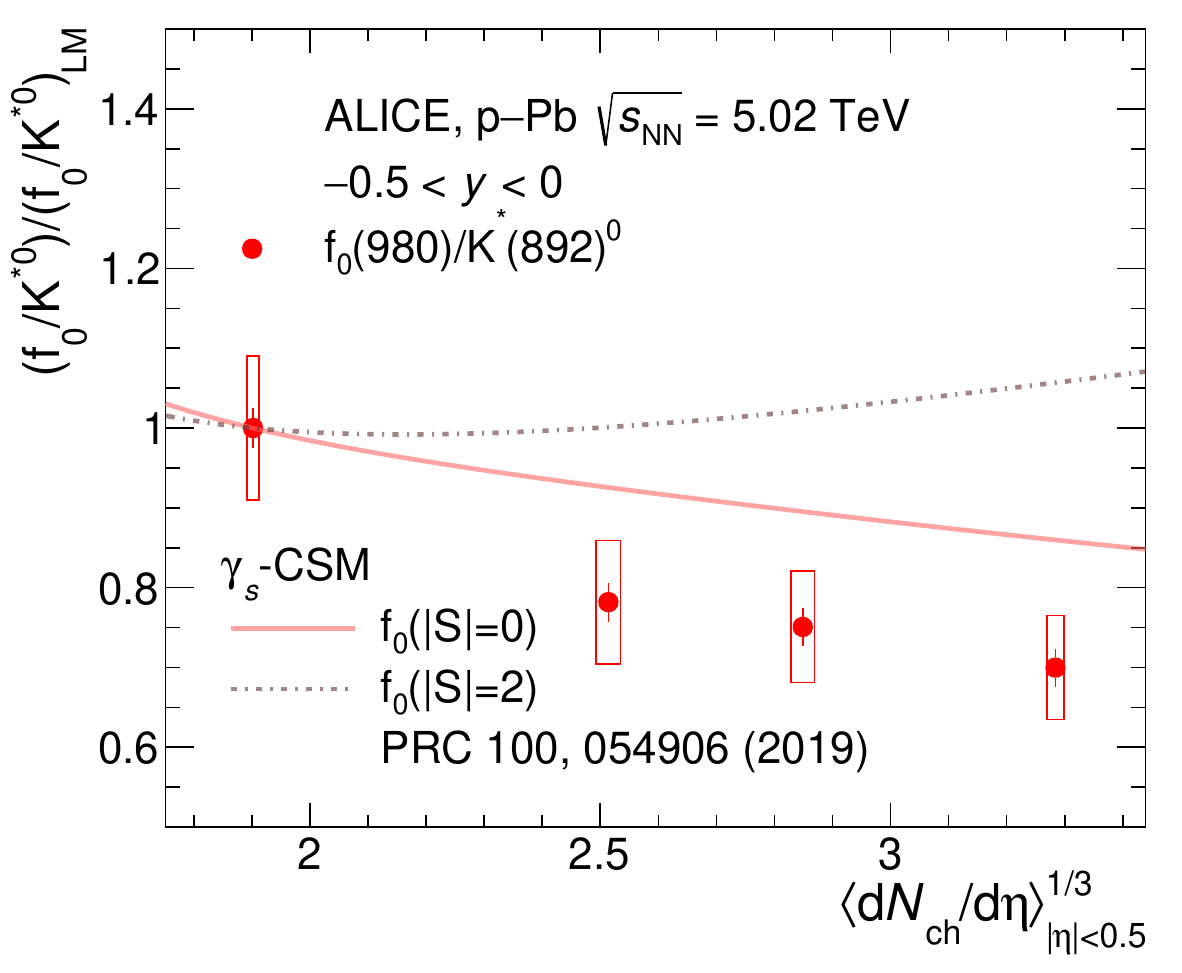}
	\caption{\label{fig:f0_ratio} Double ratios of $\phi$, $K^{*0}(892)$~\cite{ALICE:2016sak}, and $f_{0}(980)$ to $\pi$~\cite{ALICE:2016dei} (left), and $f_{0}(980)$ to $K^{}(892)^{0}$ (right), shown as a function of charged-particle multiplicity. The double ratios are raised to the power of $1/3$. Predictions from the canonical statistical model ($\gamma_{s}$-CSM)~\cite{Vovchenko:2019kes} are shown as lines.}
\end{figure}

Recently, the ALICE experiment at the LHC measured the yield of the exotic $f_{0}(980)$ resonance in p--Pb collisions at
$\sqrt{s_{\mathrm{NN}}}=5.02$~TeV and studied its multiplicity dependence by comparing it with the yields of various other hadrons.
Figure~\ref{fig:f0_ratio} (left panel) shows the double ratios (defined as the hadron-to-pion yield ratios normalized to their values in the lowest multiplicity interval (60–100\%), $(h/\pi)/(h/\pi)_{\mathrm{LM}}$) of different resonances ($\phi$, $K^{*0}$~\cite{ALICE:2016sak}, and $f_{0}(980)$~\cite{ALICE:2023cxn}) to charged pions~\cite{ALICE:2016dei} as a function of charged-particle multiplicity in p--Pb collisions at $\sqrt{s_{\mathrm{NN}}}=5.02$~TeV. The $\phi/\pi$ double ratio increases with increasing multiplicity, consistent with strangeness enhancement~\cite{ALICE:2016fzo}. Due to the relatively long lifetime of the $\phi$ meson, its yield is expected to be only weakly affected by interactions in the hadronic phase.

In contrast, the $K^{*0}(892)/\pi$ double ratio shows no significant multiplicity dependence within the reported uncertainties. This behavior can be understood as the result of competing effects between strangeness enhancement and hadronic rescattering. As discussed in the previous section, the latter is expected to dominate due to the short lifetime of the $K^{*0}(892)$. The $f_{0}(980)/\pi$ double ratio exhibits a decreasing trend with increasing multiplicity, suggesting a dominant role of rescattering effects. This is consistent with the short lifetime of the $f_{0}(980)$ ($\approx$~3--5~fm/$c$)~\cite{ALICE:2023cxn}, which is comparable to that of the $K^{*0}(892)$.

The measurements are compared with predictions from the $\gamma_{s}$-canonical statistical model (CSM)~\cite{Vovchenko:2019kes} under different assumptions for the hidden strangeness content. The model qualitatively reproduces the rising trend of the $\phi/\pi$ ratio when the $\phi$ meson is assigned two hidden strange quarks. However, it overestimates the $K^{*0}(892)/\pi$ ratio at high multiplicity, as hadronic rescattering effects are not implemented in the model, while strangeness enhancement for the $K^{*0}(892)$ is included. For the $f_{0}(980)$, the CSM assuming two hidden strange quarks predicts an increasing trend with multiplicity, in contradiction to the experimental data. The zero-hidden-strangeness scenario, although exhibiting a decreasing trend, still overestimates the observed suppression.

The right panel of Fig.~\ref{fig:f0_ratio} shows the $f_{0}(980)/K^{*0}(892)$ double ratio as a function of multiplicity. Given the comparable and short lifetimes~\cite{ParticleDataGroup:2022pth} of the two resonances, this ratio is expected to be weakly affected by hadronic interactions. The CSM prediction assuming two hidden strange quarks for the $f_{0}(980)$ shows a mild increase with increasing multiplicity, whereas the measured ratio decreases. The decreasing trend is qualitatively reproduced by the CSM with zero hidden strangeness for the $f_{0}(980)$ and can be understood as a consequence of strangeness enhancement in the $K^{*0}(892)$yield. Overall, the results indicate no clear evidence for effective strangeness enhancement of the $f_{0}(980)$ meson. The internal structure of $f_{0}(980)$ can also be probed by measuring its azimuthal momentum anisotropy, which has been discussed in section~\ref{chap3:sub:f0}.

\subsection{Hadronic Phase Time Scale}\label{chap3:sub:second}
Resonance to non-resonance yield ratio serves as a sensitive probe to estimate a lower bound on the hadronic phase lifetime, defined as the time interval $\tau$ between chemical and kinetic freeze-out. Taking an example of \NKSshort/K, the yield ratio at kinetic freeze-out can be approximated by
\begin{equation}
\left[\mathrm{K}^{*0}/\mathrm{K}\right]_{\mathrm{kinetic}} = \left[\mathrm{K}^{*0}/\mathrm{K}\right]_{\mathrm{chemical}} \times e^{-\tau/\tau_{K^{*0}}},
\end{equation}
where $\tau_{K^{*0}} = 4.16$ fm$/c$ is the vacuum lifetime of the \NKSshort resonance. The $[\mathrm{K}^{*0}/\mathrm{K}]$ ratio measured in the 70--100\% multiplicity class of pp collisions at $\sqrt{s} = 13$ TeV is used as a proxy for the chemical freeze-out value, assuming negligible hadronic rescattering in such dilute systems. Yield ratios in larger systems or higher multiplicity classes are interpreted as values after hadronic rescattering at kinetic freeze-out. Because the extraction neglects any regeneration of \NKSshort during the hadronic phase—which would partially compensate for suppression—the resulting lifetime should be regarded strictly as a conservative lower limit.

\begin{figure}[!hbt]
	\centering
	\includegraphics[height=0.5\textwidth]{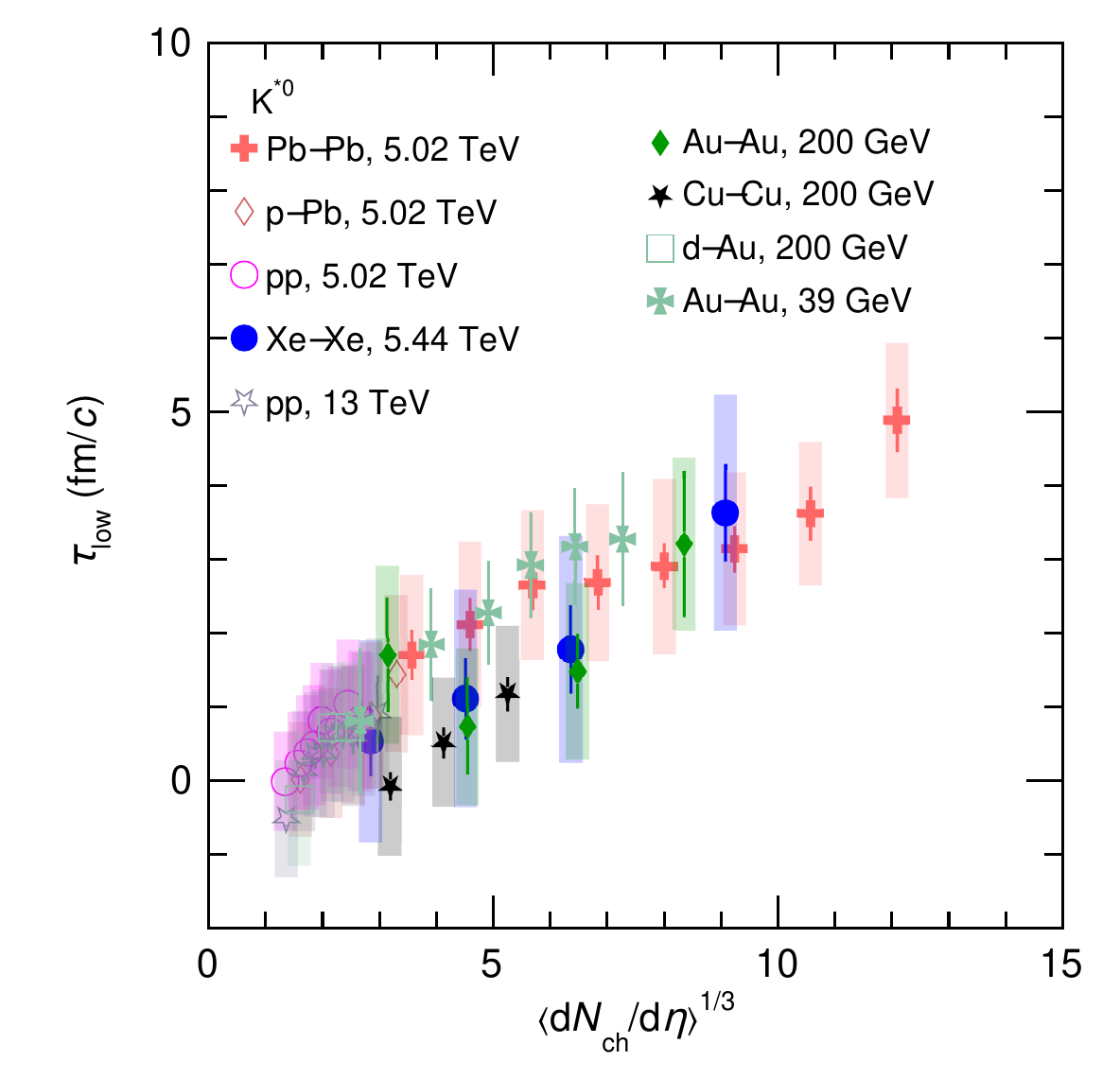}
	\caption{\label{fig:fig5} Lower bound on the hadronic phase lifetime $\tau$ as a function of \dndetaqube for various collision systems and energies. Vertical bars indicate statistical uncertainties, while shaded boxes represent systematic uncertainties.}
\end{figure}

To account for relativistic effects, the estimated $\tau$ values are scaled by a Lorentz factor,
\begin{equation}
\sqrt{1 + \left(\frac{\langle p_{\mathrm{T}} \rangle}{m_{K^{*0}}} \right)^2},
\end{equation}
where $\langle p_{\mathrm{T}} \rangle$ is the average transverse momentum of the \NKSshort. The extracted values of $\tau$ are shown in Fig.~\ref{fig:fig5} as a function of the charged-particle multiplicity density, $\langle \mathrm{d}N_{\mathrm{ch}}/\mathrm{d}\eta \rangle^{1/3}$. As expected, the hadronic phase lifetime increases with system size. In the most central Pb--Pb collisions at $\sqrt{s_{\mathrm{NN}}} = 5.02$ TeV, the estimated lower bound on the hadronic lifetime reaches 4--7~fm$/c$, comparable to the \NKSshort  lifetime and significantly shorter than that of the $\phi$ meson.

A smooth evolution of $\tau$ from small to large systems is observed, highlighting the continuity of hadronic rescattering effects. The EPOS3 model coupled with UrQMD qualitatively reproduces this trend~\cite{Knospe:2015nva}. Assuming a constant chemical freeze-out temperature, the observed increase in $\tau$ with multiplicity implies a decreasing kinetic freeze-out temperature in more central collisions—consistent with blast-wave analyses of identified particle $p_{\mathrm{T}}$ spectra.

A complementary perspective on the extracted hadronic phase lifetime can be obtained by comparing these results with femtoscopic measurements based on Hanbury Brown--Twiss (HBT) correlations. HBT analyses probe the space--time extent of the particle-emitting source at kinetic freeze-out, providing access to characteristic source radii and emission durations that reflect the final decoupling stage of the system. In Pb--Pb collisions at LHC energies, ALICE HBT measurements report system lifetimes and emission durations of the order of 5--10~fm$/c$ in central collisions, with a clear dependence on collision centrality and charged-particle multiplicity~\cite{ALICE:2017iga}.
While HBT femtoscopy is primarily sensitive to the conditions at kinetic freeze-out, the suppression of short-lived resonances such as the \NKSshort reflects the cumulative effect of hadronic interactions occurring throughout the entire hadronic phase between chemical and kinetic freeze-out. The consistency between these two independent observables provides an important cross-validation of the extracted timescales and supports a coherent picture of a prolonged hadronic rescattering phase in central heavy-ion collisions. Moreover, the smooth increase of the \NKSshort-derived lifetime with charged-particle multiplicity mirrors the system-size dependence observed in HBT radii and emission durations, indicating that both observables are governed by the same underlying dynamics of system expansion and cooling. This correspondence reinforces a unified space--time description of the hadronic stage, in which larger and more central collision systems undergo a longer-lived hadronic evolution before kinetic freeze-out.

\subsection{Kinetic Freeze-out and Thermal Decoupling}\label{chap3:sub:third}

The duration of the hadronic phase is also reflected in the separation between chemical and kinetic freeze-out temperatures. In most heavy-ion studies, the kinetic freeze-out temperature $T_{\mathrm{kin}}$ and the average transverse expansion velocity $\langle \beta_{\mathrm{T}} \rangle$ are extracted from transverse-momentum spectra using blast-wave--type parameterizations, while the chemical freeze-out temperature $T_{\mathrm{ch}}$ is independently constrained from the yields of long-lived hadrons and found to be approximately 155~MeV at LHC energies. Although successful, this approach relies on assumptions regarding the collective flow profile and the freeze-out hypersurface.
An alternative and conceptually distinct method to extract the kinetic freeze-out temperature has been developed within the framework of a Hadron Resonance Gas with Partial Chemical Equilibrium (HRG-PCE)~\cite{Motornenko:2019jha}. In this approach, chemical freeze-out fixes the total abundances of long-lived hadrons, while short-lived resonances remain in pseudo-elastic equilibrium with their decay products during the subsequent hadronic evolution. As the system cools below $T_{\mathrm{ch}}$, resonance yields continue to evolve due to ongoing regeneration and rescattering processes until kinetic freeze-out is reached at a lower temperature $T_{\mathrm{kin}}$.
Within the HRG-PCE framework, the kinetic freeze-out temperature is determined by a simultaneous fit to the measured yields of stable hadrons ($\pi^{\pm}$, K$^{\pm}$, p($\overline{\mathrm{p}}$)) and resonances such as the $\phi$ and \NKSshort, without invoking assumptions about the transverse flow velocity or the space--time structure of the freeze-out hypersurface. This allows $T_{\mathrm{ch}}$ and $T_{\mathrm{kin}}$ to be constrained directly from hadron and resonance yields alone, providing a self-consistent thermodynamic description of the hadronic phase.

\begin{figure}[!hbt]
	\centering
	\includegraphics[height=0.4\textwidth]{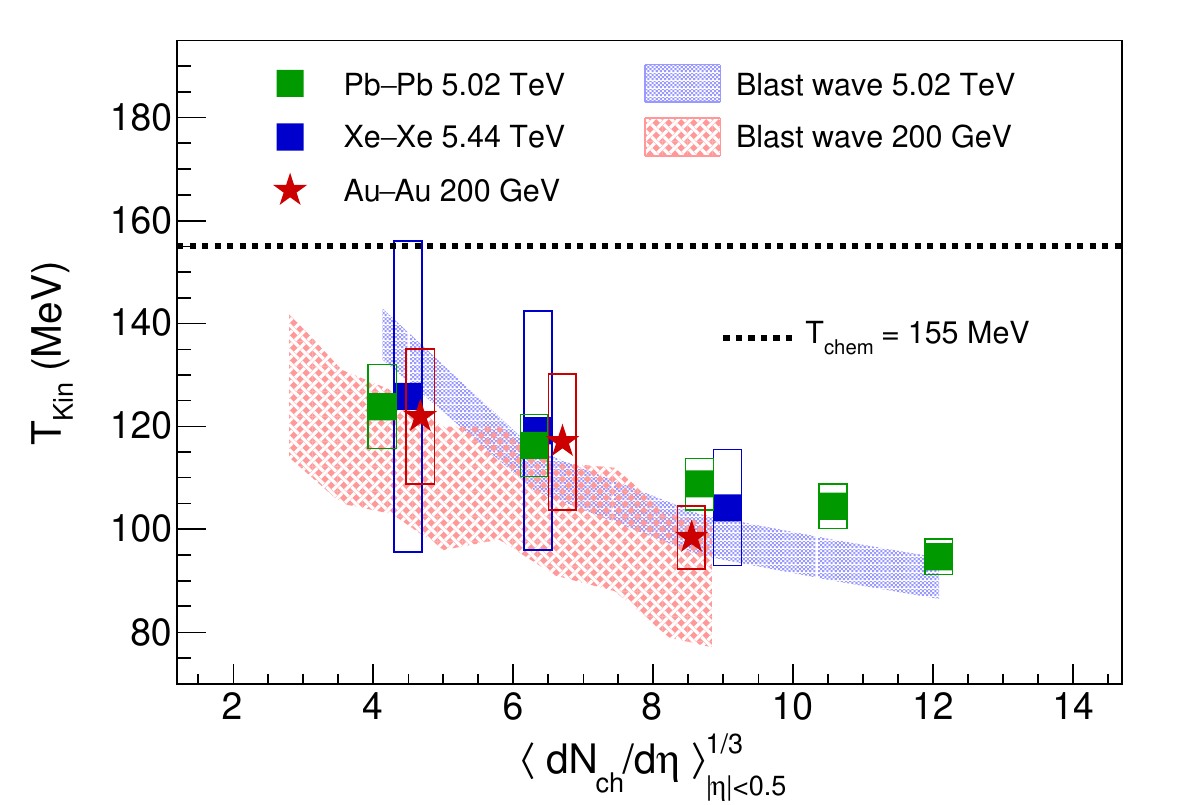}
	\caption{\label{fig:fig6a} Kinetic freeze-out temperature estimated using HRG-PCE fits to the yields of $\pi^{\pm}$, K$^{\pm}$, p($\overline{\mathrm{p}}$), $\phi$, and \NKSshort in different centrality classes of Pb--Pb, Xe--Xe, and Au--Au collisions. Vertical bars indicate statistical uncertainties, while boxes represent systematic uncertainties. Shaded bands show predictions from blast-wave model calculations.}
\end{figure}

Figure~\ref{fig:fig6a} shows the extracted kinetic freeze-out temperatures as a function of the charged-particle multiplicity. A clear decrease of $T_{\mathrm{kin}}$ from peripheral to central collisions is observed, consistent with a longer-lived hadronic phase in larger systems. While the uncertainties are larger for Xe--Xe and Au--Au collisions, the extracted $T_{\mathrm{kin}}$ values across different collision systems are compatible within errors when compared at similar charged-particle multiplicities.

The separation between chemical and kinetic freeze-out temperatures provides independent evidence for the existence of a hadronic phase with finite lifetime. In central Pb--Pb collisions, the extracted $T_{\mathrm{kin}}$ values reach temperatures of the order of 90--100~MeV, significantly below $T_{\mathrm{ch}}$, indicating substantial cooling and prolonged hadronic rescattering prior to kinetic freeze-out. These results are in fair agreement, within systematic uncertainties, with kinetic freeze-out temperatures obtained from blast-wave fits to transverse-momentum spectra, demonstrating consistency between yield-based and spectrum-based approaches.

\subsection{Energy Loss and Resonance Suppression}\label{chap3:sub:fourth}

\subsubsection{Nuclear modification in large systems}

\begin{figure}[!hbt]
\centering
\includegraphics[height=0.38\textwidth]{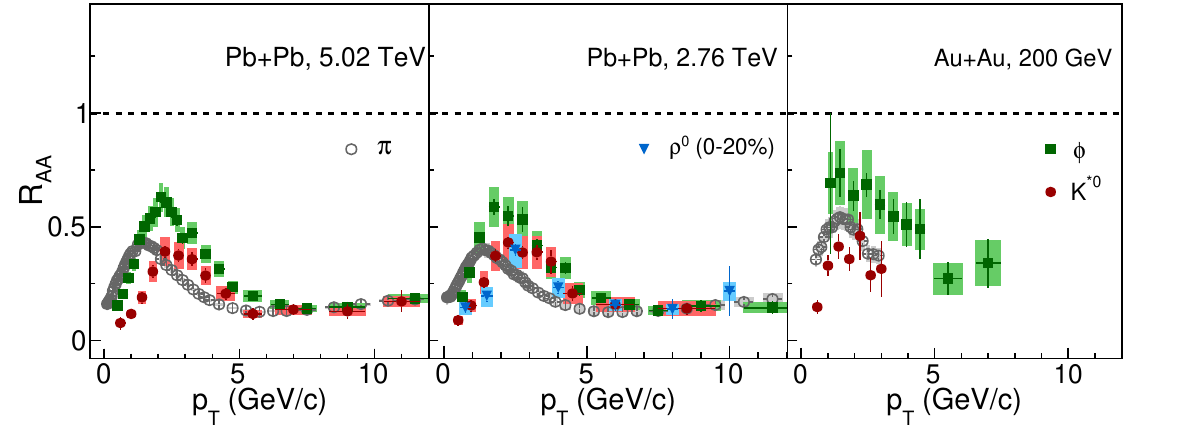}
\caption{\label{fig:fig6b} Nuclear modification factor ($R_{\rm AA}$) of various hadrons measured by ALICE~\cite{ALICE:2019hno,ALICE:2014juv,ALICE:2016dei,ALICE:2018qdv,ALICE:2021ptz,
ALICE:2017ban,ALICE:2021rpa} and STAR~\cite{STAR:2004bgh,PHENIX:2013kod,STAR:2008twt} collaborations at different heavy-ion (Pb--Pb and Au--Au) systems and energies. Vertical bars indicate statistical uncertainties, while shaded boxes represent systematic uncertainties.}
\end{figure}

The nuclear modification factor of resonances ($R_{\rm AA}$) in heavy-ion collisions has been extensively studied at RHIC \cite{STAR:2004bgh,PHENIX:2013kod,PHENIX:2010bqp,STAR:2008twt} and the LHC
\cite{ALICE:2019hno,ALICE:2014juv,ALICE:2016dei,ALICE:2018qdv,ALICE:2021ptz,
ALICE:2017ban,ALICE:2021rpa}, and is summarized in Fig.~\ref{fig:fig6b} with measurements from the ALICE and STAR experiments in central Pb--Pb and Au--Au collisions. At high transverse momentum ($p_{\mathrm{T}} \gtrsim 8$--10~GeV/$c$), the nuclear modification factors of short-lived resonances such as the $\rho^{0}$, \NKSshort\ and the $\phi$ meson are found to be comparable, within uncertainties, to those of stable hadrons (e.g.\ $\pi$ shown here). This similarity indicates that, once produced in initial hard scatterings, the parent partons of resonances and long-lived hadrons
experience a comparable amount of medium-induced energy loss before hadronization. Within current uncertainties, no additional resonance-specific medium modification is observed at high $p_{\mathrm{T}}$, and the dominant physics mechanism is partonic energy loss in the quark--gluon plasma, which is largely insensitive to the lifetime or quantum numbers of the final-state hadron. These measurements therefore suggest
that, at high $p_{\mathrm{T}}$, the suppression of resonances is largely indistinguishable from that of stable particles with similar masses and
quark content.

However, at low transverse momentum ($p_{\mathrm{T}} \lesssim 2$~GeV/$c$), the suppression of short-lived resonances is observed to be stronger than that of longer-lived particles such as the $\phi$ meson, particularly at LHC energies. This behavior is commonly interpreted as being consistent with the effects of hadronic re-scattering in the late hadronic phase, which can reduce the reconstructed yields of resonances that decay early, while longer-lived hadrons are expected to be less affected. These observations suggest that resonance-specific modifications at low $p_{\mathrm{T}}$ may predominantly arise during the hadronic phase, where
decay and regeneration processes take place within the medium.

\subsubsection{Nuclear modification in small systems}

\begin{figure}[!hbt]
\centering
\includegraphics[height=0.55\textwidth]{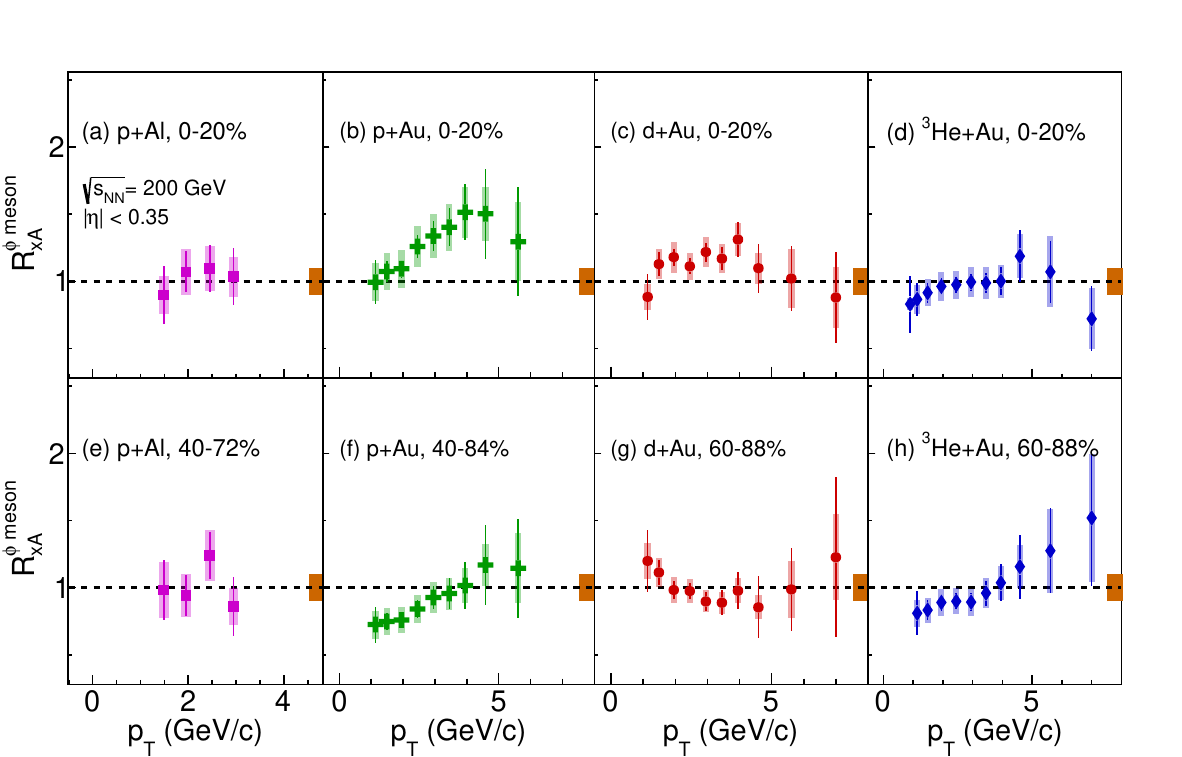}
\caption{\label{fig:phenixrda} Nuclear modification factors $R_{\mathrm{xA}}(p_{\mathrm{T}})$ for $\phi$ mesons in $p$--Al, $p$--Au, $d$--Au, and $^3$He--Au collisions at
$\sqrt{s_{\mathrm{NN}}}=200$~GeV measured by the PHENIX Collaboration~\cite{PHENIX:2022hku}. Vertical bars indicate statistical uncertainties, while shaded boxes represent systematic uncertainties.}
\end{figure}

In contrast to large A--A collisions, small collision systems are not expected to produce an extended, long-lived deconfined medium, and therefore strong partonic energy loss effects are generally absent. Instead, nuclear modification in small systems is commonly interpreted in terms of cold-nuclear-matter effects, such as nuclear PDFs, initial-state multiple scattering, and Cronin-type transverse-momentum broadening. The Fig.~\ref{fig:phenixrda} presents the nuclear modification of $\phi$ mesons in small collision systems (denoted by $R_{\mathrm{xA}}$) in $p$--Al, $p$--Au, $d$--Au, and $^3$He--Au collisions at $\sqrt{s_{\mathrm{NN}}}=200$~GeV at midrapidity ($|\eta|<0.35$) from the PHENIX Collaboration~\cite{PHENIX:2022hku}. In general, $R_{\mathrm{xA}}$ remains close to unity within uncertainties, indicating the absence of strong suppression in these small system collisions. However, in the most central collisions at intermediate $p_{\mathrm{T}}$, a weak system ordering is observed, with a tendency for $R_{^3\mathrm{HeAu}} < R_{\mathrm{dAu}} < R_{\mathrm{pAu}}$, which may reflect differences in event activity, collision geometry, and underlying multiplicity rather than strong partonic energy-loss effects. 

\begin{figure}[!hbt]
\centering
\includegraphics[height=0.45\textwidth]{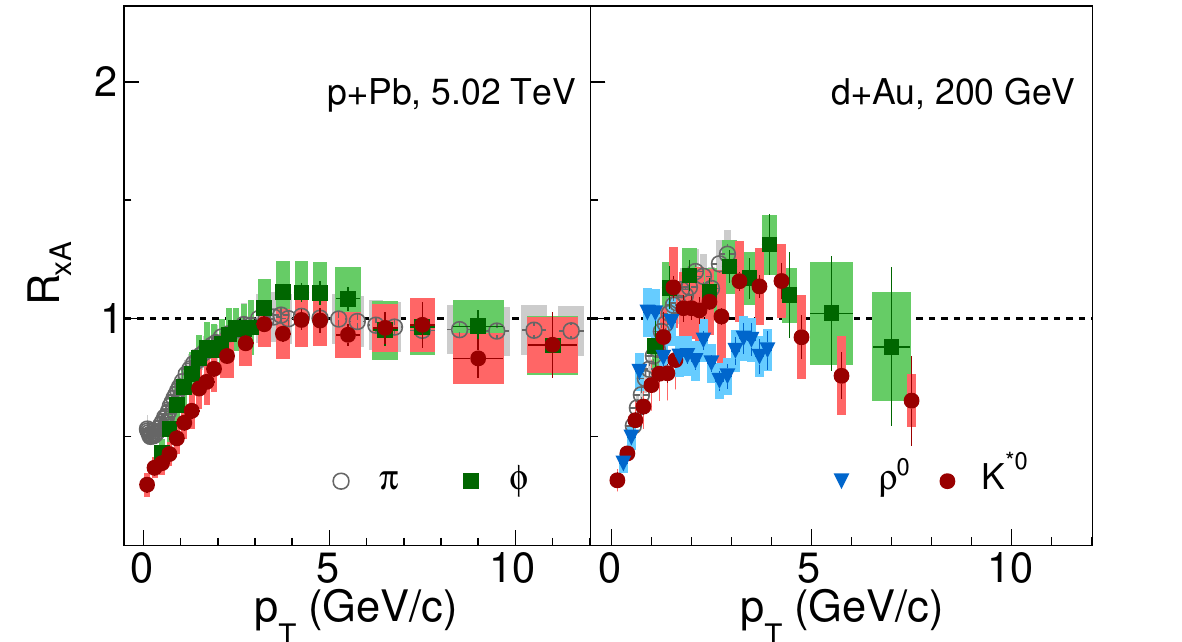}
\caption{\label{fig:RxApPbdAu} Nuclear modification factors $R_{\mathrm{xA}}(p_{\mathrm{T}})$ for $\phi$ and $K^{*0}$ mesons in $p$-Pb collisions at $\sqrt{s_{\mathrm{NN}}}=5.02$~TeV measured by the ALICE Collaboration~\cite{ALICE:2021rpa}, and
$R_{\mathrm{xA}}(p_{\mathrm{T}})$ for $\phi$, $K^{*0}$, and $\rho^{0}$ mesons in $d$+Au collisions at $\sqrt{s_{\mathrm{NN}}}=200$~GeV measured by the STAR~\cite{STAR:2008twt} and PHENIX~\cite{PHENIX:2014kia} Collaborations. Resonance measurements are compared with those of pions from the respective experiments and collision systems. Vertical bars indicate statistical uncertainties, while shaded boxes represent systematic uncertainties.}
\end{figure}

Figure~\ref{fig:RxApPbdAu} extends the discussion of nuclear modification in small systems by presenting $R_{\mathrm{xA}}(p_{\mathrm{T}})$ for resonance production measured in $p$--Pb and $d$--Au collisions, with particular emphasis on the low- and intermediate-$p_{\mathrm{T}}$ region. Shown are $R_{\mathrm{xA}}$ for $\phi$ and $K^{*0}$ mesons measured by the ALICE Collaboration in $p$--Pb collisions at $\sqrt{s_{\mathrm{NN}}}=5.02$~TeV~\cite{ALICE:2021rpa}, together with corresponding measurements of $\phi$, $K^{*0}$, and $\rho^{0}$ mesons by the STAR and PHENIX Collaborations in $d$--Au collisions at $\sqrt{s_{\mathrm{NN}}}=200$~GeV~\cite{STAR:2008twt,PHENIX:2014kia}. Furthermore, the $R_{\mathrm{xA}}$ of short-lived resonances are found to be comparable to those of pions within uncertainties. At low transverse momentum, the $R_{\mathrm{xA}}$ values of the different resonance species are found to be compatible within uncertainties, indicating no significant resonance-dependent modification in this kinematic regime. At intermediate and high $p_{\mathrm{T}}$, the measured $R_{\mathrm{xA}}$ values remain consistent with unity within uncertainties for all resonance species, further supporting the absence of strong final-state partonic energy loss in these small collision systems.

\begin{figure}[!hbt]
\centering
\includegraphics[height=0.45\textwidth]{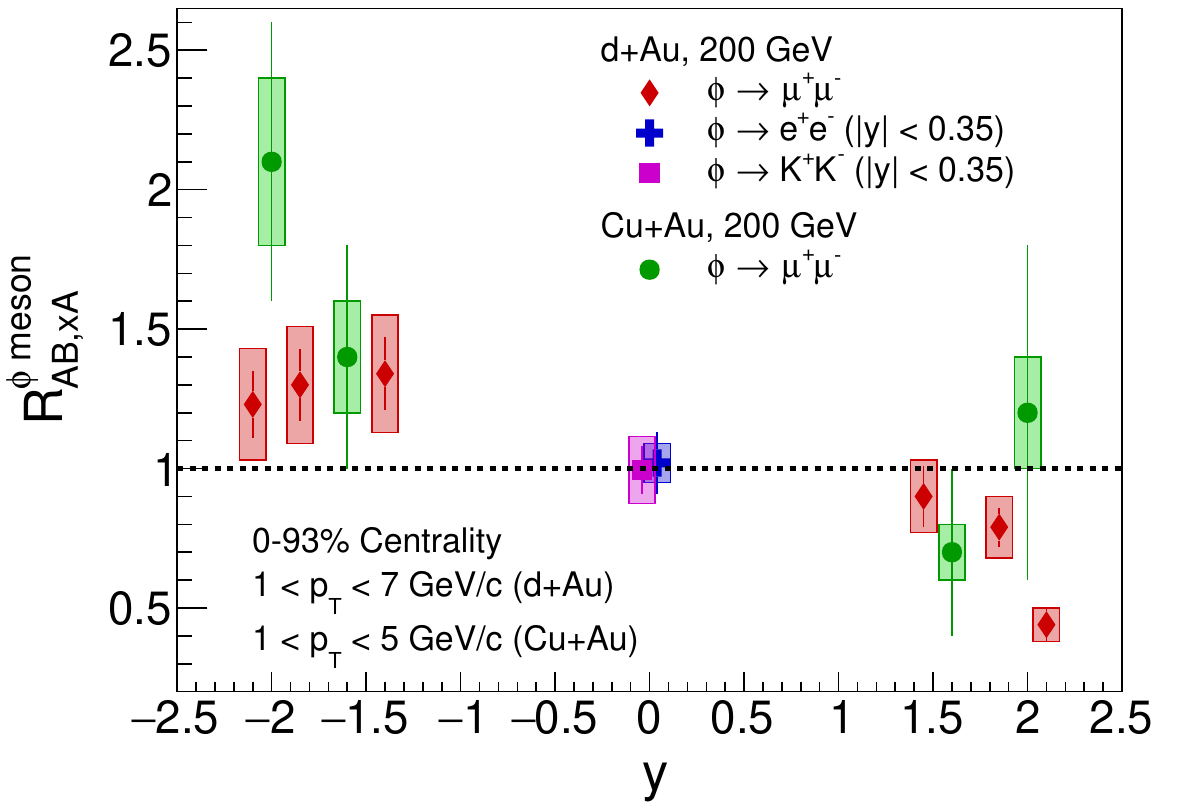}
\caption{\label{fig:figRdasy} Nuclear modification factor ($R_{\rm AB, xA}$) versus rapidity for $\phi$ meson via dimuon, dielectron and hadronic channels in $d$--Au and Cu--Au collisions at $\sqrt{s_{\mathrm{NN}}}=200$~GeV measured by PHENIX collaboration~\cite{PHENIX:2015spp,PHENIX:2015cea}. Vertical bars indicate statistical uncertainties, while shaded boxes represent systematic uncertainties.}
\end{figure}

Figure~\ref{fig:figRdasy} shows the nuclear modification factors for $\phi$ mesons measured by PHENIX in $d$--Au and Cu--Au collisions at $\sqrt{s_{\mathrm{NN}}}=200$~GeV as a function of rapidity~\cite{PHENIX:2015spp,PHENIX:2015cea}. In $d$--Au collisions, $R_{d\mathrm{Au}}$ is shown for $\phi$ mesons reconstructed via the dielectron and hadronic decay channels within $|y|<0.35$. The dimuon channel measurements are performed within $1.2<|y|<2.2$ in both $d$--Au and Cu--Au collisions. A clear rapidity asymmetry is observed in both systems in the dimuon channel. In $d$--Au collisions, $\phi$ meson production is suppressed in the light-ion--going direction and enhanced in the heavy-ion--going direction, with $R_{d\mathrm{Au}}$ at midrapidity consistent with binary-collision--scaled $p$+$p$ expectations. A similar asymmetry is observed in Cu--Au collisions, with enhanced $\phi$ production in the Au-going direction and values consistent with unity, within uncertainties, in the Cu-going direction. The rapidity-dependent modification of $\phi$ meson production in these asymmetric systems is qualitatively similar to that observed for open heavy-flavor hadrons at RHIC~\cite{PHENIX:2013txu}.
However, $J/\psi$ production exhibits suppression at both forward and backward rapidities ($1.2<|y|<2.2$)~\cite{PHENIX:2015spp,PHENIX:2015cea}. Comparable enhancement and suppression
patterns have also been reported by the ALICE Collaboration for $\phi$ mesons in $p$--Pb collisions at $\sqrt{s_{\mathrm{NN}}}=5.02$~TeV~\cite{ALICE:2015cql}, indicating that such rapidity-dependent nuclear effects persist across collision energies and provide essential baselines for interpreting resonance modification in larger symmetric A--A collisions.

Taken together, nuclear modification measurements of resonances reveal distinct modification mechanisms in large and small collision systems. In large A--A collisions, resonance suppression at high $p_{\mathrm{T}}$ is dominated by partonic energy loss in the quark--gluon plasma, while resonance-specific modifications at low $p_{\mathrm{T}}$ arise from hadronic interactions in the late stage of the collision. In contrast, small collision systems exhibit nuclear modification factors close to unity over a broad $p_{\mathrm{T}}$ range, indicating the absence of strong partonic energy loss, while short-lived resonances may remain sensitive to the density and lifetime of the hadronic medium in high-multiplicity $p$--A events. In this context, the $\phi$ meson serves as a useful long-lived reference for comparisons with shorter-lived resonances such as the $K^{*0}$. Overall, these measurements place important constraints on the onset of final-state interactions and provide quantitative baselines for interpreting collective-like phenomena observed in small collision systems.


\subsection{Spin Alignment}\label{chap3:sub:third}
\subsubsection{Introduction}
Resonances, particularly vector mesons such as $K^{*}(892)$ and $\phi(1020)$, are invaluable probes in the study of the QCD medium created in high-energy heavy-ion collisions. Owing to their short lifetimes and sensitivity to the surrounding environment, resonances serve as temporally resolved messengers of the medium's evolution and its collective dynamics. Among the diverse roles played by resonances, their potential sensitivity to the system’s angular momentum and magnetic field—particularly through spin alignment phenomena—has attracted considerable attention in recent years.

High-energy heavy-ion collisions generate a transient state of deconfined quarks and gluons, known as the quark--gluon plasma (QGP), characterized by extreme temperature and energy density~\cite{ALICE:2022wpn}. Extensive experimental and theoretical studies have demonstrated that the QGP behaves as a strongly coupled fluid, exhibiting a low shear viscosity-to-entropy ratio and collective flow patterns~\cite{ALICE:2022wpn}. Non-central collisions, where the impact parameter is non-zero, introduce additional structure: an enormous angular momentum on the order of $O(10^{7}\hbar)$~\cite{THAM}, and intense magnetic fields reaching $O(10^{15}~\mathrm{T})$~\cite{THMF}, both oriented perpendicular to the reaction plane~\cite{THAM}.

These macroscopic quantities—angular momentum and magnetic field—are not directly measurable, but their influence may be imprinted on final-state observables. This is where resonances play a pivotal role. Their decay topologies, sensitivity to hadronic rescattering, and—crucially—their spin degrees of freedom allow for unique access to early-time dynamics. In particular, vector mesons with spin $J=1$ are ideal candidates to probe the local polarisation structure of the medium via measurements of spin alignment. Deviations in the spin density matrix element $\rho_{00}$ from the isotropic value of $1/3$ carry imprints of the interplay between parton polarisation, hadronisation dynamics, and possible medium-induced effects such as vorticity or magnetic field-induced polarisation.

During the collision process, part of the system’s initial angular momentum is transferred to the QGP fluid, generating a vorticity field that can polarise quarks. Similarly, strong electromagnetic fields created by the colliding nuclei can polarise quarks through magnetic moment coupling. Both effects are expected to be short-lived and spatially non-uniform, requiring sensitive probes that can reflect early-time conditions. Vector mesons, particularly those reconstructed from hadronic decay channels, offer precisely this sensitivity, as their spin alignment can capture subtle spin-dependent dynamics during hadronisation.

Measuring spin alignment involves studying the angular distribution of decay products in the vector meson’s rest frame relative to a polarisation axis—typically chosen perpendicular to the reaction plane in heavy-ion collisions. The angular distribution for the decay of a vector meson into two spin-zero particles is given by~\cite{Schilling:1969um}:
\begin{equation}
\frac{\mathrm{d}N}{\mathrm{d}\cos{\vartheta^{*}}} \propto \left[ 1 - \rho_{00} + (3\rho_{00} - 1 )\cos^{2}{\vartheta^{*}} \right],
\label{eq:dndcost}
\end{equation}
where $\vartheta^{*}$ is the angle between one of the decay products and the quantisation axis. An isotropic spin orientation corresponds to $\rho_{00} = 1/3$; any deviation signals spin alignment and, by extension, sensitivity to medium vorticity or other spin-polarising effects.

In proton--proton collisions, the quantisation axis is typically the helicity axis aligned with the particle’s momentum in the lab frame, whereas in heavy-ion collisions, the direction defined by the system’s angular momentum or magnetic field is more appropriate. These choices are further informed by the anisotropic expansion of the QGP, characterised by the Fourier decomposition of azimuthal particle distributions, notably the elliptic flow coefficient $v_2$~\cite{Snellings:2011sz}. Spin alignment measurements must therefore disentangle flow-induced correlations from genuine polarisation effects.

Various theoretical models have explored how spin alignment in vector mesons could arise from mechanisms such as quark polarisation in the early-stage medium, followed by recombination into mesons. Coalescence models predict a suppression of the $m=0$ spin projection state (i.e., $\rho_{00} < 1/3$) for mesons formed from polarised quarks~\cite{Liang:2004xn, Yang:2017sdk}, while fragmentation processes may lead to $\rho_{00} > 1/3$ depending on the spin transfer mechanisms.

Crucially, vector meson spin alignment complements global polarisation studies in hyperons. Whereas hyperon polarisation reflects the net spin orientation of their constituent quarks, vector mesons are more sensitive to local spin correlations in quark--antiquark pairs, enabling access to finer spatial and temporal resolution of spin dynamics~\cite{Lv:2024uev}. This distinction reinforces the importance of including resonances—especially vector mesons—in any comprehensive study of spin phenomena in heavy-ion collisions.

Beyond coalescence, a range of other mechanisms have been proposed:
\begin{itemize}
    \item Local polarisation from anisotropic medium expansion~\cite{local1, local2}
    \item Spin fluctuations in turbulent colour fields, driven by axial charge currents~\cite{Muller:2021hpe}
    \item Initial-state spin correlations in the glasma phase due to strong chromo-electromagnetic fields~\cite{Kumar:2023ghs}
\end{itemize}

Additional field-induced scenarios include polarisation through meson field fluctuations, as in the case of the $\phi$ meson polarising strange quarks via pseudoscalar currents~\cite{Sheng:2019kmk, Sheng:2022wsy}, or spin-dependent modifications to spectral functions in holographic models~\cite{Sheng:2024kgg}.

Despite a diversity of mechanisms, a consistent, unified theoretical framework capable of quantitatively describing spin alignment across vector meson species, collision energies, and system sizes remains an open challenge. Nevertheless, resonances continue to provide indispensable insights into the spin structure of the QGP, highlighting their central role in mapping the angular momentum and magnetic properties of the strongly interacting medium.

\subsubsection{Experimental Techniques and Analysis Details}
\subsection*{Coordinate System}

In spin alignment studies of vector mesons, three reference frames are commonly used to define the quantization axis: (1) the helicity frame, (2) the production plane frame, and (3) the reaction plane frame. Each frame provides a distinct orientation against which the angular distribution of the decay daughters is analyzed.

\begin{figure}[H]
\centering
\includegraphics[scale=0.5]{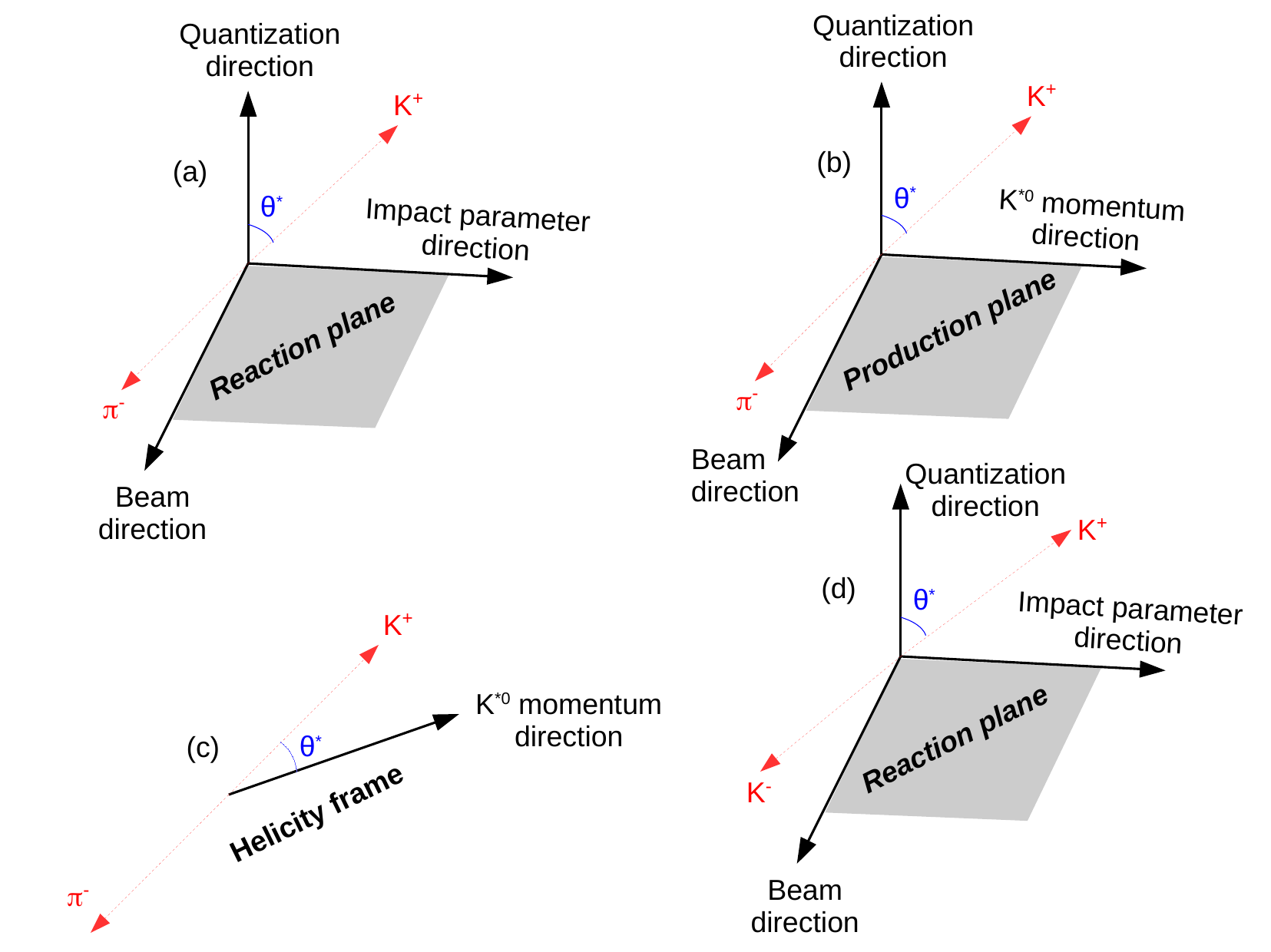}
\caption{Reference frames used in spin alignment measurements of vector mesons.}
\label{coordinate}
\vspace{-3mm}
\end{figure}

In the helicity frame, the quantization axis is aligned with the meson's momentum in the laboratory frame. The angle $\theta^*$ is then measured between this axis and the momentum of a decay daughter in the meson's rest frame, as shown in panel (c) of Fig.~\ref{coordinate}.

In the production plane frame, the quantization axis is defined as the normal  vector to the plane formed by the beam direction and the meson's momentum vector. This setup is illustrated in panel (b) of Fig.~\ref{coordinate}.

The reaction plane frame uses a quantization axis perpendicular to the reaction plane, which is defined by the beam direction and the impact parameter vector. This axis corresponds to the direction of the system’s global angular momentum in non-central collisions. Panels (a) and (d) of Fig.~\ref{coordinate} depict this configuration for the K$^*$ and $\phi$ mesons, respectively.

Since the true impact parameter direction is experimentally inaccessible, the event plane—constructed from the azimuthal distribution of produced charged particles—is used as an experimental approximation. To account for the finite resolution of the event plane reconstruction, the measured spin alignment parameter $\rho_{00}^{\mathrm{EP}}$ is corrected using~\cite{analytic_accep_corr}:
\begin{equation}
\rho^{\mathrm{RP}}_{00} - \frac{1}{3} = \left(\rho^{\mathrm{EP}}_{00} - \frac{1}{3}\right)\frac{4}{1 + 3R},
\label{res_eq11}
\end{equation}
where $R$ is the second-order event plane resolution.

\section*{Analysis Technique}

Vector mesons such as K$^*$ decay rapidily, while the $\phi$ has significantly longer lifetime, and must be reconstructed from their decay products. The decay channels used are: K$^* \rightarrow K^{+}\pi^{-}$, and $\phi \rightarrow K^{+}K^{-}$. Charged daughters are identified via particle identification techniques. At ALICE and STAR, identification primarily uses energy loss in the Time Projection Chamber (TPC) and timing from the Time-of-Flight (TOF) detector.

The invariant mass spectrum of the daughter pairs includes both true signal and a combinatorial background. Background estimation methods include:
\begin{itemize}
    \item \textbf{Mixed Event:} Pairs formed from tracks in different events.
    \item \textbf{Like-Sign:} Using same-charge pairs to model background.
    \item \textbf{Rotation Method:} Rotating one daughter’s momentum to break correlations.
\end{itemize}

\begin{figure}[H]
\centering
\includegraphics[scale=0.3]{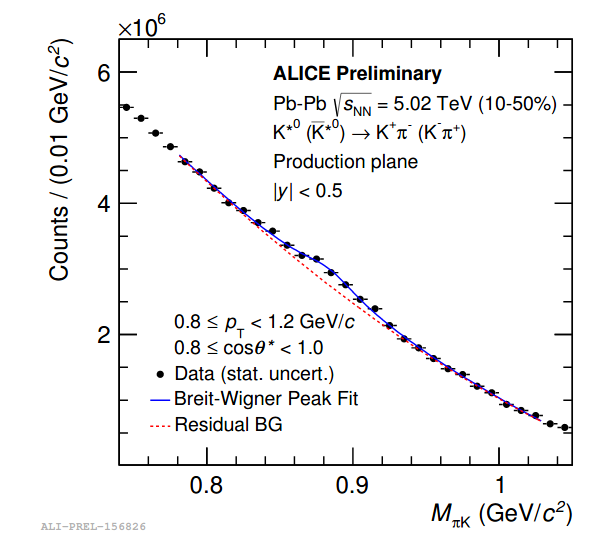}
\caption{Background-subtracted invariant mass distribution of $K\pi$ pairs in Pb--Pb collisions at $\sqrt{s_{NN}} = 2.76$~TeV. The signal is fitted with a Breit--Wigner plus polynomial. This example is for 0.8 $<$ $p_T$ $<$ 1.2 GeV/$c$ and 0.8 $<$ $\cos\theta^*$ $<$ 1.0. Figure from~\cite{rsingh}.}
\label{invmass}
\vspace{-3mm}
\end{figure}

After subtraction, residual background (from correlated sources like jets) remains. The signal is modeled with a Breit--Wigner (or Voigtian) function, and the residual background with a second-order polynomial as shown in Fig.~\ref{invmass}..

\begin{figure}[H]
\begin{center}
\includegraphics[scale=0.5]{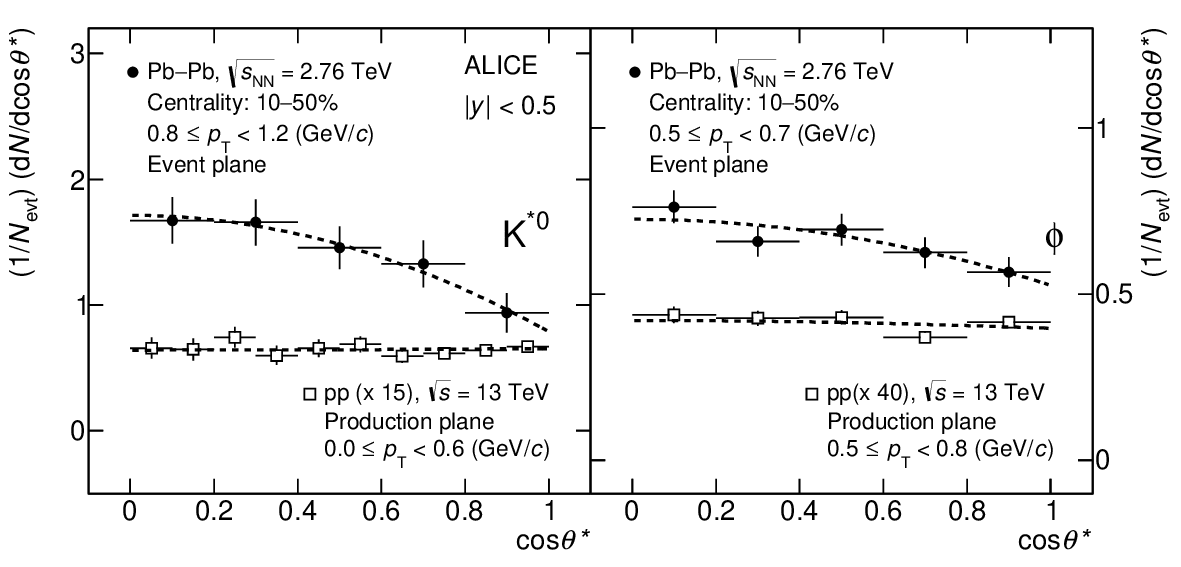}
\caption{Corrected angular distributions for K$^*$ and $\phi$ mesons in Pb--Pb at $\sqrt{s_{NN}} = 2.76$ TeV and pp at $\sqrt{s} = 13$ TeV. Fits with Eq.~(\ref{res_eq11}) yield $\rho_{00}$ values.}
\label{ctdist}
\end{center}
\vspace{-3mm}
\end{figure}

Extracted signals in $\cos\theta^*$ bins are further corrected for detector effects. Acceptance $\times$ efficiency corrections are evaluated using Monte Carlo simulations with full detector modelling via GEANT. The corrected distributions are fitted with the angular function:
\[
\frac{dN}{d\cos\theta^*} \propto 1 - \rho_{00} + (3\rho_{00} - 1)\cos^2\theta^*,
\]
to extract $\rho_{00}$ values. Figure~\ref{ctdist} shows example of acceptance $\times$ efficiency corrected angular distribution fitting. Additional details are in~\cite{alice,Abelev:2008ag}, with an alternative analytic correction method proposed in~\cite{analytic_accep_corr}.

To complement the traditional yield-based approach for extracting spin alignment, measurement of $\langle \cos^2 \theta^* \rangle$ can be used to improve statistical robustness and mitigate background-related and fit-related uncertainties. Instead of binning the invariant mass distribution in $\cos\theta^*$, the method relies on calculating the average value $\langle \cos^2 \theta^* \rangle$ in fine bins of invariant mass ($m_{\mathrm{inv}}$). The relation between this average and the spin alignment observable is given by:
\begin{equation}
\Delta\rho\{\theta^*\} = \rho_{00} - \frac{1}{3} = \frac{5}{2} \left( \langle \cos^2 \theta^* \rangle - \frac{1}{3} \right).
\end{equation}

A profile of $\langle \cos^2 \theta^* \rangle$ as a function of $m_{\mathrm{inv}}$ is constructed and converted into a $\Delta\rho\{\theta^*\}$ profile. This is then fitted using the model:
\begin{equation}
\Delta\rho\{\theta^*\}(m_{\mathrm{inv}}) = \Delta\rho_{\mathrm{sig}} \cdot r(m_{\mathrm{inv}}) + \Delta\rho_{\mathrm{bkg}} \cdot (1 - r(m_{\mathrm{inv}})),
\end{equation}
where $r(m_{\mathrm{inv}})$ is the signal fraction extracted from a fit to the invariant mass distribution using a Breit--Wigner plus polynomial background.

As an alternative method to determine the spin-alignment parameter $\rho_{00}$, one can employ the observable $\Delta\rho\{\phi^*\}$, defined using the azimuthal angle $\phi^*$ of the decay daughter relative to the event-plane angle $\Psi_{\mathrm{RP}}$~\cite{Niida:2024ntm}:
\begin{equation}
\Delta\rho\{\phi^*\} = -\frac{4}{3} \langle \cos 2(\phi^* - \Psi_{\mathrm{RP}}) \rangle.
\end{equation}

A variant of this method uses the parent meson's azimuthal angle $\phi_p$ in place of $\Psi_{\mathrm{RP}}$, defining a helicity-inspired frame and yielding:
\begin{equation}
\Delta\rho_p\{\phi^*\} = -\frac{4}{3} \langle \cos 2(\phi^* - \phi_p) \rangle.
\end{equation}

\section*{Data-Driven Corrections via Meson-Level Mixing}

In spin-alignment measurements, detector acceptance and efficiency effects must be carefully accounted for to avoid biasing the extracted $\rho_{00}$. One common strategy relies on detailed Monte Carlo simulations with embedded vector-meson decays in a realistic detector model. Several complementary analytical techniques~\cite{analytic_accep_corr, Niida:2024ntm} have also been proposed.

Alternatively, a data-driven meson-level mixing technique can be used. Reconstructed vector mesons from one event are replaced by mesons from different events with matching kinematics ($p_T$, $\eta$, $\phi$). This preserves phase-space population and acceptance, while eliminating genuine physical spin alignment correlations. Polarization is encoded in daughter–axis correlations in the meson rest frame, which are destroyed by mixing.

The resulting $\cos\theta^*$ distribution from the mixed-event sample reflects only detector effects. A corrected distribution is then obtained as:
\[
\left( \frac{dN}{d\cos\theta^*} \right)_{\mathrm{corr}} = \frac{ \left( \frac{dN}{d\cos\theta^*} \right)_{\mathrm{same}} }{ \left( \frac{dN}{d\cos\theta^*} \right)_{\mathrm{mixed}} }.
\]

From this, the corrected angular moment is computed:
\[
\Delta\rho\{\theta^*\} = \frac{3}{2} \left( \langle \cos^2\theta^* \rangle - \frac{1}{3} \right).
\]

\begin{figure}[H]
  \centering
  \includegraphics[width=1.0\textwidth]{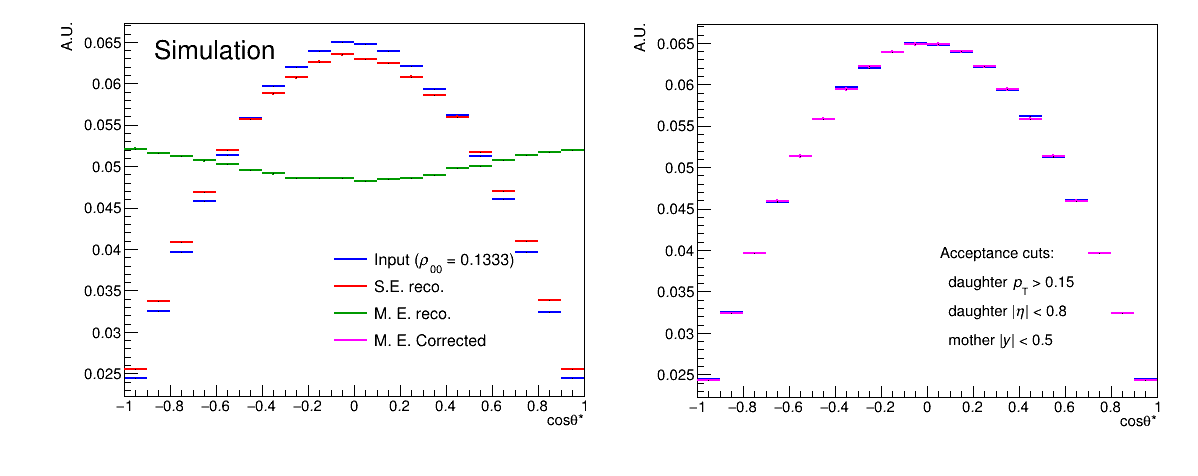}
  \caption{Two-panel validation of the meson-level mixing approach.
  Left: input angular distribution (blue), same-event reconstructed with detector effects (red), and mixed-event reconstructed (green).
  Right: input (blue) versus mixed-event corrected result (magenta).}
  \label{fig:meson_mixing}
\end{figure}

This meson-level mixing approach is fully data-driven, model-independent, and can be applied to both $\Delta\rho\{\theta^*\}$ and $\Delta\rho\{\phi^*\}$. It requires careful kinematic matching and 3D reweighting for acceptance correction. Figure~\ref{fig:meson_mixing} presents a two-panel validation of the meson-level mixing procedure using a Toy MC study, in which the input, same-event reconstructed, mixed-event acceptance reference, and corrected distributions are compared.

\subsubsection{Results}
\paragraph{Small Systems} 
Spin alignment studies provide a unique window into the hadronization and production mechanisms of vector mesons. Investigations of light-flavor vector meson spin alignment—such as for \( \rho \), \( K^{*0} \), and \( \phi \)—has been studied in a range of collision systems, from \( e^{+}e^{-} \) to \( pp \), spanning energies from fixed-target experiments to the LHC. These measurements, which use different choices of polarization reference frames, quantify the degree of spin alignment via the spin density matrix element \( \rho_{00} \). In the helicity frame, a value of \( \rho_{00} = 1/3 \) corresponds to no alignment, while deviations from this value signify spin alignment. These small systems serve not only as a benchmark for heavy-ion collisions but also test the underlying QCD production mechanisms.

Several theoretical models have been proposed to explain the observed spin alignment patterns in small systems. These include:

\begin{itemize}
  \item \textbf{Statistical spin counting}~\cite{Donoghue:1978yb}, which assumes equal probabilities for quark–antiquark pair creation in all helicity states. In this model, if the quark and antiquark spins are parallel, the resulting vector meson has helicity \( \lambda = \pm 1 \); if antiparallel, the system can yield either a pseudoscalar meson with probability \( f \), or a vector meson with probability \( 1 - f \). The spin density matrix element \( \rho_{00} \) is then expressed as:
  \[
      \rho_{00} = \frac{1}{2}(1 - \mathrm{P}/\mathrm{V}) = \frac{1 - f}{1 - 2f},
  \]
  where P/V is the pseudoscalar-to-vector meson production ratio. This model predicts \( \rho_{00} \in [0, 0.5] \).

  \item \textbf{QCD-inspired soft gluon fragmentation}~\cite{Augustin:1978wf,Farrar:1975yb} describes a scenario where a fast quark emits soft gluons that subsequently form quark–antiquark pairs. When the soft antiquark shares the same helicity as the original fast quark, their recombination favors mesons with helicity \( \lambda = \pm 1 \), leading to a prediction of \( \rho_{00} = 0 \).

  \item \textbf{Helicity conservation models}~\cite{Vainshtein:1977db} consider processes like \( q \rightarrow qV \), where the quark couples to the vector meson as a conserved current. In this case, helicity conservation leads predominantly to mesons with \( \lambda = 0 \), corresponding to \( \rho_{00} = 1 \).
\end{itemize}

\begin{figure}[h]
    \centering
    \includegraphics[scale=0.7]{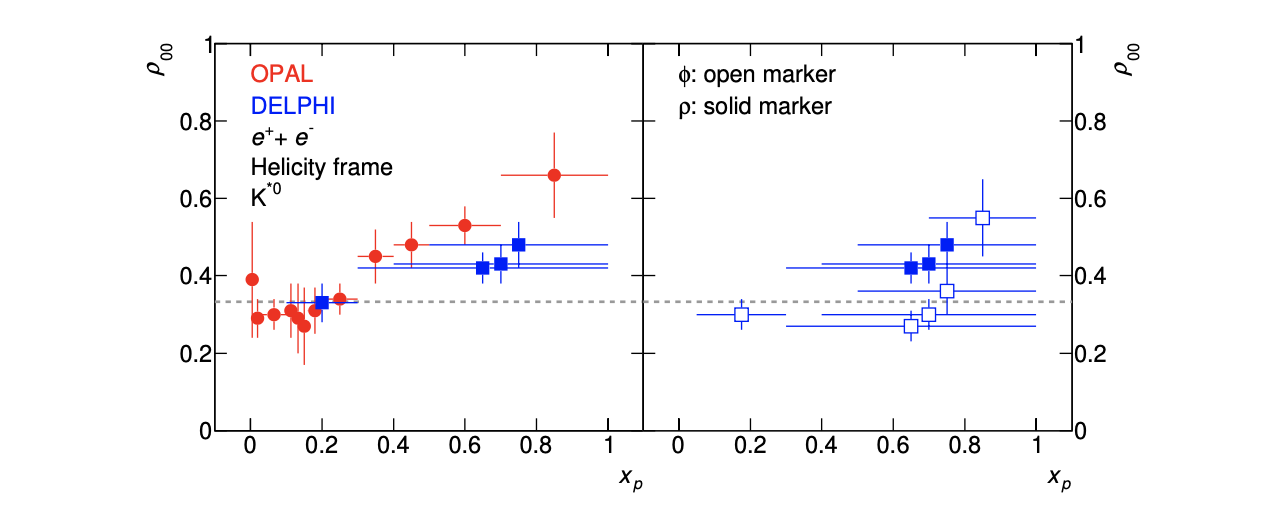}
    \caption{Spin density matrix element \( \rho_{00} \) as a function of $x_\mathrm{p}$ for $\rho$, K$^{*0}$ and $\phi$  meson in $e^{+}e^{-}$ collisions. Measurements were carried out in the helicity frame. Uncertainties on data points are the quadrature sum of statistical and systematics uncertainties.}
    \label{fig:rho_ee}
    \end{figure}

Experimental results in \( e^{+}e^{-} \) collisions from the OPAL~\cite{OPAL:1997vmw,OPAL:1997nwj} and DELPHI~\cite{DELPHI:1997ruo} collaborations provide detailed insight. As shown in Fig.~\ref{fig:rho_ee}, the measured \( \rho_{00} \) values for \( \rho \), \( K^{*0} \), and \( \phi \) mesons in the helicity frame exceed the unpolarized value of \( 1/3 \) at high scaled momenta \( x_p = p/p_\mathrm{beam} \), whereas at low \( x_p \), the values are consistent with \( \rho_{00} = 1/3 \). Specifically, deviations are seen for:
\begin{itemize}
    \item \( \rho \) and \( K^{*0} \) when \( x_p > 0.3 \),
    \item \( \phi \) mesons when \( x_p > 0.7 \).
\end{itemize}

Furthermore, DELPHI observed \( \rho_{00} > 0.5 \) for \( \phi \) mesons when \( x_E = E/E_{\mathrm{lab}} > 0.7 \), which violates the upper bound of the statistical spin counting model. Such high values of \( \rho_{00} \) imply that pseudoscalar mesons containing the primary quark are suppressed. These findings point toward a production mechanism favoring helicity-zero states, consistent with QCD-based models~\cite{Augustin:1978wf,Farrar:1975yb}.

Both OPAL and DELPHI also measured the off-diagonal matrix element \( \rho_{1,-1} \), which was found to be consistent with zero. This disfavors models involving coherence effects in meson formation. While no such coherence effects were observed within current experimental uncertainties, future precision measurements may help further probe this regime.

\begin{figure}[h]
    \centering
    \includegraphics[scale=0.7]{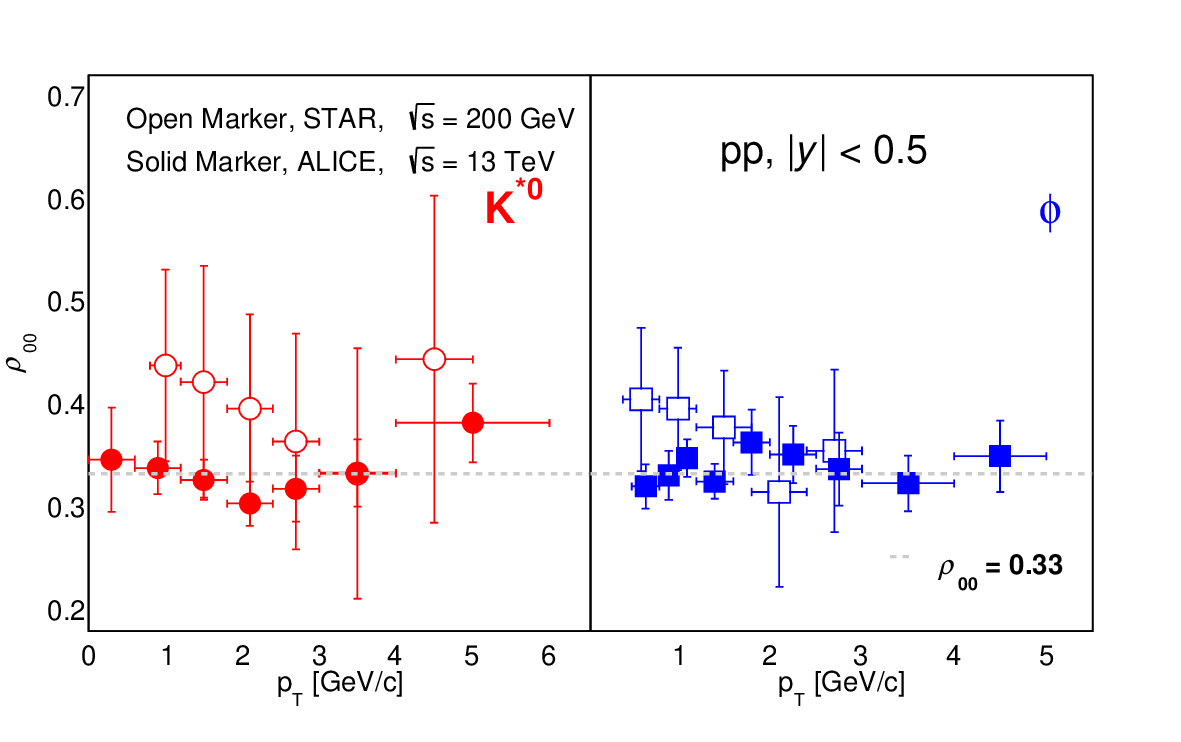}
    \caption{Spin density matrix element \( \rho_{00} \)  as a function of $p_\mathrm{T}$ for K$^{*0}$, $\phi$ in pp collisions at RHIC and LHC energies. Measurements were carried out in the production plane. Uncertainties on data points are the quadrature sum of statistical and systematics uncertainties.}
    \label{fig:rho_pp}
    \end{figure}

 The \( \rho_{00} \)  for K$^{*0,\pm}$ has also been measured in Kp and nC interactions~\cite{Chliapnikov:1972ei,Paler:1975qf,French-Soviet:1980hsq,Brussels-Genoa-Mons-Nijmegen-Serpukhov-CERN:1982pxs}, with respect to the direction perpendicular to the production plane.  The observed \( \rho_{00} \)  values are significantly higher than the expected value for the case with no spin alignment.
 This can be explained by the parton recombination model~\cite{Ayala:2009ab}, which attributes the spin alignment of the vector mesons to Thomas' precession of the quark spin during the hadronization process. Spin alignment measurements of K$^{*0}$ and $\phi$ vector mesons with respect to the production plane have also been performed in high-energy pp collisions at RHIC and the LHC energies as shown in Fig.~\ref{fig:rho_pp}. The \(\rho_{00}\)  consistent with 1/3 within current uncertainties in the measured $p_\mathrm{T}$ for both mesons at RHIC and LHC energies~\cite{Abelev:2008ag,Zhou:2019lun}.

Recent theoretical studies presented in Ref.~\cite{Chen:2020pty} the possibility of spin alignment for vector mesons such as $\rho$ and K$^{*0}$ in pp collisions using the helicity frame. The model predicts possible spin alignment in the fragmentation region. This region corresponds to high transverse momentum ($p_\mathrm{T}$) or large Feynman-$x$ ($x_\mathrm{F} = 2p_{z}/\sqrt{s}$) . The effect arises from the spin-dependent fragmentation function $D_\mathrm{1LL}$. Predictions for vector mesons at LHC energies are proposed using measured $\rho_{00}$ values from $e^{+}e^{-}$ collisions as input.
These theoretical expectations can be further tested with data from RHIC and LHC to improve our understanding of vector meson production mechanisms in the fragmentation region via spin-alignment studies. 

The observed spin alignment in these collisions challenges the conventional expectations for heavy-ion collisions. Furthermore, small collision systems can serve as an essential baseline to investigate whether local spin alignment effects or other local phenomena contribute to the observed spin alignment in heavy-ion collisions.

\paragraph{Heavy-ion collisions}
Spin alignment measurements in heavy-ion collisions at RHIC and LHC energies have opened a new avenue for probing spin-orbit interactions in QCD. In non-central collisions (impact parameter $\sim$ 3–10 fm), large angular momentum and strong magnetic fields are generated. While the magnetic field is transient, lasting only a few fm/$c$, the angular momentum persists and can influence the system evolution throughout its lifetime. Through spin-orbit coupling, this initial angular momentum may polarize quarks, resulting in a net spin alignment of produced vector mesons such as \( K^{*0} \) and \( \phi \). These measurements offer a valuable experimental probe of the early-stage dynamics and spin phenomena in the QCD medium created in heavy-ion collisions.

Several theoretical frameworks—ranging from quark recombination to hydrodynamic models—have been developed to interpret the spin alignment phenomenon. The quark recombination model predicts \( \rho_{00} < 1/3 \) when vector mesons are formed from two polarized quarks, particularly at low transverse momentum and midrapidity where recombination dominates. In contrast, the quark polarization model, which becomes relevant at high \( p_T \) and forward rapidity, expects \( \rho_{00} > 1/3 \) due to fragmentation of polarized quarks. Additionally, it predicts a stronger spin alignment effect for lighter mesons (e.g., \( K^{*0} \)) than for heavier ones (e.g., \( \phi \)) due to an  inverse mass dependence of quark polarization.

An alternative mechanism involves the formation of a coherent \( \phi \)-meson field, which can also result in \( \rho_{00} > 1/3 \), with a predicted dependence on the center-of-mass energy. This mechanism has been invoked to explain the enhanced \( \phi \)-meson spin alignment at RHIC. Furthermore, magnetic fields may modify spin alignment: increasing \( \rho_{00} \) for neutral mesons and decreasing it for charged ones. However, recent hydrodynamic calculations suggest \( \rho_{00} < 1/3 \), indicating a possible contribution from vorticity and collective effects rather than quark-level mechanisms alone.

\begin{figure}[h]
    \centering
    \includegraphics[scale=0.99]{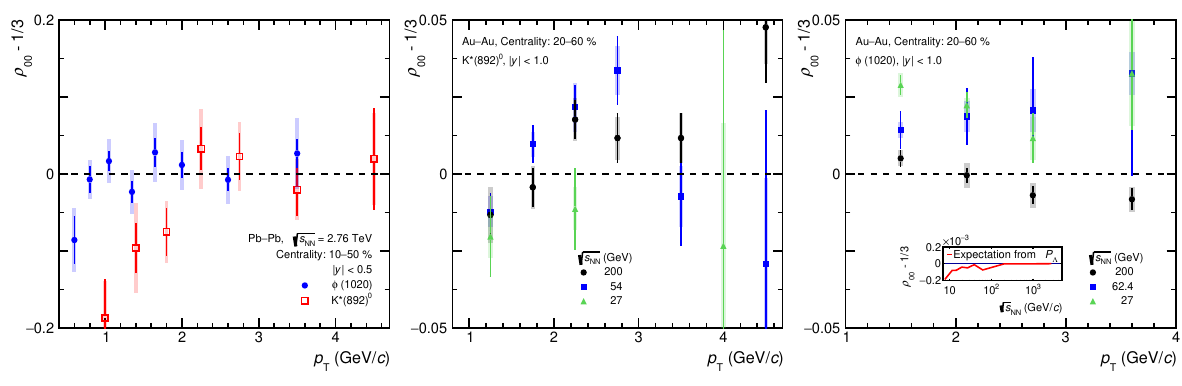}
    \caption{Spin density matrix element \( \rho_{00} \)  as a function of $p_\mathrm{T}$ for K$^{*0}$, $\phi$ in heavy-ion collisions (Au--Au, Pb-Pb) at RHIC and LHC energies. Measurements were carried out with respect to the normal to the event plane. Uncertainties on data points are the quadrature sum of statistical and systematics uncertainties.}
    \label{fig:rho_pt_AA}
    \end{figure}

Figure~\ref{fig:rho_pt_AA} shows \( \rho_{00} - 1/3 \) as a function of \( p_T \) for \( K^{*0} \) and \( \phi \) mesons in semi-central (20--40\% and 10--50\%) Au--Au and Pb--Pb collisions at RHIC and LHC. The spin alignment is evaluated with respect to the event plane, and the measurements span \( |y| < 1.0 \) (RHIC) and \( |y| < 0.5 \) (LHC). At low \( p_T \), both mesons show \( \rho_{00} < 1/3 \), with the effect stronger for \( K^{*0} \). At high \( p_T \), the values approach 1/3, indicating a lack of alignment. RHIC results for \( \phi \) mesons show an opposite trend \( \rho_{00} > 1/3 \) compared to the LHC.

These trends are consistent with the recombination model, where spin alignment is strongest at low \( p_T \), and with the expectation that \( K^{*0} \) mesons exhibit larger effects than \( \phi \) mesons due to mass and quark content. The significantly larger spin alignment observed for vector mesons compared to global \( \Lambda \) hyperon polarization suggests different underlying mechanisms.

\begin{figure}[htbp]
    \centering
    \includegraphics[scale=0.8]{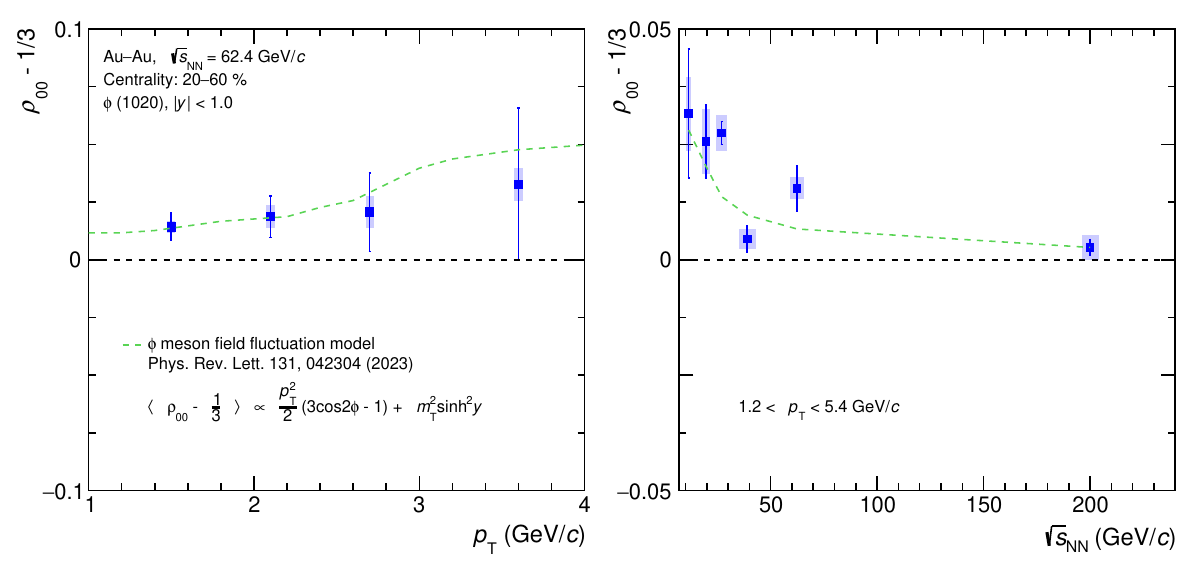}
    \caption{$\rho_{00} \ -\frac{1}{3}$  as a function of $p_\mathrm{T}$ (left) and  center-of mass energy \snn (right)  are shown for the $\phi$-meson in  Au--Au collisions. Measurements are compared  to coherent $\phi$- meson field calculations. Uncertainties on data points are the quadrature sum of statistical and systematic uncertainties.}
    \label{fig:rho_AA_model}
    \end{figure}

A more detailed comparison of \( \phi \)-meson spin alignment with predictions from the coherent field model is shown in Fig.~\ref{fig:rho_AA_model}~\cite{Sheng:2022wsy,Sheng:2023urn}. The model provides a good description of the \( p_T \)-differential and energy-dependent behavior of \( \rho_{00} \) in the range \( 1.2 < p_T < 5 \) GeV/\emph{c}. Despite small differences in rapidity coverage, the model captures the trends observed at both RHIC and LHC, suggesting that the coherent field scenario offers a unifying explanation for \( \phi \)-meson spin alignment across collision energies.

It is important to clearly distinguish between the experimentally established observations and their theoretical interpretations in the context of vector-meson spin alignment. Experimentally, measurements in heavy-ion collisions unambiguously demonstrate deviations of the spin-density matrix element \( \rho_{00} \) from the unpolarized value of 1/3 for several vector mesons, within specific kinematic regions and collision centralities. These deviations constitute robust evidence that the spin degrees of freedom of produced hadrons retain sensitivity to the medium created in relativistic nuclear collisions.
However, the underlying physical mechanisms responsible for the observed spin alignment remain model dependent. Interpretations invoking medium vorticity, strong electromagnetic fields, spin–orbit interactions during hadronization, or polarization transfer in the hadronic phase rely on different theoretical assumptions and dynamical inputs, and no single mechanism has yet been shown to provide a quantitative and unified description of all existing measurements. At present, the available data do not allow an unambiguous separation of these effects, nor a precise determination of their relative contributions. Future measurements with improved precision, extended kinematic coverage, and complementary observables, together with fully dynamical theoretical calculations, will be essential to establish the dominant origin of spin alignment in high-energy nuclear collisions.

\subsubsection{Discussion and Future Prospects}

\begin{itemize}
\item \textbf{Rapidity-, azimuth-, and species-differential spin-alignment measurements}  
A broad range of recent theoretical approaches predict that vector-meson spin alignment is predominantly governed by local mechanisms rather than global vorticity. These include  
(i) strong-force field or ``glasma'' scenarios, where fluctuating chromoelectric and chromomagnetic
fields generate rapidity-dependent quark spin correlations at very early times~\cite{KumarGlasma23,ShengReview25};  
(ii) gradient-driven or second-order spin-hydrodynamic frameworks, in which anisotropic shear and expansion gradients induce polarization of spin-1 fields and generate characteristic $\cos(2\phi)$ modulations in $\rho_{00}$~\cite{KumarGradient24,GoncalvesTorrieri22}; and  
(iii) holographic models, where rotation or electromagnetic fields modify the longitudinal and transverse spectral components of vector mesons in a polarization-dependent manner~\cite{HolographySpin24,WangHeavyHolo24}.  

Despite their conceptual differences, all these models share a robust qualitative prediction:  
\emph{vector meson spin alignment should exhibit strong rapidity dependence and pronounced azimuthal modulation relative to the event plane.}  
Such dependencies arise naturally from local spin–correlation fluctuations, longitudinal color-field structures, and polarization-sensitive spectral anisotropies.  

High-statistics measurements of $\rho_{00}(y,p_{T})$ and $\rho_{00}(\phi-\Psi_{n})$ therefore provide a direct experimental handle on the space--time localization of these effects. The extended rapidity coverage and multi-differential capabilities of upcoming detector upgrades (LHC Run~3/4, RHIC sPHENIX) are ideally suited to discriminate among strong-field-driven, gradient-driven, and holographic mechanisms.

A particularly sensitive test comes from \emph{species dependence}.  
Holographic QCD models predict that vector mesons with different quark content experience \emph{different polarization-dependent in-medium modifications} even under the same local conditions. The ratio of longitudinal to transverse spectral strength---which directly determines $\rho_{00}$---differs for strange ($\phi$) and charm (J/$\psi$) mesons due to their distinct holographic flavor embeddings~\cite{HolographySpin24,WangHeavyHolo24,FujitaSoftWall09}.  
Consequently, $\phi$ and J/$\psi$ are expected to exhibit \emph{different spin-alignment patterns at the same rapidity and centrality}.  
Observation of such differences would strongly support a picture in which polarization-dependent spectral modifications dominate over a universal spin–vorticity coupling.

Simultaneous measurements of $\rho_{00}$ for $\phi$ and J/$\psi$ in matched $(y,p_{T})$ bins thus offer a clean and experimentally tractable way to test holographic predictions. With improved strange-hadron reconstruction at RHIC and high-statistics quarkonium samples expected from LHC Run~3/4, these differential species comparisons are now experimentally within reach.

\item \textbf{Prospect of $\rho$ meson spin alignment}
The short-lived $\rho$ meson plays a central role in connecting early-time quark spin correlations to late-stage hadronic dynamics. Owing to its strong coupling to pions and lifetime comparable to the hadronic rescattering timescale, the $\rho$ is highly sensitive to \emph{spin relaxation} and \emph{hadronic interactions}. Kinetic-theory studies show that the spin-density matrix of the $\rho$ can evolve significantly during the hadronic stage, with the final value of $\rho_{00}$ depending on the interplay between initial polarization and pion--$\rho$ spin-dependent scattering~\cite{RhoKinetic24}. Transport simulations further demonstrate that the energy dependence of rescattering alters the measured spin alignment of $\rho$, $\phi$, and $K^{*0}$ differently across beam energies~\cite{Rescattering25}.

Low-energy programs (RHIC BES-II, FAIR, NICA) provide a unique opportunity to measure the energy dependence of $\rho_{00}$ for the $\rho$ meson, in parallel with $\phi$ and $K^{*0}$. Because $\rho$ is more sensitive to hadronic evolution while $\phi$ is more reflective of partonic spin structure, comparing these species across $\sqrt{s_{NN}}$ offers a powerful method to disentangle hadronic and partonic contributions to spin alignment. The $\rho$ is therefore a key observable for exploring spin transport in the hadronic phase and for isolating early-time spin correlations.

\item \textbf{Probing the QCD phase diagram with spin alignment}
Spin alignment of vector mesons may provide sensitivity to the QCD phase structure complementary to fluctuation and flow observables. Spin-transport coefficients and relaxation times are expected to change across the hadron-gas to QGP transition, and may exhibit enhanced fluctuations near a possible critical point~\cite{RischkeReview24,ShengReview25}. In addition, strong-field models predict that the magnitude and character of field-induced quark spin correlations vary with the degree of deconfinement and baryon density~\cite{KumarGlasma23}.

A systematic beam-energy scan of $\rho_{00}$ for multiple meson species---particularly $\rho$, $\phi$, and J/$\psi$---can:  
(i) search for non-monotonic behavior associated with the critical region;  
(ii) identify the transition from hadronic to partonic spin transport; and  
(iii) correlate spin alignment with global hyperon polarization to obtain a multi-observable handle on the QCD phase diagram.  
Future facilities with enhanced PID and rapidity coverage will enable spin alignment to become a quantitative probe of spin transport across the phase boundary.
\end{itemize}

\subsection{Collectivity}\label{chap3:sub:fourth}
The azimuthal anisotropy coefficients ($v_{_n}$) are essential tools for investigating the collective behavior of matter produced in relativistic heavy-ion collisions~\cite{Snellings:2011sz}. These coefficients capture the azimuthal modulation of particle emission with respect to the reaction plane and serve as the clearest experimental signatures of collective flow. Nonzero values of $v_{n}$ indicate significant interactions among the system constituents, which drive the system toward thermal equilibrium. Thus, the magnitude and pattern of anisotropic flow reflect the degree of thermalization achieved in the early stages of the collision and provide indirect access to the properties of the QGP. Short-lived resonances such as $K^{*0}$ and $\phi$ mesons are particularly sensitive probes of the hadronic phase, as their decay products can be affected by rescattering and regeneration processes. The measurement of resonance flow coefficients thus offers unique insights into the time evolution of the system and the interplay between partonic and hadronic degrees of freedom~\cite{ALICE:2017ban,ALICE:2014jbq}. $v_n$ measurements of resonances also enable separation of partonic collectivity from hadronic distortions and sharpen inferences about QGP transport and thermalization  timescales~\cite{PhysRevC.107.064910}.


\subsubsection{Azimuthal anisotropy of $\phi$ meson: A probe of partonic collectivity}\label{chap3:subsub:one}
Although elliptic flow ($v_{_n}$) originates early in the system evolution, its final strength can be altered by late-stage hadronic interactions. The $\phi$ meson, however, with its small hadronic cross section and early freeze-out time, retains information from the partonic phase, making it a clean probe of early dynamics. Owing to its mass, comparable to that of protons and $\Lambda$ baryons, the $\phi$ meson's $v_{_n}$ measurements allow a meaningful comparison with other hadron species, facilitating the separation of mass effects from baryon-meson differences in $v_{n}(p_{T})$. Therefore, the $\phi$ meson $v_{_n}$ can be considered as a reliable probe for studying the early conditions of the system formed in nucleus-nucleus collisions~\cite{Schenke:2010rr,PHENIX:2022rvg,PHENIX:2007tef}.

\subsubsection{Flow measurement method for $\phi$  meson}\label{chap3:subsub:two}
The azimuthal distribution of particles relative to the event plane is expanded in a Fourier series~\cite{Poskanzer:1998yz}:
\begin{equation}
\frac{dN}{d(\phi - \Psi_n)} \propto 1 + 2 \sum_{n \geq 1} v_n \cos[n(\phi - \Psi_n)] + 2 \sum_{n \geq 1} v_n \sin[n(\phi - \Psi_n)],
\label{fPsi_phi}
\end{equation}
where $\phi$ is the azimuthal angle of the particle, $\Psi_n$ is the event plane angle, and $v_n = \langle \cos[n(\phi - \Psi_n)] \rangle$.

Two principal methods are used to extract $v_n$: the event plane method and the invariant mass method. In the event plane method, particles are first identified, and their yields are analyzed as a function of $\phi - \Psi_n$. In contrast, the invariant mass method evaluates $\langle \cos[n(\phi - \Psi_n)] \rangle$ as a function of invariant mass, extracting the signal $v_n$ after background subtraction.

\begin{figure}[h]
\begin{center}
\includegraphics[scale=0.5]{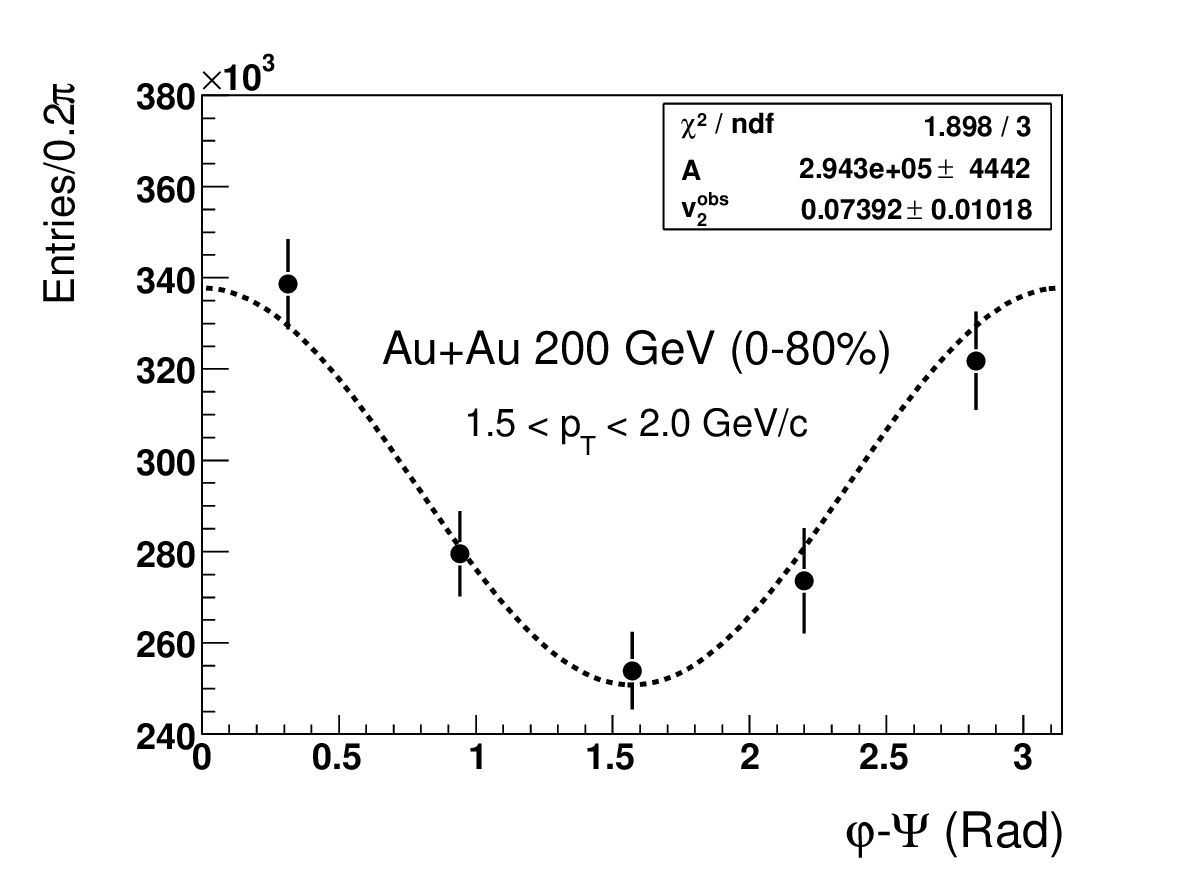}
\caption{(Color online) $\phi$-meson yields as a function of $\phi - \Psi_n$ bins. The distribution issourav.kundu@cern.ch fitted with Eq.~~(\ref{fPsi_phi}) }
\label{v2_phi_method_1}
\end{center}
\end{figure}

In the event plane method, $v_n$ is obtained by fitting the yield distributions with Eq.~~(\ref{fPsi_phi}). An example fit is shown in Fig.~~\ref{v2_phi_method_1}. Particle yields are determined in invariant mass bins, typically using Breit--Wigner fits for the $\phi$ meson.

In the invariant mass method, $v_2$ is calculated as a function of the invariant mass of the daughter particles ($K^+$ and $K^-$). Since signal and background cannot be separated event-by-event, the total $v_2^{\text{Sig+Bg}}$ includes both components:

\begin{eqnarray}
v_2^{\rm Sig+Bg}(M_{\rm inv}) & = & \left\langle \cos[2(\phi - \Psi_2)] \right\rangle{M_{\rm inv}}, \label{finvmass_v2} \\
v_2^{\rm Sig+Bg}(M_{\rm inv}) & = & v_2^{\rm Sig} \cdot \frac{\rm Sig}{\rm Sig+Bg}(M_{\rm inv}) + v_2^{\rm Bg}(M_{\rm inv}) \cdot \frac{\rm Bg}{\rm Sig+Bg}(M_{\rm inv}), \label{finvmass_v2_fit_B} \\
v_2^{\rm Bg}(M_{\rm inv}) & = & p_0 + p_1 M_{\rm inv} + p_2 M_{\rm inv}^2. \label{finvmass_v2_fit_C}
\end{eqnarray}

\begin{figure}[h]
\begin{center}
\includegraphics[scale=0.52]{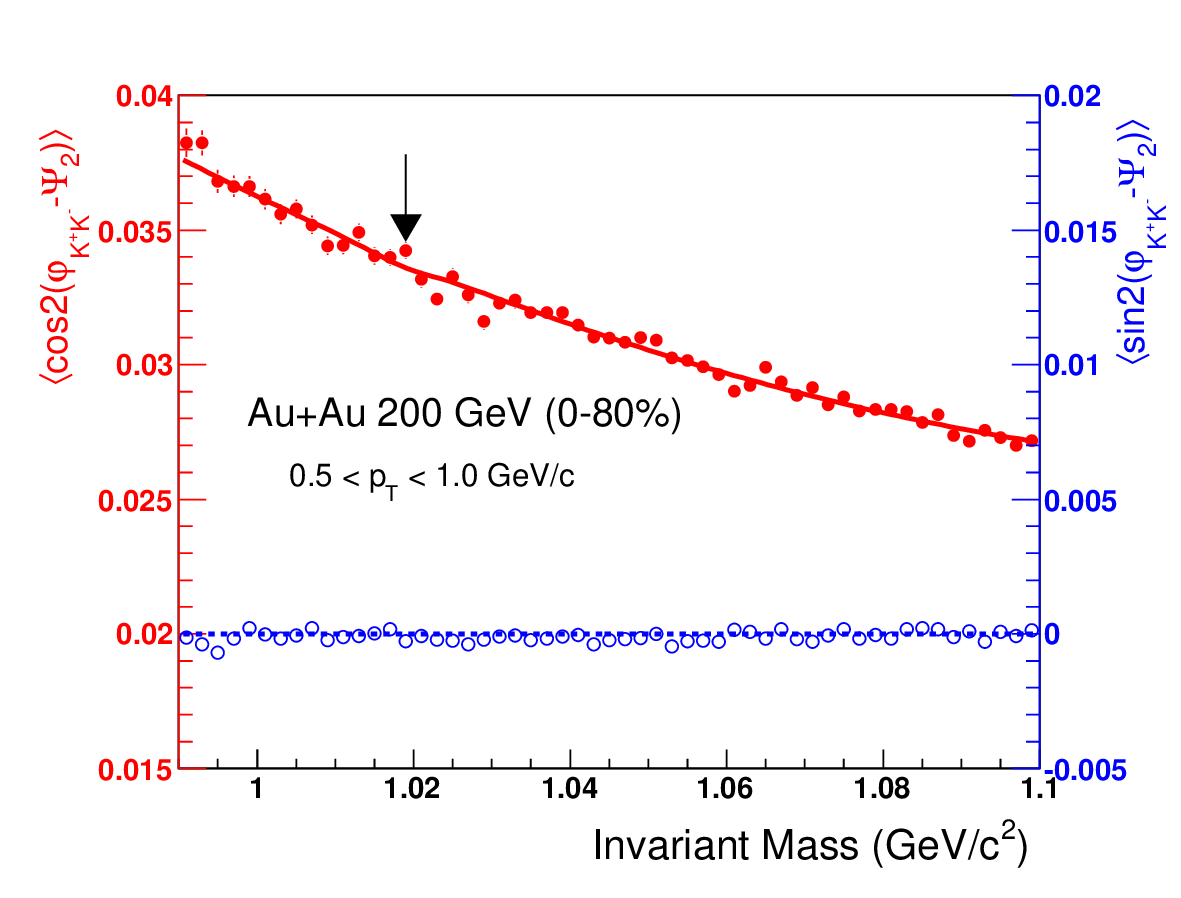}
\caption{(Color online) $\left\langle \cos\left[2(\phi_{K^{+}K^{-}}-\Psi_2)\right] \right\rangle$ and $\left\langle \sin\left[2(\phi_{K^{+}K^{-}}-\Psi_2)\right] \right\rangle$  as a function of invariant mass of the daughter particles pairs ($K^+$ and $K^-$) in Au+Au collisions at $\sqrt{s_{NN}}$ = 200 GeV.  }
\label{v2_phi_method_2}
\end{center}
\end{figure}

Figure~\ref{v2_phi_method_2} shows an example fit using this method. The cosine term reflects the elliptic flow signal, while the sine term remains near zero as expected from the symmetry of the collision geometry. The final $v_2$ is corrected for event plane resolution.

\subsubsection{Directed flow (first order flow harmonic) of $\phi$ meson}\label{chap3:subsub:three}
\begin{figure}[h]
\begin{center}
\includegraphics[scale=0.42]{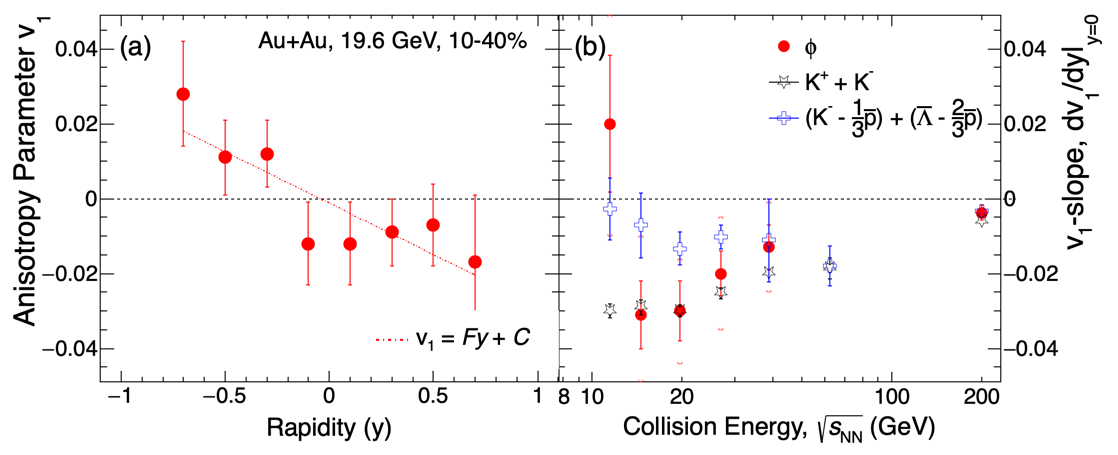}
\caption{(Color online) Rapidity (y) dependent of $v_{1}$ for  $\phi$ meson in 10-40$\%$ centrality, Au+Au collisions at $\sqrt{s_{NN}}$ = 19.6 GeV (left panel). $v_1$-slope as a function of collision energy is shown for $\phi$, $K^{+}+K^{-}$ and ($K^{-}-\frac{1}{3}\bar{p}) + (\bar{\Lambda}-\frac{2}{3}\bar{p})$ (right panel)~\cite{Adamczyk_2018,STAR:PRL.112.162301,NAYAK2021121855}.}
\label{v1_phi_snn}
\end{center}
\end{figure}

The first-order flow coefficient, known as directed flow ($v_{1}$), characterizes the collective sideward motion of produced particles and nuclear fragments in heavy-ion collisions. 
The directed flow is also sensitive to the early stage of the collision. It is believed to be generated during the nuclear passage time, and therefore it probes the pre-equilibrium stage in a heavy-ion collision. Directed flow of identified hadrons has been measured extensively in the heavy-ion collision, particularly at RHIC. As an example,  $v_{1}$ of $\phi$ mesons as a function of rapidity is shown in Fig.~\ref{v1_phi_snn}. A hydrodynamic calculation with a tilted initial QGP source can explain the observed negative $v_{1}$ slope or “anti-flow”  near mid-rapidity~\cite{PhysRevC.81.054902}. A linear fit to the data was used to extract the slope $dv_{1}/dy$.

The directed flows ($v_{1}$) of identified hadrons are expected to follow the coalescence sum rule when the created matter is initially in degrees of freedom of parton and then hadronizes through quark coalescence. As per coalescence sum rule, $v_{1}$ of the mesons or baryons is the sum of $v_{1}$ of their constituent quarks. This sum rule has been tested on hadrons carrying light quarks. In this review paper, we have tested this sum rule on $\phi$ mesons as shown in Fig.~\ref{v1_phi_snn}. Figure~\ref{v1_phi_snn} shows the slope $dv_{1}/dy$ as a function of the energy of the center of mass in Au+Au collisions. The red marker represents $dv_{1}/dy$ of $\phi$ mesons calculated by fitting $v_{1}(y)$ of $\phi$ mesons as shown in Fig.~\ref{v1_phi_snn}. The $dv_{1}/dy$ of $s$ and $\bar{s}$ quarks are calculated using measured $v_{1}$ of $K^{-}$, $\bar{p}$ and $\bar{\Lambda}$ as shown in equations~\ref{coal_s} and \ref{coal_s_bar}~\cite{Adamczyk_2018}. If $\phi$ mesons are produced from coalescence of $s$ and $\bar{s}$ quarks, then $v_{1}$ of $\phi$ mesons will be the simple sum of $v_{1}$ of $s$ and $\bar{s}$.

The blue marker represents the sum of $s$ and $\bar{s}$ $dv_{1}/dy$ as a function of $\sqrt{s_{NN}}$ and compared to $dv_{1}/dy$ of $\phi$ mesons. For energies greater than 39 GeV, coalescence sum rule predicted  $dv_{1}/dy$ of $\phi$ mesons from $s$ and $\bar{s}$ coalescence is consistent with the measured $dv_{1}/dy$ of $\phi$ mesons. For energies greater lower than 39 GeV, consistency is not very good, although we can not make a conclusion due to large uncertainties on measured  $dv_{1}/dy$ of $\phi$ mesons. In the hadronic medium, $\phi$ mesons can be produced through the coalescence of $K^{+}$ and $K^{-}$. Hence, we have calculated $dv_{1}/dy$ of $\phi$ mesons  using $dv_{1}/dy$ of $K^{+}$ and $K^{-}$ under the assumption of coalescence sum rule. It is found that $dv_{1}/dy$ of $\phi$ mesons calculated from $K^{+}$ and $K^{-}$ is also consistent with measured $dv_{1}/dy$ of $\phi$ mesons, except at 11.5 GeV~\cite{Adamczyk_2018,NAYAK2021121855}. This is very striking observation, as we know from other measurements at RHIC that $\phi$ meson are primarily produced from the coalescence of $s$ and $\bar{s}$ quarks. At lower energies (< 11.5 GeV), there is a hint of sign change of the $\phi$ meson $v_1$-slope and a similar prediction is shown by the UrQMD models [will add the ref.].


\begin{equation}
v_{1}(s) =v_{1}(K^{-})-\frac{v_{1}(\bar{p})}{3}.
\label{coal_s}
\end{equation}
\begin{equation}
v_{1}(\bar{s}) =v_{1}(\bar{\Lambda})-\frac{2v_{1}(\bar{p})}{3}.
\label{coal_s_bar}
\end{equation}

\subsubsection{Elliptic flow (second order flow harmonic) of $\phi$ meson}\label{chap3:subsub:four}

\begin{figure}[h]
\begin{center}
\includegraphics[scale=0.9]{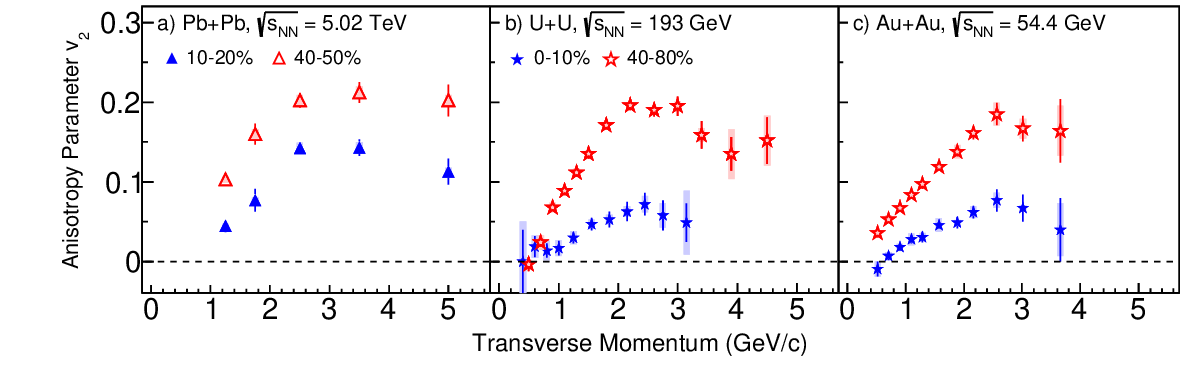}
\caption{(Color online) Elliptic flow $v_{2}$ as a function of $p_{T}$ for different centrality classes in A+A collisions at different center of mass energies~\cite{ALICE:2018yph,STAR:107.024912,STAR:103.064907}.}
\label{v2_cen}
\end{center}
\end{figure}

\begin{figure}[h]
\begin{center}
\includegraphics[scale=0.5]{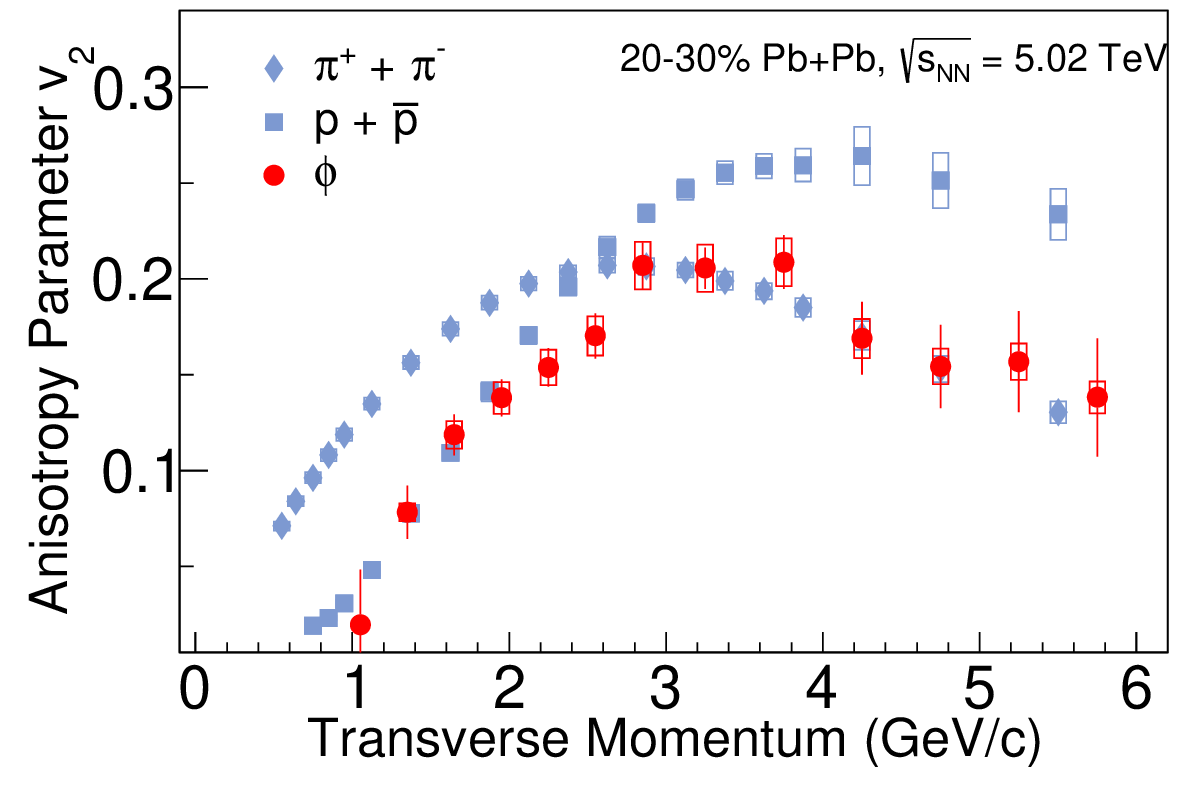}
\caption{(Color online) Elliptic flow $v_{2}$ of pions, protons and $\phi$ mesons as a function of transverse momentum $p_{T}$ in 20-30$\%$ Pb+Pb collisions at $\sqrt{s_{NN}}$ = 5.02 TeV~\cite{ALICE:2018yph}.}
\label{v2_phi_pt}
\end{center}
\end{figure}

Figure~\ref{v2_cen} shows $v_{2}$ as a function of $p_{T}$ for different centrality classes in A+A collisions at different center of mass energies. A clear centrality dependence of $v_{2}(p_{T})$ is observed for $\phi$ mesons similar to that of other identified light and strange hadrons previously measured by the STAR experiment~\cite{Adamczyk_2013}. The values of $v_{2}$ are found to be larger in peripheral collisions compared to those in central collisions. This observation is consistent with an interpretation in which the final momentum anisotropy is driven by the initial spatial anisotropy.\\
Figure~\ref{v2_phi_pt} shows $v_{2}$ of pions, protons and $\phi$ mesons as a function of transverse momentum $p_{T}$. The measurement was performed in mid-rapidity in 20-30$\%$ Pb+Pb collisions at $\sqrt{s_{NN}}$ = 5.02 TeV~\cite{ALICE:2018yph}. Figure~\ref{v2_phi_pt} shows that at low transverse momentum ($p_{T}$ $<$ 2 to 2.5 GeV/c), there is a clear mass ordering of $v_{2}$ among the measured particles, $v_{2}$ of pions ($m_{\pi}$$\approx$139 MeV) is higher than that of protons($m_{p}$$\approx$938 MeV) and $\phi$ mesons ($m_{\phi}$$\approx$1019 MeV). The observed mass ordering at low $p_{T}$ region can be explained by the hydrodynamic model calculations. It is observed that in the intermediate-pT region (2.5 $< p_{T} < $6.0 GeV/c), the $p_{T}$-differential $v_{2}$ of pions and $\phi$ mesons are the same, while protons show larger $v_{2}$. This observation in the intermediate $p_{T}$ was successfully reproduced by models in which a quark-coalescence mechanism is considered to be the dominant process for hadronization.

\begin{figure}[h]
\begin{center}
\includegraphics[scale=0.8]{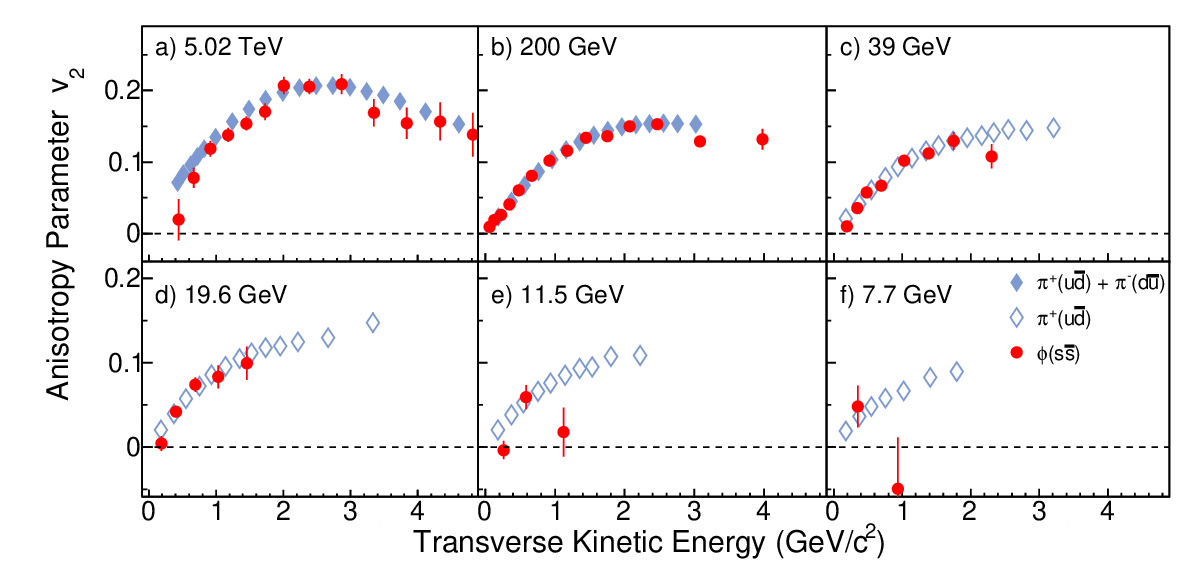}
\caption{(Color online) Elliptic flow $v_{2}$ is plotted as a function of transverse kinetic energy, $m_{T}$-$m_{0}$, for $\phi$ and $\pi$ mesons in collisions of A+A in $\sqrt{s_{NN}}$ = 5.02 TeV, 200 GeV, 39 GeV, 19.6 GeV, 11.5 GeV and 7.7 GeV~\cite{ALICE:2018yph,STAR:107.024912,Adamczyk_2013,PHENIX:2007tef}. }
\label{v2_phi_ncq}
\end{center}
\end{figure}

In Figure~\ref{v2_phi_ncq}, the $v_{2}$ is plotted as a function of transverse kinetic energy, $m_{T}$-$m_{0}$, for $\phi$ and $\pi$ mesons in collisions of Pb + Pb and Au + Au in $\sqrt{s_{NN}}$ = 5.02 TeV, 200 GeV, 39 GeV, 19.6 GeV, 11.5 GeV and 7.7 GeV~\cite{ALICE:2018yph,Adamczyk_2013,PHENIX:2007tef}. It is clear from Fig.~\ref{v2_phi_ncq} that the $v_{2}$ of hadrons (like $\phi$) consisting only of strange quarks is similar to that of $\pi$ consisting of light quarks. However, unlike $\pi$, the $\phi$ mesons do not participate strongly in hadronic interactions, suggesting that the major part of collectivity develops during the partonic phase in A+A collisions for $\sqrt{s_{NN}}$ $\geq$ 7.7 GeV. In the latest BES-II data, we expect improved $\phi$ meson flow measurement which can give an idea about the scaling.

\subsubsection{Triangular Flow ($v_3$): Sensitivity to Initial-State Fluctuations}
\begin{figure}[h]
\begin{center}
\includegraphics[scale=0.8]{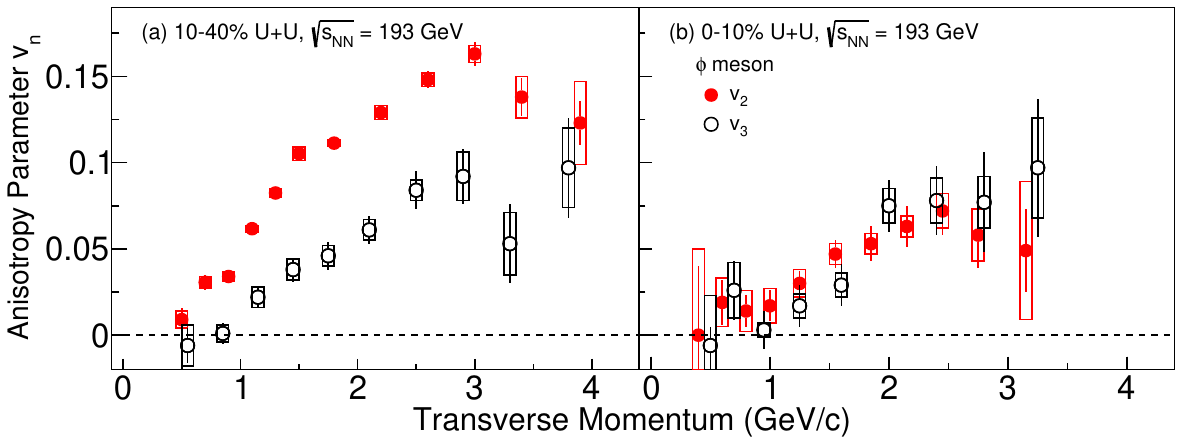}
\caption{(Color online) $v_{2}$ and $v_{3}$ in U+U collisions at $\sqrt{s_{NN}}$ = 193 GeV for central collisions of 10-40\% and 0-10\%~\cite{STAR:103.064907}}
\label{v2_v3_comp}
\end{center}
\end{figure}

\begin{figure}[h]
\begin{center}
\includegraphics[scale=0.8]{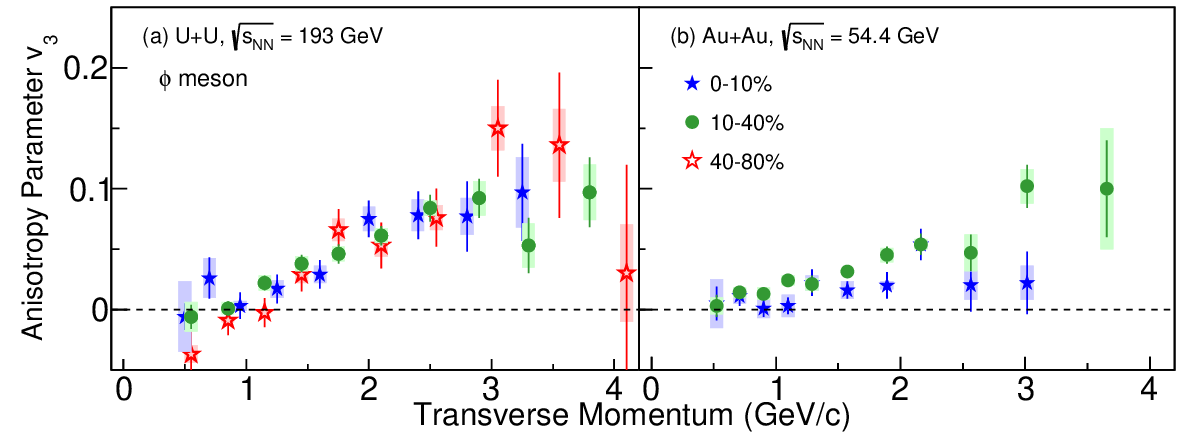}
\caption{(Color online)  The centrality dependence of $\phi$ meson $v_{3}$ in U+U collisions at $\sqrt{s_{NN}}$ =193 GeV and Au+Au collisions at $\sqrt{s_{NN}}$ = 54.4 GeV~\cite{STAR:103.064907,STAR:107.024912}.}
\label{v3_cen}
\end{center}
\end{figure}

The triangular flow coefficient ($v_3$), driven by initial-state fluctuations in the nucleon positions and gluon fields, provides additional insight into the initial conditions and viscosity of the medium. Unlike $v_2$, $v_3$ is not directly correlated with the overall collision geometry and is less sensitive to the impact parameter. The $\phi$ meson's $v_3$, as shown in Figures~\ref{v2_v3_comp} and \ref{v3_cen}, is comparable to that of other hadrons in central collisions, where fluctuations dominates~\cite{STAR:103.064907}. These $v_{n}$ measurements indicates most of the observed $\phi$ flow is generated before hadronization, with hadronic effects providing, at most, modest late-stage modifications.

\subsubsection{Probing quark composition of $f_{0}(980)$ through elliptic anisotropy measurement} \label{chap3:sub:f0}
\begin{figure}[h]
\begin{center}
\includegraphics[scale=0.7]{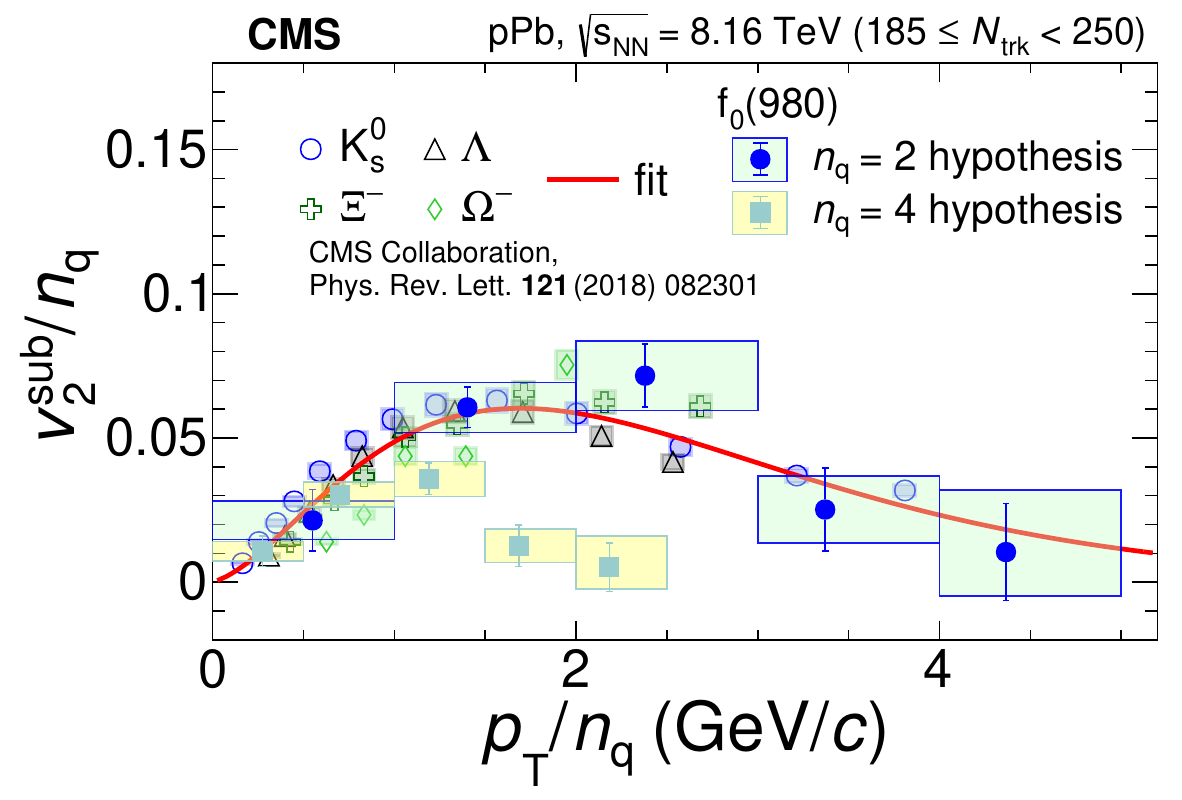}
\caption{(Color online)  The scaled elliptic flow, $v_{2}/n_q$, of the $f_0(980)$ state, evaluated under the $n_q=2$ and $n_q=4$ hypotheses, is shown as a function of $p_T/n_q$ and compared with that of the strange hadrons $K^0_S$, $\Lambda$, $\Xi^{-}$, and $\Omega$ in high-multiplicity p-Pb collisions~\cite{CMS:f0}. Error bars represent statistical uncertainties, while the shaded bands indicate systematic uncertainties. The red curve corresponds to the NCQ-scaling parameterization obtained from a simultaneous fit to the $K^0_S$, $\Lambda$, $\Xi^{-}$, and $\Omega$ data.}
\label{v2_ncq_f0}
\end{center}
\end{figure}

Hadrons are conventionally classified as quark-antiquark ($q\bar q$) mesons or three-quark ($qqq$) baryons; however, quantum chromodynamics also allows for more exotic configurations, including tetraquarks and meson-molecule states ($qq\bar q\bar q$), pentaquarks ($qqqq\bar q$), and dibaryons ($qqqqqq$)~\cite{f0_into:1,f0_into:2,f0_into:3}. Among these candidates, the scalar meson $f_0(980)$, discovered over five decades ago, remains one of the most debated states, with proposed interpretations ranging from a conventional $q\bar q$ meson to a compact tetraquark, a loosely bound $K\bar K$ molecule, or a hybrid configuration with explicit gluonic degrees of freedom. Relativistic heavy-ion collisions offer a unique opportunity to discriminate among these scenarios by probing the internal structure of the $f_0(980)$ through measurements of its elliptic anisotropy in momentum space.
Within the quark coalescence framework, the elliptic anisotropy of hadrons formed at hadronization is directly related to the collective flow of their constituent quarks. To leading order, the hadron elliptic flow can be approximated as
\begin{equation}
v_{2}^{\mathrm{had}}(p_T) \simeq n_q \, v_{2}^{q}\!\left(\frac{p_T}{n_q}\right),
\end{equation}
where $n_q$ denotes the number of constituent quarks and $v_{2}^{q}$ represents the partonic elliptic flow. This relation gives rise to the empirical number-of-constituent-quark (NCQ) scaling observed for a wide range of identified hadrons.
In this context, the elliptic flow of the $f_0(980)$ offers a sensitive probe of its internal quark structure. A conventional $q\bar q$ configuration would be expected to exhibit meson-like scaling behavior with $n_q=2$, whereas a compact tetraquark interpretation would naturally lead to a larger elliptic flow consistent with an effective $n_q=4$.
Fig.~\ref{v2_ncq_f0} shows measurement of $v_{2}$ of the $f_0(980)$ in p-Pb collisions at  $\sqrt{s_{NN}}$ = 8.16 TeV by CMS collaboration~\cite{CMS:f0}.  The scaled elliptic flow, $v_{2}/n_q$, of the $f_0(980)$ state, evaluated under the $n_q=2$ and $n_q=4$ hypotheses, is plotted against $p_T/n_q$ and compared with that of the strange hadrons $K^0_S$, $\Lambda$, $\Xi^{-}$, and $\Omega$. It is found that the $n_q=2$ ($q\bar q$) hypothesis is favored over the $n_q=4$ ($q\bar q q\bar q$ or $K\bar K$) scenarios by $7.7$, $6.3$, and $3.1$ standard deviations in the $p_T < 10$, $8$, and $6~\mathrm{GeV}/c$ ranges, respectively. The $n_q=2$ hypothesis is also preferred over the $n_q=3$ ($q \bar qg$ hybrid) interpretation by $3.5$ standard deviations in the $p_T < 8~\mathrm{GeV}/c$ range.\\
Like the $f_{0}(980)$, the $X(3872)$ is an exotic hadron candidate that is widely interpreted as a loosely bound $D^{0}\bar{D}^{*0}$ molecular state, a compact tetraquark, or a superposition of different configurations~\cite{X(3872)_into:1,X(3872)_into:2,X(3872)_into:3}. The first evidence for $X(3872)$ production in relativistic heavy-ion collisions has been reported by the CMS Collaboration~\cite{CMS:X(3872)}. Future measurements of its anisotropic flow in relativistic heavy-ion collisions could provide valuable insight into its internal structure.
 
\section{Hadronic Resonance Production in Ultra-Peripheral Collisions}\label{sec:upc_hadreso}
Ultra-peripheral collisions (UPCs) occur when two heavy nuclei interact at impact parameters ($b$) larger than the sum of their nuclear radii, thus avoiding direct hadronic overlap while enabling electromagnetic interactions through their intense Lorentz-contracted fields. These interactions facilitate photon-photon and photonuclear processes, offering a clean environment to study hadronic resonance production in the absence of the hot and dense QCD medium typically formed in central heavy-ion collisions~\cite{Bertulani:2005ru}.
\begin{figure}[h]
\centering
\includegraphics[scale=0.8]{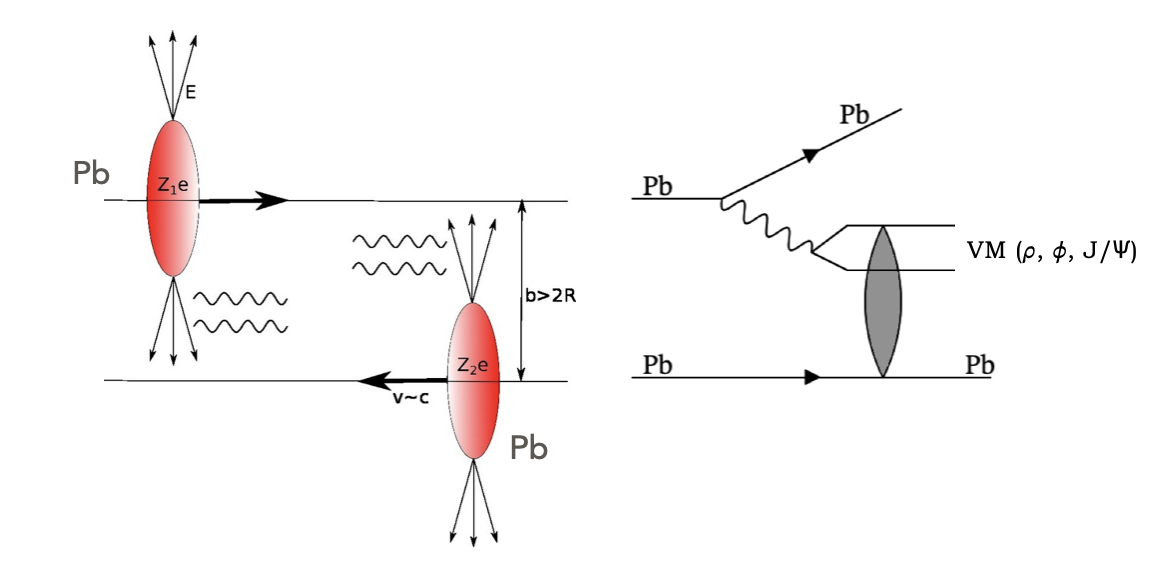}
\caption{(Color online): Schematic view of ultra-peripheral collisions (UPCs) and production of vector mesons (VMs).}
\label{upc_VM}
\end{figure} 

This section reviews experimental results on photo-produced $\rho$ mesons in UPCs and, where available, other light-flavor vector mesons. A comparative discussion with hadronic collisions highlights the influence of the medium on resonance properties.

In UPCs, the electromagnetic field of a relativistically moving nucleus is treated as a flux of quasi-real photons. These photons can interact with the opposing nucleus through photonuclear processes, producing vector mesons such as the $\rho$, $\phi$, and $J/\psi$, as illustrated in Fig.~\ref{upc_VM}. The production mechanisms can be classified as either coherent or incoherent, depending on the photon’s coupling to the entire nucleus or to individual nucleons. Coherent production dominates at low transverse momentum ($\langle p_\mathrm{T} \rangle \sim 60$ MeV/$c$), while incoherent processes produce broader $p_\mathrm{T}$ spectra ($\langle p_\mathrm{T} \rangle \sim 500$ MeV/$c$) due to nucleon-level interactions~\cite{Bertulani:2005ru}.
\begin{figure}[h]
\centering
\includegraphics[scale=0.7]{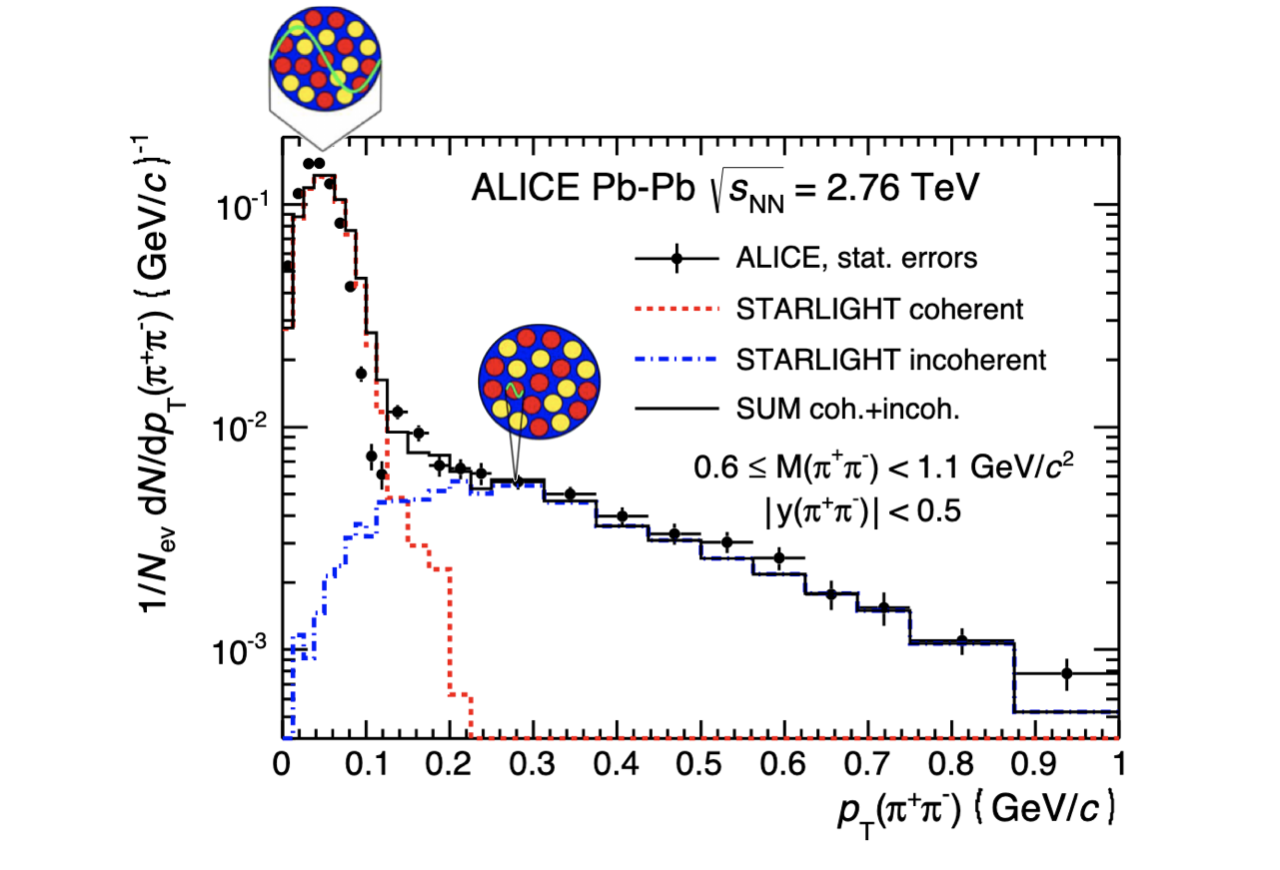}
\caption{(Color online): Transverse momentum spectra of $\rho$ mesons in UPCs, showing coherent and incoherent contributions, compared with STARLIGHT model predictions~\cite{Klein:2016yzr}.}
\label{upc_ptspectra}
\end{figure} 
Figure~\ref{upc_ptspectra} shows the $p_\mathrm{T}$ spectrum of $\rho$ mesons in UPC Pb--Pb collisions at $\sqrt{s_\mathrm{NN}} = 2.76$ TeV~\cite{ALICE:2015nbw}, compared with STARLIGHT model predictions~\cite{Klein:2016yzr}. The model accurately reproduces the spectrum, with coherent production dominating at low $p_\mathrm{T}$ and incoherent contributions becoming significant at higher $p_\mathrm{T}$, consistent with expectations.

\begin{figure}[h]
\centering
\includegraphics[scale=0.6]{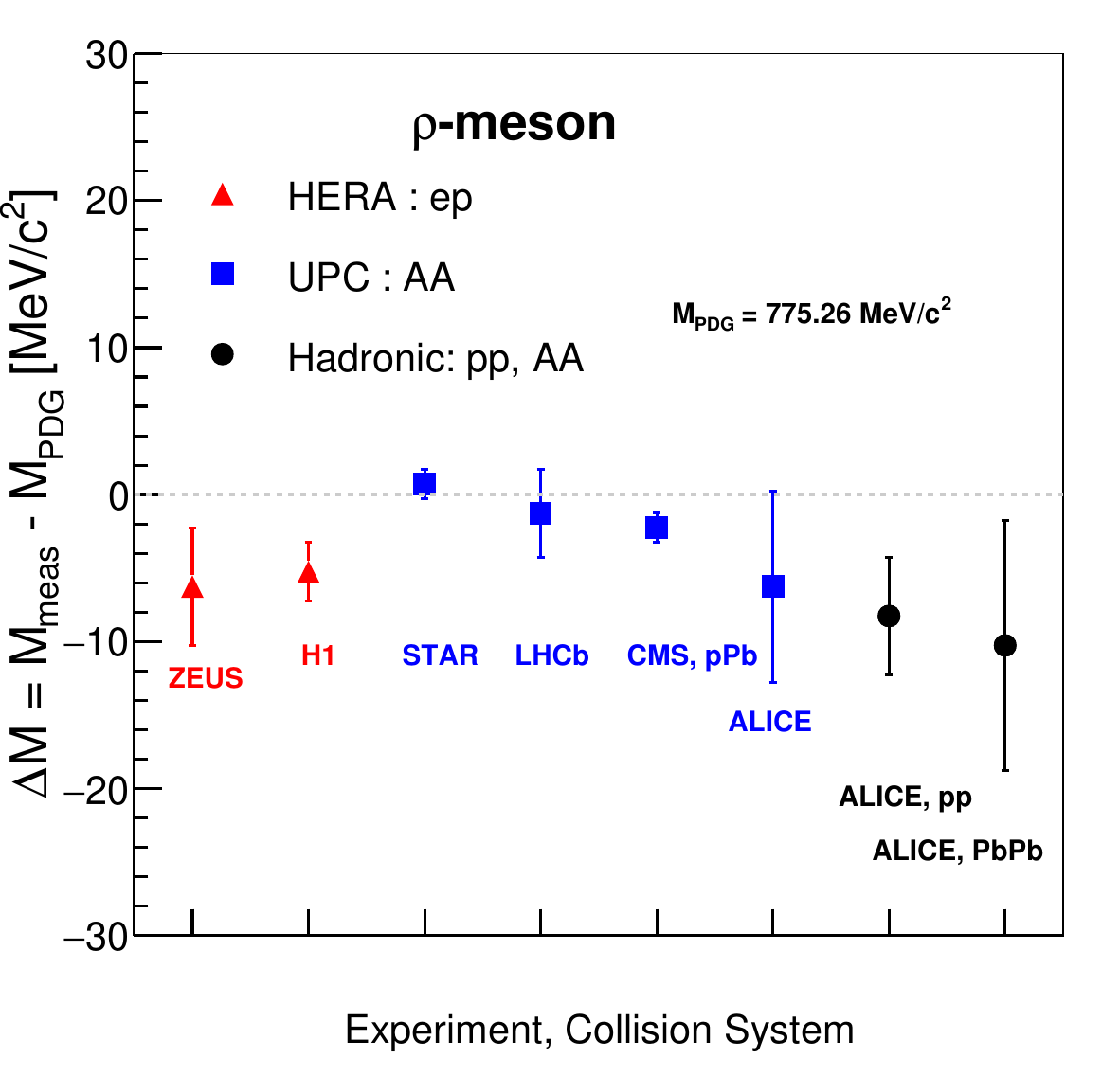}
\caption{(Color online): Comparison of the $\rho$-meson mass shift ($\Delta M$) between UPCs, electroproduction, and hadronic collisions~\cite{H1:2020lzc,H1:2009cml,ZEUS:1997rof,STAR:2007elq,ALICE:2020ugp,CMS:2019awk,LHCb:2025fzk}.}
\label{upc_rhomass}
\end{figure}

Vector mesons are valuable probes of the medium in high-energy collisions. In-medium modifications to their spectral functions—such as mass shifts and width broadening—can signal changes in hadronic interactions and partial chiral symmetry restoration~\cite{GREEN198399}. Because UPCs do not create a hot medium, they serve as a baseline to quantify such effects observed in hadronic collisions.

Among the vector mesons, the $\rho$ meson is particularly sensitive due to its short lifetime and dominant decay to $\pi^+\pi^-$. This makes it highly susceptible to in-medium modifications, including rescattering and regeneration in the hadronic phase.

\begin{figure}[h]
\centering
\includegraphics[scale=0.6]{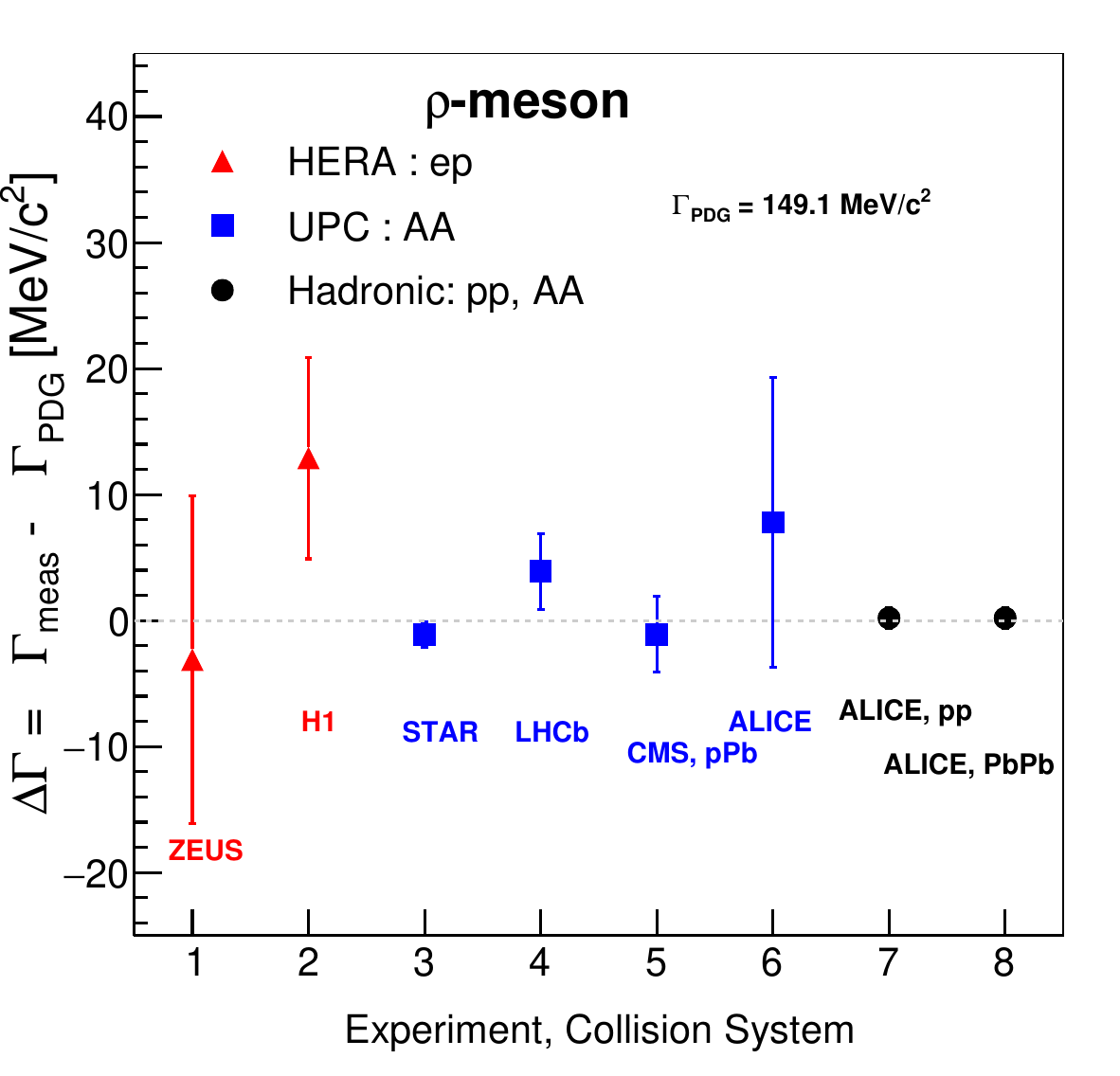}
\caption{(Color online): Comparison of the $\rho$-meson width shift ($\Delta\Gamma$) across various systems~\cite{H1:2020lzc,H1:2009cml,ZEUS:1997rof,STAR:2007elq,ALICE:2020ugp,CMS:2019awk,LHCb:2025fzk}.}
\label{upc_rhowidth}
\end{figure}

Figure~\ref{upc_rhomass} compares the $\rho$-meson mass shift, defined as $\Delta M = M_\mathrm{measurement} - M_\mathrm{PDG}$, across various systems including electroproduction (HERA), UPCs, and hadronic collisions. While UPC and electroproduction results are consistent with the vacuum mass, measurements in hadronic collisions exhibit deviations, indicative of in-medium effects. Figure~\ref{upc_rhowidth} shows similar comparisons for the width shift ($\Delta\Gamma = \Gamma_\mathrm{measurement} - \Gamma_\mathrm{PDG}$). In most cases, measured widths are consistent with PDG values within uncertainties. Notably, in hadronic collisions the width is often fixed during signal extraction, limiting sensitivity to broadening effects.

\begin{figure}[h]
\centering
\includegraphics[scale=0.6]{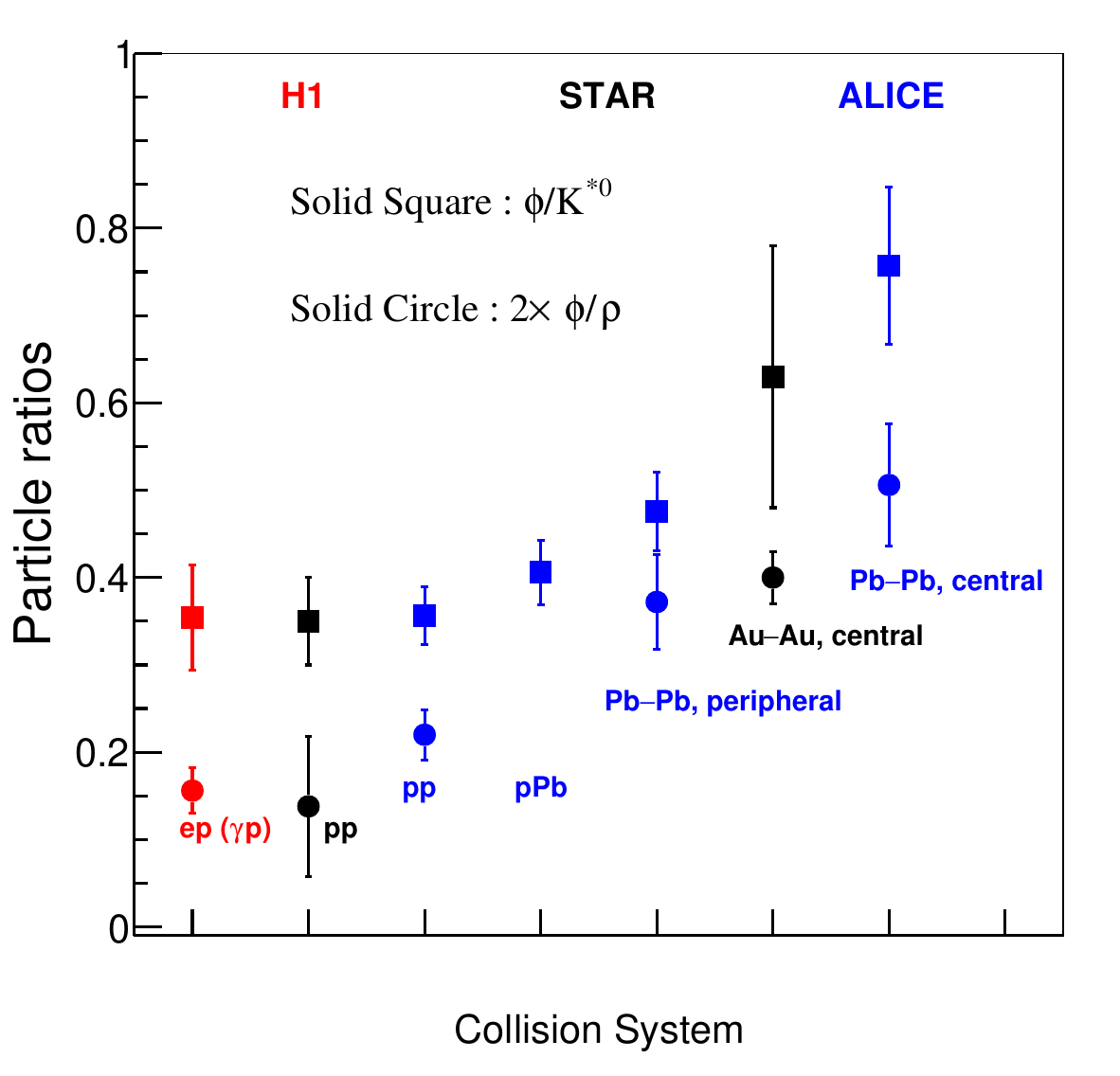}
\caption{(Color online): Particle yield ratios $\phi$/K$^{*0}$ (solid squares) and $\phi/\rho \times 2$ (solid circles) across different systems, from $ep$ (H1)~\cite{H1:2009bga}, pp (STAR, ALICE)~\cite{STAR:2004bgh,ALICE:2019etb}, p--Pb~\cite{ALICE:2021rpa}, to peripheral and central Pb--Pb collisions~\cite{ALICE:2023ifn}.}
\label{upc_ratios_mtom}
\end{figure}

In addition to spectral modifications, yield ratios of resonances to stable hadrons offer complementary insight. While such ratios are extensively studied in hadronic and small collision systems, analogous measurements in UPCs are still lacking. Figure~\ref{upc_ratios_mtom} presents the meson-to-meson yield ratios $\phi$/K$^{*0}$ and $\phi/\rho$ across several systems, including ep (HERA), pp, p--Pb, and Pb--Pb collisions.

An increasing trend in both ratios is observed with system size, particularly in central heavy-ion collisions. This enhancement is attributed to the suppression of short-lived resonances like $\rho$ and K$^{*0}$ due to re-scattering in the hadronic phase, while the longer-lived $\phi$ meson remains largely unaffected. In small systems and in $ep$ collisions, the ratios remain constant, indicating negligible final-state interactions in the absence of a dense medium.

Ultra-peripheral collisions provide more than a baseline reference for resonance production in the absence of a hadronic medium. Owing to the dominance of coherent and incoherent photonuclear interactions, UPC measurements effectively isolate vacuum spectral functions of vector mesons, free from final-state hadronic rescattering and regeneration effects. This makes UPCs uniquely sensitive to intrinsic line-shape properties, interference effects, and photoproduction amplitudes, which are otherwise difficult to disentangle in hadronic collisions.
In this context, resonance measurements across UPCs, hadronic heavy-ion collisions, and dilepton channels offer complementary and hierarchical sensitivity to medium effects. UPCs probe vacuum and initial-state photonuclear dynamics, hadronic decay channels in nucleus–nucleus collisions are sensitive to late-stage hadronic interactions and re-scattering, while dilepton measurements access in-medium spectral functions throughout the space–time evolution of the fireball with minimal final-state interactions. Together, these approaches form a coherent experimental strategy to disentangle genuine in-medium modifications from hadronic phase effects and detector-related distortions.

\noindent\textbf{Outlook:} 
\\
UPCs provide an essential reference for studying resonance production without final-state hadronic interactions. Precise measurements of spectral properties and yields of resonances—especially with upcoming LHC Run 3 data and the BES-II program at RHIC—will enable more detailed comparisons with hadronic collisions. In particular, photonuclear processes with nuclear break-up, but without medium formation, could offer intermediate benchmarks to study the onset of in-medium effects. Future measurements of identified hadron yields and correlations in UPCs will play a pivotal role in disentangling initial- and final-state effects, thereby deepening our understanding of hadronic resonance behavior in QCD matter.

\newpage

\section{Prospects for Charm Resonance Studies}\label{chap2:sub:charm}

While the reconstruction and phenomenology of light and strange hadronic resonances have been extensively studied in pp, p--A, and AA collisions, analogous studies in the charm sector remain comparatively underexplored. With recent and upcoming experimental upgrades at RHIC and the LHC, the study of short-lived charm resonances is becoming increasingly feasible. These resonances provide a natural extension of resonance-based probes to the heavy-flavor sector, opening new windows into the mechanisms of hadronization, in-medium modification, and charm transport in high-energy nuclear collisions.

Table~\ref{tab:charm_resonances} summarizes a selection of prominent charm meson and baryon resonances that are considered the counterparts of $K^{*}$, $\rho$, and $\Lambda^{*}$ in the charm sector.

\begin{table}[H]
\centering
\begin{tabular}{|l|c|c|c|c|}
\hline
\textbf{Resonance} & \textbf{Quark Content} & \textbf{Mass (MeV/$c^2$)} & \textbf{Width (MeV)} & \textbf{Dominant Decay Channel} \\
\hline
$D_0^*(2300)^0$     & $c\bar{u}$     & $\sim$2300 & $\sim$200 & $D\pi$ \\
$D_1(2420)^0$       & $c\bar{u}$     & $\sim$2420 & $\sim$25  & $D^*\pi$ \\
$D_2^*(2460)^0$     & $c\bar{u}$     & $\sim$2460 & $\sim$50  & $D\pi$, $D^*\pi$ \\
$D_s^*(2317)^+$     & $c\bar{s}$     & $\sim$2317 & $<3.8$    & $D_s^+\pi^0$ \\
$D_{s1}(2460)^+$    & $c\bar{s}$     & $\sim$2460 & $<3.5$    & $D_s^{*+}\pi^0$ \\
\hline
$\Sigma_c(2455)^{++,+,0}$ & $qqc$ ($q = u,d$) & $\sim$2455 & $\sim$2–3 & $\Lambda_c^+\pi$ \\
$\Sigma_c(2520)^{++,+,0}$ & $qqc$             & $\sim$2520 & $\sim$15  & $\Lambda_c^+\pi$ \\
$\Lambda_c(2595)^+$       & $udc$             & $\sim$2595 & $\sim$2.6 & $\Lambda_c^+\pi\pi$ \\
$\Lambda_c(2625)^+$       & $udc$             & $\sim$2625 & $<1.9$    & $\Lambda_c^+\pi\pi$ \\
$\Xi_c^*(2645)^+$         & $usc$             & $\sim$2645 & $\sim$2   & $\Xi_c\pi$ \\
\hline
\end{tabular}
\caption{Examples of charm resonances analogous to light and strange sector resonances. Properties taken from PDG~\cite{ParticleDataGroup:2022pth}.}
\label{tab:charm_resonances}
\end{table}

These states are generally broad (except a few narrow ones), with decay topologies involving ground-state $D$ mesons or charmed baryons such as $\Lambda_c^+$. Their lifetimes range from $\sim 10^{-24}$ to $10^{-23}$~s, making them experimentally accessible through invariant mass reconstruction of their decay daughters, much like their light-quark counterparts.

Reconstruction of these resonances in heavy-ion collisions is challenging due to:
\begin{itemize}
    \item Low production cross-sections of charm hadrons,
    \item Complex decay chains and possible feed-down contributions,
    \item Large combinatorial backgrounds in high-multiplicity environments.
\end{itemize}

Nonetheless, recent and future detector upgrades provide an opportunity to study these states in both proton–nucleus and nucleus–nucleus environments. Table~\ref{tab:charm_expt} summarizes the experimental prospects.

\begin{table}[H]
\centering
\begin{tabular}{|l|l|c|c|}
\hline
\textbf{Experiment} & \textbf{Access to Charm Resonances} & \textbf{Collision Systems} & \textbf{Timeline} \\
\hline
ALICE (LHC, Run 3/4) & $D_0^*$, $D_1$, $D_2^*$, $\Sigma_c$, $\Lambda_c^*$ & pp, p–Pb, Pb–Pb & 2022–2030 \\
LHCb (Run 3/4)       & Precision spectroscopy in pp; limited in p–A fixed target & pp, p–Ar & Ongoing \\
STAR (RHIC)          & $\Lambda_c^+$, possibly $\Sigma_c$, $\Lambda_c^*$ with upgrades & Au–Au & 2025 onward \\
CBM (FAIR)           & Near-threshold charm baryons; possible resonance access & Au–Au (10–40 AGeV) & $\sim$2028+ \\
MPD (NICA)           & Open charm program in later phases & Au–Au (4–11 AGeV) & Late 2020s \\
\hline
\end{tabular}
\caption{Experiments with current or future capability to access charm resonances.}
\label{tab:charm_expt}
\end{table}

The study of charm resonances is expected to shed light on:
\begin{itemize}
    \item The mechanism of charm quark hadronization (fragmentation vs coalescence),
    \item Potential in-medium modifications to charm resonance masses or widths,
    \item Freeze-out hierarchy across different hadron species,
    \item Possible restoration of chiral symmetry in the charm sector.
\end{itemize}

These measurements, though experimentally demanding, represent a critical step toward extending resonance-based probes to heavy-flavor QCD.
Charm resonances differ qualitatively from their light-flavour counterparts in that hadronic rescattering and regeneration are expected to play a significantly reduced role, owing to their smaller hadronic interaction cross sections and the dominance of charm-quark–level hadronization processes. As a consequence, late-stage hadronic effects that strongly modify light-flavour resonances are expected to be subleading for charm resonances.

\section{Resonances as probes of the chiral transition}

According to quantum chromodynamics (QCD), strongly interacting matter is expected to undergo a transition from a hadronic phase to deconfined quark matter when subjected to sufficiently high temperatures or baryon densities. In this regime, chiral symmetry, which is spontaneously broken in hadronic matter, is anticipated to be restored. High-energy nucleus–nucleus collisions offer a unique opportunity to study these phenomena under controlled laboratory conditions.

Among the available probes, the $\rho(770)$ meson is particularly sensitive to in-medium effects due to its short lifetime of about 1.3~fm/$c$. Changes in the mass and width of the $\rho$ meson have long been proposed as possible signals of the chiral transition~\cite{rho_chiral:th1,rho_chiral:th2,rho_chiral:th3}. While most theoretical approaches predict a substantial broadening of the $\rho$ spectral function near the transition, the behavior of its mass remains unsettled, with models suggesting a decrease, little change, or even an increase.

\begin{figure}
\begin{center}
\includegraphics[scale=0.4]{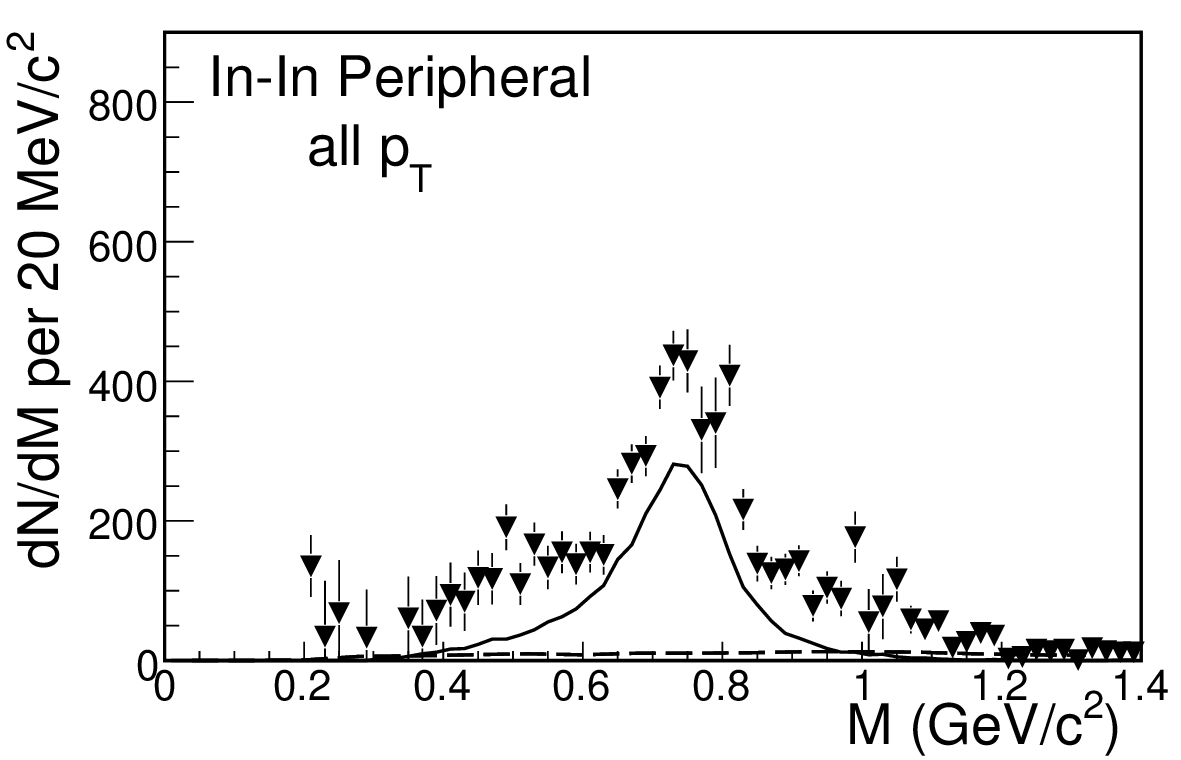}
\includegraphics[scale=0.4]{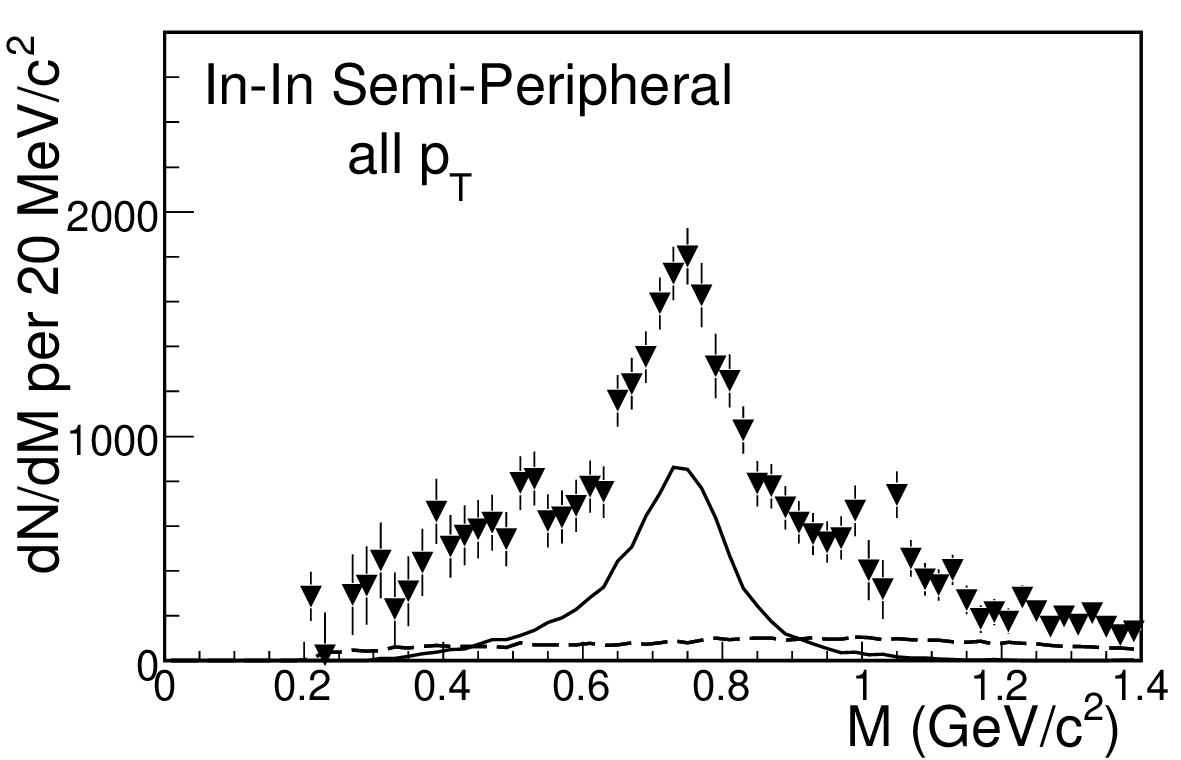}
\includegraphics[scale=0.4]{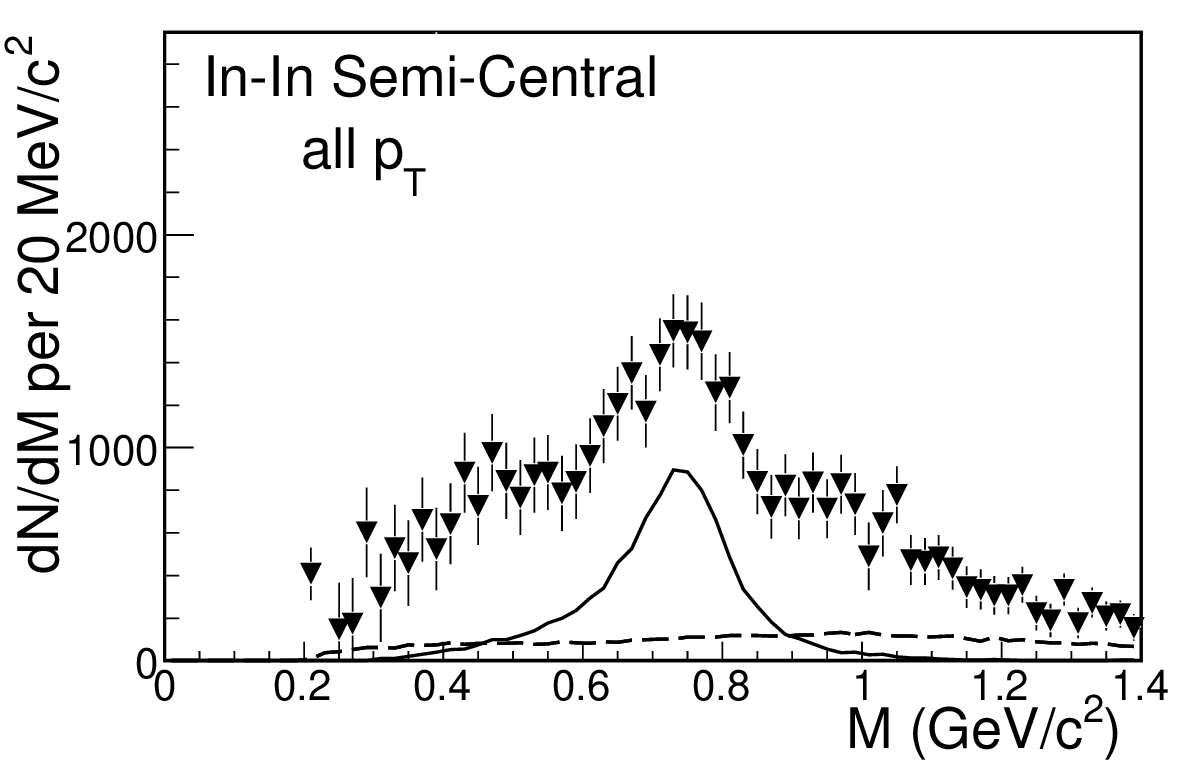}
\includegraphics[scale=0.4]{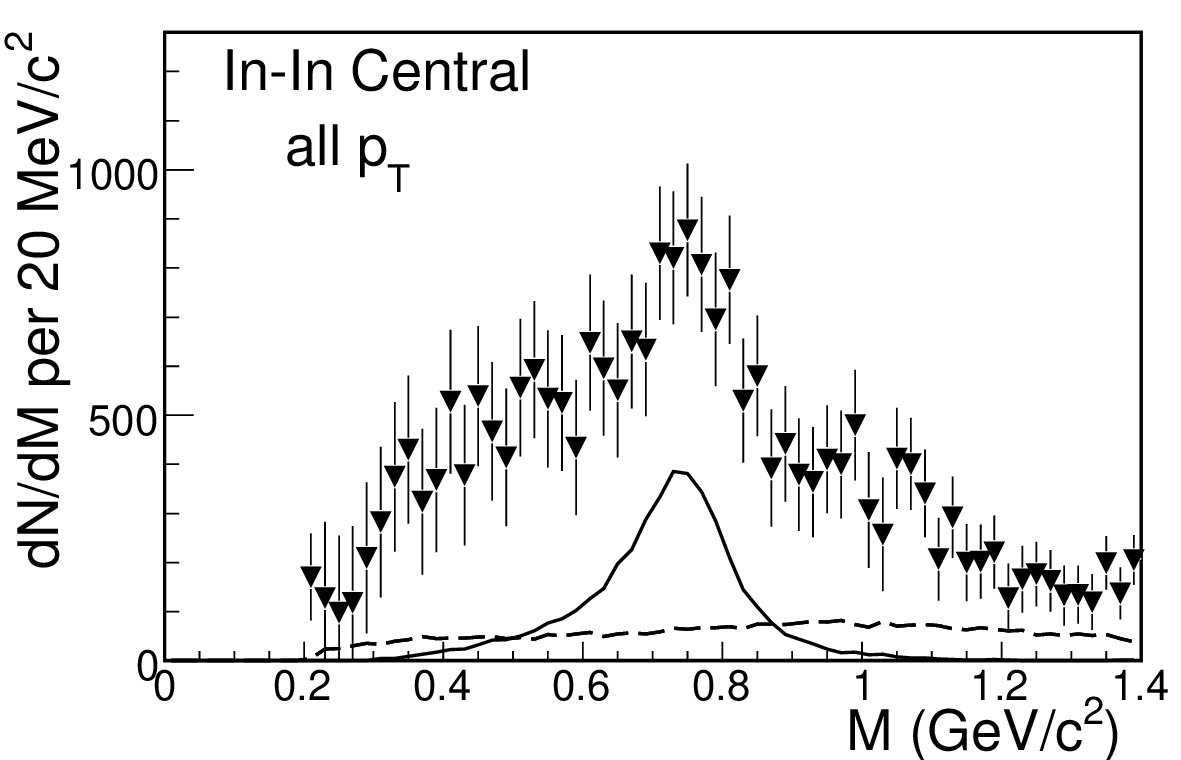}
\vspace*{-0.3cm}
   \caption{Excess mass spectra of dimuons. The cocktail $\rho$
   (solid) and the level of uncorrelated charm decays
   (dashed) are shown for comparison. The errors are purely
   statistical. Figure is taken from ref.~\cite{rho_chiral:data}}
   \label{fig:chiral}
\end{center}
\vspace*{-0.6cm}
\end{figure}

The NA60 experiment at the CERN SPS has performed a high-precision study of in-medium modifications of the $\rho(770)$ meson using low-mass dimuon spectra measured in 158~$A$GeV In–In collisions~\cite{rho_chiral:data}. These measurements reveal a pronounced excess of dileptons over the yield expected from electromagnetic decays of neutral mesons. The excess is generally attributed to thermal radiation from the hot and dense fireball, dominated by pion annihilation, $\pi^+ \pi^- \to \rho \to \ell^+ \ell^-$, involving an intermediate $\rho$ meson that is strongly modified by the medium. 
The excess mass spectra for  four centrality bins are shown in Fig.~\ref{fig:chiral}. The spectra exhibit striking qualitative features: in all cases a pronounced peak is observed that broadens significantly with increasing centrality, but without a significant shift from the position of the nominal $\rho$ mass. Further studies, including precise $p_T$-dependent analyses, are required to consolidate these findings. Current hadronic resonance measurements provide only indirect and model-dependent constraints on chiral symmetry restoration.\\
A crucial missing element in this topic has been the lack of direct experimental constraints on the vacuum $\rho$--nucleon interaction, which enters theoretical descriptions through the in-medium $\rho$ self-energy. This gap has recently been addressed by the ALICE Collaboration~\cite{ALICE:rho2025}, which reported the first measurement of the $\rho^{0}$--proton interaction using femtoscopic two-particle correlations in high-multiplicity $pp$ collisions. By quantitatively constraining the strength and range of the $\rho$--baryon interaction in vacuum, this result could provide a key baseline for interpreting in-medium spectral modifications and for linking the NA60 observations to microscopic mechanisms of hadronic broadening and chiral symmetry restoration.

\section{Summary and outlook}\label{sec:sum}

Over the past two decades, measurements of light-flavour hadronic resonances have become an indispensable part of the experimental programme aimed at characterizing the space–time evolution of strongly interacting matter created in high-energy hadronic and heavy-ion collisions. This review has summarized the state-of-the-art reconstruction techniques, measurements of masses, widths, yields, mean transverse momenta, and flow coefficients, as well as the nuclear modification factors obtained across pp, p--A, d--A, A--A and ultra-peripheral collisions at SPS, RHIC and the LHC. Together, these observables provide a detailed and internally consistent picture of how resonances probe the hadronic phase, freeze-out dynamics, and the collective behaviour of the medium.

A global pattern emerges from the systematic comparison of resonance yields and ratios. Short-lived resonances, notably $\rho(770)$, $K^{*}(892)$, and $\Delta(1232)$, exhibit pronounced suppression in central heavy-ion collisions relative to expectations from statistical hadronization, most prominently at low transverse momentum. This suppression, absent in small systems, reflects the strong influence of hadronic rescattering on their decay daughters, while the behaviour of longer-lived resonances such as $\phi(1020)$ serves as a control measurement insensitive to late-stage interactions. The smooth evolution of resonance-to-stable particle ratios across systems of varying size and multiplicity highlights the continuity of final-state hadronic effects and reinforces the interpretation that the charged-particle multiplicity is the dominant scaling variable governing resonance survival. Complementary studies of masses and widths reveal that, after accounting for detector effects, no large in-medium modifications have yet been observed at RHIC or LHC energies, although small downward shifts for certain species persist at low $p_T$.

Resonance observables also offer quantitative access to the timescales of the hadronic stage. Lower bounds on the hadronic lifetime, extracted from the multiplicity dependence of $K^*/K$ and related ratios, indicate durations of 4--7 fm/$c$ in the most central Pb–Pb collisions, comparable to the lifetime of short-lived resonances themselves. Comparisons with model calculations that incorporate partial chemical equilibrium or hadronic transport show that regeneration and re-scattering must both be accounted for to describe the data. Consistency between the hadronic lifetimes inferred from resonances and kinetic freeze-out temperatures obtained from HRG-PCE fits provides a coherent chronology of hadronic evolution across energies and system sizes. Nuclear modification factors for resonances at intermediate and high $p_T$ extend these insights to earlier times, where parton energy loss, shadowing and medium response influence resonance production, complementing $R_{AA}$ measurements for stable hadrons and jets. Complementary information is provided by rapidity-dependent measurements in asymmetric collision systems, where resonance nuclear modification exhibits a clear forward--backward asymmetry, reflecting the interplay of longitudinal geometry and nuclear effects beyond those accessible at midrapidity.

Beyond yields and spectral modifications, resonances also provide essential information on collective dynamics. Measurements of anisotropic flow for the $\phi$ meson, including directed ($v_1$), elliptic ($v_2$), and triangular ($v_3$) flow, demonstrate that the $\phi$ participates in the collective expansion of the medium, consistent with its expected early decoupling and small hadronic interaction cross section. In central Pb–Pb collisions, the $v_2$ of the $\phi$ follows the mass-ordering pattern characteristic of hydrodynamic behaviour, supporting the interpretation that partonic collectivity dominates the early stages of the evolution. In contrast, systematic comparisons of $\langle p_T\rangle$ and flow observables across p--p, p--Pb, and Pb--Pb collisions reveal that this mass ordering is violated in small systems, where the $\phi$ and $K^*$ exhibit harder spectra and higher $\langle p_T\rangle$ than protons despite having comparable masses. These deviations point to a different interplay of geometry, hadronization mechanisms, and limited final-state interactions in small systems, and highlight the ability of resonance measurements to discriminate between genuine collectivity and non-hydrodynamic spectral hardening.

Spin-alignment measurements of vector mesons, particularly the $K^{*0}$ and $\phi$, have opened a new experimental window on spin degrees of freedom in high-energy collisions. In p--p collisions, the spin-density matrix element $\rho_{00}$ remains consistent with the unpolarized value of 1/3 across the measured $p_t$ range, providing a clean baseline. In heavy-ion collisions, however, clear system- and energy-dependent structures emerge. At LHC energies, both $K^{*0}$ and $\phi$ exhibit $\rho_{00}$ < 1/3 at low $p_t$, approaching 1/3 at higher $p_t$, with a stronger effect for $K^{*0}$. In contrast, BES measurements at RHIC show the opposite trend for the $\phi$ meson, where $\rho_{00}$ > 1/3 is observed and the deviation increases at lower collision energies. This striking energy dependence places new constraints on theoretical interpretations involving vorticity, magnetic fields, quark polarization, coherent $\phi$-field formation, and spin–orbit interactions during hadronization. Although a unified framework is still lacking, the emerging patterns across $\sqrt{s_{NN}}$, $p_t$, and species provide essential input to developing a systematic understanding of spin transport in QCD matter. Ultra-peripheral collisions, in turn, supply near-vacuum baselines for resonance production and simultaneously probe the nuclear gluon density at low Bjorken-x.

Looking ahead, several open issues highlighted throughout this review point to clear directions for future work. Measurements of multiple resonances with different lifetimes and interaction strengths remain essential for disentangling re-scattering and regeneration effects, as current transport and hybrid models do not simultaneously describe all observed species and system-size trends. The extraction of the hadronic-phase duration from resonance suppression still carries model-dependent uncertainties, motivating more precise measurements across collision energies and centralities. The behaviour of resonances in small systems, where suppression patterns and $\langle p_T\rangle$ systematics differ markedly from those in Pb--Pb collisions, also requires further study to clarify the role of final-state interactions at low multiplicity. In addition, the non-trivial $p_T$ and energy dependences observed in vector-meson spin-alignment measurements, particularly the contrasting trends between LHC and RHIC BES energies—remain incompletely understood, underscoring the need for higher-precision data and improved theoretical frameworks. Ultra-peripheral collisions will continue to provide important baselines for line-shape and photo-production studies, enabling a cleaner separation between vacuum and in-medium effects.

At higher energies and luminosities, upcoming UPC measurements at the LHC Run 3 and Run 4, as well as future EIC-like facilities for photon–nucleus interactions, will allow systematic studies of vector mesons over a broad kinematic range and provide improved constraints on the nuclear gluon distribution. Finally, the emergence of charm-resonance reconstruction, enabled by upgraded tracking, PID, and high-rate DAQ systems at RHIC, LHC, FAIR and NICA, opens a new domain in which in-medium modifications and hadronization dynamics can be probed for charmed excitations. These measurements have the potential to transform our understanding of heavy-flavour interaction in both partonic and hadronic phases.

Taken together, the measurements reviewed here demonstrate that resonance production has become a central component of the experimental effort to characterize the properties and evolution of the medium created in high-energy collisions. Through systematic studies of yields, spectral shapes, suppression patterns, flow observables, and spin alignment across a wide range of collision energies and system sizes, resonances provide complementary information on rescattering and regeneration, the duration of the hadronic phase, and the emergence of collective behavior. Ultra-peripheral collisions further supply a clean baseline for comparing vacuum and in-medium production. As higher-precision data and additional resonance species become accessible at RHIC, the LHC, and upcoming facilities, these measurements will continue to refine our understanding of the hadronic stage and the mechanisms that govern the production and evolution of strongly interacting matter.
A brief summary of established experimental observations on resonance production and the corresponding open physics questions is given in Table~\ref{tab:resonance_summary_longtable}.

\renewcommand{\arraystretch}{1.32}

\begin{longtable}{p{0.47\linewidth} p{0.47\linewidth}}
\caption{Summary of established experimental observations on resonance production and the corresponding open physics questions.}
\label{tab:resonance_summary_longtable} \\
\hline
\textbf{Established experimental observations} &
\textbf{Open physics questions} \\
\hline
\endfirsthead

\multicolumn{2}{c}%
{{\tablename\ \thetable{} -- continued from previous page}} \\
\hline
\textbf{Established Experimental Observations} &
\textbf{Open Physics Questions} \\
\hline
\endhead

\hline
\multicolumn{2}{r}{{Continued on next page}} \\
\endfoot

\hline
\endlastfoot

\textbf{(1) Rescattering systematics} \\
Measurements at RHIC and the LHC show that short-lived resonances such as 
K$^{*}(892)$, $\rho(770)$, and $\Lambda(1520)$ exhibit decreasing yields 
from pp to p--Pb to A--A collisions and with increasing centrality, while the 
$\phi(1020)$ yield remains approximately independent of centrality. 
These trends are consistent with hadronic rescattering being the dominant effect shaping the reconstructable yields of short-lived resonances.
&
From yield suppression alone, it is not possible to isolate the relative roles of rescattering, regeneration, and any possible in-medium spectral-function changes of resonances. 
Direct measurements of in-medium resonance spectral shapes at collider energies are not yet available, and extracting hadronic-phase lifetimes requires dynamical modeling including relevant scattering rates, cross sections, regeneration channels, and the space--time evolution of the hadron gas. \\

\textbf{(2) Modeling and cross-section uncertainties} \\
Statistical-hadronization calculations with partial chemical equilibration and hadronic transport models qualitatively reproduce the observed centrality dependence of the K$^{*}$/K and $\Lambda(1520)$/\,$\Lambda$ yield ratios.
&
The quantitative interpretation of these ratios remains limited because key hadronic scattering cross sections ($\pi K$, $\pi N$, $K\bar K$, $\pi\Lambda$) in the relevant temperature and momentum ranges of the hadronic phase are insufficiently constrained by existing experimental data. \\

\textbf{(3) Beam energy dependence} \\
STAR Beam Energy Scan results show that the K$^{*}$/K ratio decreases with decreasing $\sqrt{s_{NN}}$, extending the suppression trend into the finite-$\mu_{B}$ region.
&
The evolution of resonance production with baryochemical potential, and its connection to changes in hadronic lifetime or to the QCD phase structure, is not yet established. \\

\textbf{(4) Multiplicity and event-shape dependence} \\
Across pp, p--Pb, and Pb--Pb collisions, resonance yields and $\langle p_{T}\rangle$ increase with charged-particle multiplicity, and event-shape selections show similar trends, in parallel with long-lived hadrons.
&
The relative contributions of collective flow, non-flow correlations, and hadronic-phase effects to the observed multiplicity and event-shape dependence of resonance observables have not been determined experimentally. \\

\textbf{(5) High-$p_{T}$ suppression} \\
At $p_{T} \gtrsim 8$--10 GeV/$c$, the nuclear modification factors ($R_{\mathrm AA}$) of K$^{*}(892)$ and $\phi(1020)$ agree with those of inclusive charged hadrons within uncertainties, indicating that high-$p_{T}$ suppression is governed by partonic energy loss rather than hadronic-phase effects.
&
Whether quenched parton fragmentation produces resonances and long-lived hadrons with identical probabilities remains unquantified, because resonance-resolved fragmentation and jet-correlation measurements are not yet available with sufficient precision. \\

\textbf{(6) Rapidity asymmetry in small systems} \\
In $p$--A and $d$--A collisions, measurements of $R_{\mathrm{xA}}$ for several resonance species indicate values close to unity at midrapidity over a broad $p_{\mathrm{T}}$ range, while in more asymmetric systems a clear rapidity
asymmetry of $\phi$ mesons is observed at forward and backward rapidity, characterized by suppression in the light-ion--going direction and enhancement in the heavy-ion--going direction.
&
At present, the most comprehensive system- and rapidity-differential measurements in small collision systems are available for the $\phi$ meson, and extending such studies to additional resonance species would be beneficial for disentangling possible initial- and final-state contributions and for achieving a more complete understanding of nuclear effects in small
systems. \\

\textbf{(7) Flow measurements} \\
Measurements of the anisotropic flow of the $\phi(1020)$ meson established the fact that there is mass ordering at low $p_{T}$ and baryon--meson grouping at intermediate $p_{T}$, and the small hadronic cross section of the $\phi$ implies that its flow is developed predominantly in the partonic phase.
&
For short-lived resonances such as K$^{*}(892)$ and $\Lambda(1520)$, which undergo significant hadronic re-scattering, it is not known how much of their measured flow reflects genuine partonic collectivity and how much is modified by hadronic interactions or reconstruction effects. Quantitative separation of these contributions remains open. \\

\textbf{(8) Spin alignment} \\
Measurements of the spin-density--matrix element $\rho_{00}$ for K$^{*}(892)$ and $\phi(1020)$ show deviations from the unpolarized value $\rho_{00}=1/3$, with differing magnitudes between RHIC and LHC energies and between the two meson species.
&
The origin of these patterns—including their contrasting energy dependence, species differences, and their apparent inconsistency with the trends of $\Lambda$ hyperon global polarization—has not been established. \\

\textbf{(9) Ultra-peripheral collisions (UPCs)} \\
In ultra-peripheral Pb--Pb collisions, coherent photoproduction of vector mesons such as $\rho^{0}$ and $\phi(1020)$ has been measured with high precision, providing vacuum-like baselines and constraints on nuclear gluon distributions at low $x$.
&
The quantitative roles of nuclear shadowing, gluon saturation, and coherent--incoherent interference in shaping UPC vector-meson photoproduction remain insufficiently constrained, as does the connection between the measured cross sections and the spatial distribution of gluons in nuclei. \\

\end{longtable}

\newpage
\section*{Acknowledgements}
The authors acknowledge support from the Department of Atomic Energy (DAE) and the Department of Science and Technology (DST), Government of India. B.M. is particularly grateful to the Anusandhan National Research Foundation (ANRF) for support through the J. C. Bose Fellowship. SS and AKS acknowledge support from the Chinese Academy of Sciences. KN acknowledges support from OSHEC, Department of Higher Education, Government of Odisha, Index No. 23EM/PH/124 under MRIP 2023. We thank the STAR Collaboration at RHIC and the ALICE Collaboration at the LHC for numerous stimulating discussions. Several of the authors are members of these collaborations and have directly contributed to a subset of the measurements reviewed in this article. We also thank Dr. Ziwei Lin for valuable discussions related to transport-model descriptions of resonance production.



\nocite{*}
\bibliography{bibliography_updated}



\newpage
\appendix
\renewcommand*{\thesection}{\Alph{section}}


\end{document}